\documentclass[aps,
 prb,
 twocolumn,
 longbibliography,
 preprintnumbers,
 amssymb,
 floatfix]{revtex4-1}

\usepackage{amsfonts}%
\usepackage{amssymb, amsmath}%
\usepackage{graphicx}
\usepackage{graphicx, , setspace, epstopdf, fancyhdr, mathrsfs, multirow}
\usepackage{subfigure}
\usepackage{dcolumn}
\usepackage{enumerate}
\usepackage{cleveref}
\usepackage{natbib}
\usepackage{xcolor}
\usepackage{soul}
\usepackage{stmaryrd}

\renewcommand{\d}{\text{d}}
\newcommand{\p}{\partial}

\newcommand{\aref}[1]{App.~\ref{#1}}
\newcommand{\eref}[1]{Eq.~\eqref{#1}}
\newcommand{\esref}[1]{Eqs.~\eqref{#1}}
\newcommand{\Eref}[1]{Equation~\eqref{#1}}
\newcommand{\Esref}[1]{Equations~\eqref{#1}}
\newcommand{\fref}[1]{Fig.~\ref{#1}}
\newcommand{\fsref}[1]{Figs.~\ref{#1}}

\newcommand{\sref}[1]{Sec.~\ref{#1}}
\newcommand{\ssref}[1]{Secs.~\ref{#1}}

\newcommand{\Rref}[1]{Ref.~\citenum{#1}}

\begin{document}

\title{Point dipole and quadrupole
 scattering approximation to collectively responding resonator systems}
\author{Derek W. Watson}
\author{Stewart D. Jenkins}
\author{Janne Ruostekoski}
\affiliation{Mathematical Sciences and Centre for Photonic
	Metamaterials, University of Southampton,
	Southampton SO17 1BJ, United Kingdom}
	
 \begin{abstract}
 We develop a theoretical formalism for
 collectively responding point scatterers
 where the radiating electromagnetic fields from each emitter are
 considered in the electric dipole, magnetic dipole,
 and electric quadrupole approximation.
 The contributions of the electric quadrupole moment
 to electromagnetically-mediated interactions
 between the scatterers are derived in detail
 for a system where each scatterer represents a linear
 $RLC$ circuit resonator, representing common metamaterial
 resonators in radiofrequency, microwave, and optical regimes.
 The resulting theory includes a closed set of
 equations for an ensemble of discrete
 resonators that are radiatively coupled to each other by
 propagating electromagnetic fields,
 incorporating potentially strong interactions
 and recurrent scattering processes.
 The effective model
 is illustrated and tested for examples
 of pairs of interacting point electric dipoles,
 where each pair can be qualitatively replaced by a model point
 emitter with different multipole radiation moments.
 \end{abstract}
 \date{\today}
 \maketitle

 %
 %
 \section{Introduction}
 \label{sec:Intro}

 Metamaterials are artificial media which, through design,
 exhibit functions not observed in natural materials. The constituent
 components of the metamaterial are
 resonators that are typically much smaller
 than the wavelength of the
 electromagnetic (EM) field. Each unit
 cell in a metamaterial array is formed
 by a metamolecule whose internal
 structure may then consist, e.g.,
 of a nontrivial configuration of circuit resonators.
 Metamolecules are closely spaced
 and they can also interact strongly by EM-field mediated coupling.
 The strong interactions result
 from a multiple scattering effect, whereby a resonators'
 charge and current oscillations, driven by the incident field and
 those EM fields emitted by other resonators, produce EM fields which,
 in turn, drive the charge and current oscillations of other resonators.
 The functionalities of the metamaterial then depend on these interactions.

 In principle, when the reaction of a resonator to an
 EM field is known, Maxwell's equations may be solved
 numerically for an ensemble of resonators, taking into
 account the constituent structure and geometry of each resonator.
 In practice, however, this is
 computationally demanding for more
 than a few single elements~\cite{5565504},
 or would require simplifications, such as adapting
 the discrete translational symmetry of an
 infinite lattice~\cite{KoschnyEtAlPRL2004,
 PendryEtAlIEEE1999,
 BelovSimovskiPRE2005,
 LiuEtAlPRE2007}.
 An alternative approach is to provide
 an effective model for the  individual circuit
 elements  as point scatterers interacting with
 the incident and scattered EM fields.
 A general formalism for such
 an approach was developed in \Rref{JenkinsLongPRB},
 where each metamolecule was assumed to
 comprise a set of pointlike
 circuit elements whose radiative properties were described by
 the lowest order electric and magnetic
 multipoles. The model was designed to capture
 the physics of each resonator, e.g.,
 its resonance frequency and radiative
 emission rate, that are relevant for
 collective radiative coupling between large numbers of metamolecules,
 without the need for a detailed model
 of the intrinsic structure of each circuit resonator.
 The approach then results in a coupled set
 of equations for the dynamics of the
 resonators and EM fields. The model is not
 limited to circuit resonators but can also be
 utilized for a variety of point scatterer~\cite{devries98}
 and nanoparticle
 systems~\cite{optical_antennas,
 Fromm,
 Evlyukhin_nanop,
 wang_nanoshells}.

 In \Rref{JenkinsLongPRB,
 JenkinsLineWidthNJP,
 CAIT,
 Jenkinssub} the general formalism
 of \Rref{JenkinsLongPRB} was applied
 to derive effective dipole point scatterer
 approaches to model split-ring
 resonators;~\cite{SmithEtAlPRL2000} each
 arc of the resonator was described by
 a point emitter that possess both
 electric and magnetic dipoles. The
 resulting metamolecule of two such arcs
 exhibits a strong electric or magnetic dipole
 excitation, but notably weaker electric
 quadrupole excitation~\cite{FedotovEtAlPRL2007}.
 The model was sufficient to qualitatively
 describe the strong collective effects
 of a planar metamaterial array and
 its correlated subradiant
 excitations~\cite{Jenkinssub}.  In
 related systems, it has also been
 used in electron-beam excitation
 studies of a metamaterial array~\cite{AdamoEtAlPRL2012},
 and the model can also incorporate
 additional features, such as inhomogeneous
 broadening~\cite{JenkinsRuostekoskiPRB2012b}.

 Many metamolecules cannot accurately
 be modeled by electric and magnetic
 dipoles alone, and treating each
 constituent separately may be
 computationally impractical. For instance,
 there is considerable interest in the studies
 of strong intrametamolecular couplings
 in the context of Fano transmission
 resonances or subradiance in the systems
 that are formed by combinations of several
 resonators~\cite{LiuEtAlNatMat2009,
 Lovera,
 FanCapasso,
 GiessenOligomers,
 Frimmer,
 Dregely,
 Watson2016}. Experiments on
 collective responses of large numbers
 of resonators in planar metamaterial
 arrays are also becoming common~\cite{FedotovEtAlPRL2010,
 papasimakis2009,
 SentenacPRL2008,
 LemoultPRL10,
 AdamoEtAlPRL2012,
 Anlageprx,
 dielectricmeta,
 Jenkinssub}.

 Here we extend the analysis of \Rref{JenkinsLongPRB}
 to point emitter descriptions that includes
 electric quadrupole radiation. The
 formalism then provides an effective
 model of a resonator system comprising single
 point scatterers radiating electric dipole,
 quadrupole, and magnetic dipole EM fields.
 The EM interactions between point emitters
 possessing also electric quadrupole
 moments  lead to mathematically more
 complicated expressions. We consider
 an ensemble of such effective emitters
 and develop a compact model for their
 interactions. The approach is illustrated
 and tested by simple examples of point
 electric dipoles. We compare the responses
 of two pairs of point electric dipoles to an
 effective description where each pair is
 replaced by a single point emitter with
 an electric dipole or a magnetic
 dipole and electric quadrupole moment.

 In \sref{sec:MADyn}, we review
 the theoretical model utilized to describe
 the interactions between resonators,
 and the point electric and
 magnetic dipole approximation of
 their interactions.
 In \sref{sec:E2}, we introduce
 the point emitter description with
an electric quadrupole moment and
 describe the scattered EM fields and the interactions with other emitters.
 In \sref{sec:point_multipole_resonators},
 we analyze in detail the point multipole interactions of
 different resonator systems, providing specific analytical examples.
 Some concluding remarks are
 made in \sref{sec:conclusions}.

 %
 %
 \section{Basic formalism for discrete resonator model}
 \label{sec:MADyn}

 Here, we introduce the basic formalism used to
 analyze the interaction of an EM field
 to closely spaced resonators. The formalism
 is derived in detail in \Rref{JenkinsLongPRB}.
 Here, we review how the scattered EM
 fields are obtained from general polarization
 and magnetization sources, before providing
 an overview of the point electric and
 magnetic dipole approximation of the scattered EM fields
 and their interactions with the
 resonators in \sref{subsec:E1M1_general}.
 
 \subsection{Radiated fields}

 In the general model of circuit resonator interactions with
 EM fields, we assume that the
 charge and current sources are initially driven
 by an incident electric displacement field
 ${\bf D}_{\text{in}}({\bf r},t)$, and magnetic
 induction ${\bf B}_{\text{in}}({\bf r},t)$
 with frequency $\Omega_0$.
 The electric ${\bf E}_{\text{sc},j}({\bf r},t)$
 and magnetic  ${\bf H}_{\text{sc},j}({\bf r},t)$
 fields scattered by resonator $j$,
 are a result of its oscillating polarization ${\bf P}_j({\bf r},t)$
 and magnetization ${\bf M}_j({\bf r},t)$ sources.
 In general, the electric and
 magnetic fields are related to the electric
 displacement and magnetic induction
 through the auxiliary equations
 \begin{align}
 {\bf D}({\bf r},t) &
 =
 \epsilon_0{\bf E}({\bf r},t) + {\bf P}({\bf r},t)
 \,\text{,}
 \label{eq:D_auxiliary}
 \\
 {\bf H}({\bf r},t) &
 =
 \frac{1}{\mu_0}{\bf B}({\bf r},t) - {\bf M}({\bf r},t)
 \,\text{.}
 \label{eq:H_auxiliary}
 \end{align}
 When analyzing the EM fields and resonators,
 we adopt the rotating wave approximation
 where the dynamics is dominated by $\Omega_0$.
 In the rest of the paper, all the EM field and resonator amplitudes refer
 to the slowly-varying versions of the positive frequency components of the
 corresponding variables, where the rapid oscillations $e^{-i\Omega_0t}$
 due to the dominant laser frequency has been factored out.
 The scattered EM fields are then given by~\cite{JenkinsLongPRB}
 \begin{align}
 {\bf E}_{\text{sc},j}({\bf r}) &
 =
 \frac{k^3}{4\pi\epsilon_0}\int\d^3r'\,
 \bigg[{\bf G}({\bf r} -{\bf r}')\cdotp{\bf P}_j({\bf r}',t)\nonumber
 \\
 &
 \quad
 +\frac{1}{c}{\bf G}_\times({\bf r} - {\bf r}')\cdotp{\bf M}_j({\bf r}',t)\bigg]
 \,\text{,}\label{eq:Esc}
 \\
 {\bf H}_{\text{sc},j}({\bf r}) &
 =
 \frac{k^3}{4\pi}\int\d^3r'\,
 \bigg[{\bf G}({\bf r} -{\bf r}')\cdotp{\bf M}_j({\bf r}',t)\nonumber
 \\
 &
 \quad
 -c{\bf G}_\times({\bf r} - {\bf r}')\cdotp{\bf P}_j({\bf r}',t)\bigg]
 \,\text{,}
 \label{eq:Bsc}
 \end{align}
 where $k=\Omega/c$.  Explicit expressions for the radiation kernels
 are~\cite{JenkinsLongPRB}
 \begin{align}
 {\bf G}({\bf r}) &=
 i \bigg[\frac{2}{3}{\bf I}h_0^{(1)}(kr) +
 \left(\frac{\bf r r}{r^2}- \frac{\bf I}{3}\right)h_2^{(1)}(kr)\bigg]
 \nonumber\\
 &
 \quad\quad
 - \frac{4\pi}{3}{\bf I}\delta(k{\bf r})
 \,\text{,}
 \label{eq:G}
 \\
 {\bf G}_\times({\bf r}) &= \frac{i}{k}\nabla\times\frac{e^{ikr}}{kr}{\bf I}
 \,\text{.}
 \label{eq:G_x}
 \end{align}
 Here: the dyadic ${\bf rr}$, is the outer product of
 ${\bf r}$ with itself; ${\bf I}$ is the identity matrix; and
 $h^{(1)}_n(x)$ are spherical Hankel functions
 of the first  kind, of order $n$, defined by
 \begin{align}
 h^{(1)}_0(x) &
 =
 - i \frac{e^{ ix}}{x}
 \,\text{,}
 \label{eq:h_0}\\
 h^{(1)}_2(x) &
 =
 i
 \left[\frac{1}{x} + i \frac{3}{x^2} - \frac{3}{x^3}\right]e^{ ix}
 \,\text{.}
 \label{eq:h_2}
 \end{align}
 The radiation kernel ${\bf G}({\bf r} - {\bf r}')$ determines
 the electric (magnetic) field at ${\bf r}$, from polarization (magnetization)
 sources at ${\bf r}'$~\cite{Jackson}.
 Similarly, the cross kernel ${\bf G}_\times({\bf r} - {\bf r}')$, determines
 the electric (magnetic) field at ${\bf r}$, from
 magnetization (polarization) sources at
 ${\bf r}'$~\cite{Jackson}.

 \Esref{eq:Esc} and~\eqref{eq:Bsc} give the total scattered
 EM fields as functions of the polarization and magnetization densities.
 In general, for sources other than point resonators, the scattered field
 equations are not readily solved for
 ${\bf P}_j({\bf r},t)$ and ${\bf M}_j({\bf r},t)$.
 When  resonators are separated by distances less than, or of the order of a
 wavelength, a strongly coupled system results.
 
  \subsection{Interacting resonators}

 In \Rref{JenkinsLongPRB}, a general theory was formulated
 to derive a coupled set of linear equations for the EM fields and strongly
 coupled resonators.
 The state of current oscillation in each resonator $j$ is described by a
 single dynamic variable with units of charge $Q_j(t)$
 and its rate of change $I_j(t)$, the current.
 The current oscillations within the $j$th resonator behave
 like an LC circuit with resonance frequency
 $\omega_j$,
 \begin{equation}
 \omega_j = \frac{1}{\sqrt{L_jC_j}}
 \,\text{,}
 \label{eq:omega}
 \end{equation}
 where $C_j$ and $L_j$ are an effective self-capacitance
 and self-inductance, respectively.
 The polarization and magnetization of a resonator can be obtained
 from $Q_j(t)$ and $I_j(t)$~\cite{JenkinsLongPRB}
 \begin{align}
 {\bf P}_j({\bf r},t) &
 =
 Q_j(t){\bf p}_j({\bf r})
 \,\text{,}
 \label{eq:P}
 \\
 {\bf M}_j({\bf r},t) &
 =
 I_j(t){\bf w}_j({\bf r})
 \,\text{.}
 \label{eq:M}
 \end{align}
 The charge profile function ${\bf p}_j({\bf r})$ and the  current
 profile function ${\bf w}_j({\bf r})$,
 in \esref{eq:P} and~\eqref{eq:M}, may be considered
 independent of time. The geometry of individual resonators
 determines the form of the respective profile functions.
 The polarization and magnetization densities
 are related to the charge and current
 densities of the resonators by~\cite{JenkinsLongPRB}
 \begin{align}
 \rho_j({\bf r},t) &
 =
 -\nabla\cdotp{\bf P}_j({\bf r},t)
 \,\text{,}
 \label{eq:rho}
 \\
 {\bf J}_j({\bf r},t) &
 =
 \frac{\p}{\p t}\big[{\bf P}_j({\bf r},t)\big]
 +
 \nabla\times {\bf M}_j({\bf r},t)
 \,\text{.}
 \label{eq:J}
 \end{align}
 The charge and current densities within each
 resonator are initially driven by the
 incident EM fields ${\bf D}_{\text{in}}({\bf r},t)$
 and  ${\bf B}_{\text{in}}({\bf r},t)$.
 The incident electric displacement and magnetic flux, with
 polarization vector ${\bf \hat e}_{\text{in}}$, are:
 \begin{align}
 {\bf D}_{\text{in}}({\bf r})  &
 =
 D_\text{in}{\bf \hat e}_{\text{in}}e^{i{\bf k}_\text{in}\cdotp{\bf r}}
 \,\text{,}
 \label{eq:Din}
 \\
 {\bf B}_{\text{in}}({\bf r})&
 =
 B_\text{in}
 \big[{\bf \hat k}_{\text{in}}\times
 {\bf \hat e}_{\text{in}}\big]e^{i{\bf k}_\text{in}\cdotp{\bf r}}
 \,\text{,}
 \label{eq:Bin}
 \end{align}
 where ${\bf \hat k}_{\text{in}}$ is the propagation
 vector of the incident EM field.
 The total EM fields external to resonator $j$ comprise
 the incident field and those fields scattered
 from all other resonators,
 \begin{align}
 {\bf E}_{\text{ext},j}({\bf r},t)& =
 \frac{1}{\epsilon_0}{\bf D}_{\text{in}}({\bf r},t)
 + \sum_{i \ne j}{\bf E}_{\text{sc},i}({\bf r},t)
 \,\text{,}
 \label{eq:Eext}
 \\
 {\bf H}_{\text{ext},j}({\bf r},t)& =
 \frac{1}{\mu_0}{\bf B}_{\text{in}}({\bf r},t) +
 \sum_{i \ne j}{\bf H}_{\text{sc},i}({\bf r},t)
 \,\text{.}
 \label{eq:Bext}
 \end{align}
 The total driving of the charge and current oscillations
 within the resonator  is  provided  by the
 external EM fields, \esref{eq:Eext} and~\eqref{eq:Bext},
 aligned along the direction of the
 source, providing a net electromagnetic
 force (emf)~\cite{JenkinsLongPRB},
 $\mathcal{E}_{\text{ext},j}$ and
 flux~\cite{JenkinsLongPRB}, $\Phi_{\text{ext},j}$.
 We define the external emf and flux as~\cite{JenkinsLongPRB}
 \begin{align}
 \mathcal{E}_{\text{ext},j}& =
 \frac{1}{\sqrt{\omega_jL_j}}\int\,\d^3r\,{\bf p}_j({\bf r})
 \cdotp{\bf E}_{\text{ext},j}({\bf r})
 \,\text{,}
 \label{eq:emf}
 \\
 \Phi_{\text{ext},j} &=
 \frac{\mu_0}{\sqrt{\omega_jL_j}}\int\d^3r\,{\bf w}_j({\bf r})
 \cdotp{\bf H}_{\text{ext},j}({\bf r})
 \,\text{.}
 \label{eq:flux}
 \end{align}
 The emf and flux can be decomposed into contributions
 from the incident and scattered
 EM fields,
 \begin{eqnarray}
 \mathcal{E}_{\text{ext},j} & = &\mathcal{E}_{\text{in},j}
 +
 \sum_{i \ne j}\mathcal{E}_{i,j}^{\text{sc}}
 \,\text{,}
 \label{eq:emf_insc}
 \\
 \Phi_{\text{ext},j} & = &\Phi_{\text{in},j}
 +
 \sum_{i \ne j}\Phi_{i,j}^{\text{sc}}
 \,\text{.}
 \label{eq:flux_insc}
 \end{eqnarray}
 The emf and flux resulting from the driving by
 the incident EM field is $\mathcal{E}_{\text{in},j}$
 and $\Phi_{\text{in},j}$, respectively. The driving of resonator $j$
 by the scattered EM fields from resonator $i$ are the emf
 $\mathcal{E}_{i,j}^{\text{sc}}$ and flux $\Phi_{i,j}^{\text{sc}}$.
 The total driving of a resonator can be summarized by
 the external driving $F_{\text{ext},j}$, the sum
 of the incident  $F_{\text{in},j}$
 and scattered $F_{\text{sc},j}$ driving contributions, respectively,
 where~\cite{JenkinsLongPRB}
 \begin{equation}
 F_{\text{ext},j}
 =
 F_{\text{in},j} + F_{\text{sc},j}
 = F_{\text{in},j} +  \sum_{i \ne j} \mathcal{C}_{ij}
 \,\text{,}
 \label{eq:F_ext}
 \end{equation}
 where the components
 \begin{align}
 F_{\text{in},j}
 &
 =
 \frac{i}{\sqrt{2}}
 \big(\mathcal{E}_{\text{in},j} + i\omega_j\Phi_{\text{in},j}\big)
 \,\text{,}
 \label{eq:fin}
 \\
 \big[\mathcal{C}\big]_{i \ne j}
 &
 =
 \frac{i}{\sqrt{2}}
 \big(\mathcal{E}_{i,j}^{\text{sc}}
 +
 i\omega_{j}\Phi_{i,j}^{\text{sc}}\big)
 \,\text{.}
 \label{eq:C_offdiagonal}
 \end{align}
 
  \subsection{Normal modes}

 In order to express the coupled equations for the
 EM fields and resonators we introduce
 the slowly varying normal mode oscillator
 amplitudes~\cite{JenkinsLongPRB}
 $b_j(t)$,
 \begin{equation}
 b_j(t) =
 \frac{1}{\sqrt{2\omega_j}}
 \left[\frac{Q_j(t)}{\sqrt{C_j}} + i \frac{\phi_j(t)}{\sqrt{L_j}}\right]
 \,\text{.}
 \label{eq:b}
 \end{equation}
 Here, the generalized coordinate for the
 current excitation in the resonator $j$ is
 the charge $Q_j(t)$ and $\phi_j(t)$ represents
 its conjugate momentum.
 In the rotating wave approximation
 the conjugate momentum  is
 linearly proportional to the
 current~\cite{JenkinsLongPRB}.
  The dynamic variable in \eref{eq:b} can be
 used to describe a general resonator
 with both polarization and magnetization sources.
 
 The normal mode amplitudes $b_j(t)$ describe the 
 current oscillations of the resonator. These current 
 oscillations are subject to radiative damping 
 due to their own emitted radiation. The driving of 
 $b_j(t)$ is achieved through the external 
 fields and resulting emf and flux. The
 equations of motion for 
 $Q$ and $\phi$, $\dot{Q}= I$ and 
 $\dot{\phi} = \mathcal{E}$, together with the
 scattered EM fields from each resonator and those 
 scattered fields from other resonators result in 
 a linear system of equations for $b_j(t)$.
 For a system which comprises $N$ resonators, 
these read as~\cite{JenkinsLongPRB}
 \begin{equation}
 {\bf \dot b} = \mathcal{C}{\bf b} + {\bf F}_{\text{in}}
 \,\text{,}
 \label{eq:equmot}
 \end{equation}
 where ${\bf b}$ is a column vector of $N$
 normal oscillator amplitudes
 \begin{equation}
 {\bf b}
 =
 \begin{bmatrix}
 b_1
 \\
 b_2
 \\
 \vdots
 \\
 b_N
 \end{bmatrix}
 \,\text{,}
 \label{eq:b_vector}
 \end{equation}
 ${\bf F}_{\text{in}}$ is a column
 vector formed by \eref{eq:fin}.
 The matrix $\mathcal{C}$ describes the interactions between the
 resonator's self-generated EM fields
 (diagonal elements) and those scattered
 from different resonators [off-diagonal
 elements; the interaction terms in
 \eref{eq:C_offdiagonal}]~\cite{JenkinsLongPRB}.

 As we will see (\ssref{subsec:E1M1_general} and~\ref{sec:E2})
 the solutions to \eref{eq:C_offdiagonal} become increasingly
 complicated  as the complexity of
 the resonators increases.
 The diagonal elements of $\mathcal{C}$
 contain the resonance frequency shift  and the total decay rate
 $\Gamma_j$~\cite{JenkinsLongPRB},
 \begin{equation}
 \big[\mathcal{C}\big]_{j,j} =
 -i(\omega_j - \Omega_0) - \frac{\Gamma_j}{2}
 \,\text{.}
 \end{equation}
 The total decay rate $\Gamma$ results
 from the radiative emission rate and
 ohmic losses.
 Although generally the emitters can
 have different resonance frequencies,
 here, for simplicity, we focus on the
 case of equal frequencies, i.e., $\omega_j = \omega_0$, for all $j$.
 
  \subsection{Collective eigenmodes}

 Strong multiple scattering results in
 collective excitation modes of the system.
 The collective modes of current oscillation within the system are described by
 the eigenvectors ${\bf v}_n$ of the interaction
 matrix $\mathcal{C}$. The corresponding eigenvalues
 $\xi_n$ have real and imaginary parts corresponding to the decay rate and
 resonance frequency shift of the mode,
 \begin{equation}
 \xi_n = - \frac{\gamma_n}{2} - i(\Omega_n - \Omega_0)
 \,\text{.}
 \label{eq:xi_n}
 \end{equation}
 The number of  resonators $N$,
 determines the number of collective modes.
  The collective eigenmodes can then exhibit different
 resonance frequencies and linewidths and line
 shifts~\cite{JenkinsLongPRB,
 Jenkinssub}.
 The different modes may have superradiant or
 subradiant characteristics. The former occurs when the emitted radiation is
 enhanced by the interactions of the resonators
 ($\gamma_n>\Gamma$). The latter occurs when the
 radiation is suppressed and confined to the metamaterial
 ($\gamma_n<\Gamma$).

 \section{Point electric and magnetic dipole approximation}
 \label{subsec:E1M1_general}

The general model of interacting
 resonators summarized above and
 introduced formally in \Rref{JenkinsLongPRB}
 is  applicable to any type of circuit element resonators.
 In practice, however, some approximations
 to the intrinsic structure of the resonators
 is required. When the size of the resonator
 is much less than the wavelength, the
 resonators' scattered EM fields are often
 approximated as those of point multipole sources.
 For split ring resonators the scattered
 fields are dominated by electric and
 magnetic dipole radiation. This motivated
 the formal theory of  the point electric
 and magnetic dipole approximation,
 introduced in \Rref{JenkinsLongPRB}, which
 we first review here. Later, in \sref{sec:E2},
 we extend the theory by deriving the point
 electric quadrupole approximation in the
 same formalism that can be used to model
 also more general resonator and emitter systems.
 
  \subsection{Radiating point dipoles}

 The  electric  ${\bf E}_{\text{sc},j}({\bf r})$
 and magnetic  ${\bf H}_{\text{sc},j}({\bf r})$
 fields scattered from the $j$th resonator located
 at ${\bf r}'$ due to its polarization and magnetization
 sources  follow from \esref{eq:Esc} and~\eqref{eq:Bsc}
 with the polarization
 density \eref{eq:P} and magnetization density \eref{eq:M}.
In the electric and magnetic dipole approximation, the
 mode functions,
 ${\bf p}_{j}({\bf r}) = {\bf p}^{\text{d}}_{j}({\bf r})$
 and ${\bf w}_{j}({\bf r})$, respectively,
 are defined as~\cite{JenkinsLongPRB}
 \begin{align}
 {\bf p}^{\text{d}}_{j}({\bf r})
 &
 = H_j{\bf \hat d}_j\delta({\bf r} - {\bf r}_j)
 \,\text{,}
 \label{eq:p_E1}
 \\
 {\bf w}_j({\bf r})
 &
 = A_{\text{M},j}{\bf \hat m}_j\delta({\bf r} - {\bf r}_j)
 \,\text{.}
 \label{eq:w_M1}
 \end{align}
 Here, the proportionality constant $H_j$ has units of
 length and the unit vector ${\bf \hat d}_j$ indicates the
 orientation of the electric dipole, whilst $A_{\text{M},j}$ has units of
 area and ${\bf \hat m}_j$ indicates the orientation of the magnetic
 dipole.
 The interaction of the resonator
 with its self-generated EM fields
 causes radiative damping to occur.
 The radiation rates of
 the electric and magnetic dipoles
 of the $j$th resonator  are
 $\Gamma_{\text{E1},j}$~\cite{JenkinsLongPRB}
 and
 $\Gamma_{\text{M1},j}$~\cite{JenkinsLongPRB},
 respectively, where
 \begin{align}
 \Gamma_{\text{E1},j} &
  =
 \frac{C_jH_j^2\omega_j^4}{6\pi\epsilon_0c^3}
 \,\text{,}
 \label{eq:GammaE1}
 \\
 \Gamma_{\text{M1},j} &
  =
 \frac{\mu_0A_{\text{M},j}^2\omega_j^4}{6\pi L_jc^3}
 \,\text{.}
 \label{eq:GammaM1_temp}
 \end{align}
 We account for nonradiative losses by adding the phenomenological decay
 rate $\Gamma_{\text{O},j}$. For simple gold or silver
 resonators, $\Gamma_{\text{O}}$ can be estimated
 by applying the Drude model of permittivity with specific
 material paramaters to the scattered cross section 
 of the resonator, see e.g.,~\Rref{Watson2016}.
 The total
 decay rate is then the sum of the radiative
 emission rate and ohmic losses.
 In the dipole approximation the total
 decay rate $\Gamma_{j}$ is~\cite{JenkinsLongPRB}
 \begin{equation}
 \Gamma_{j} = \Gamma_{\text{E1},j}
 +
 \Gamma_{\text{M1},j}
 +
 \Gamma_{\text{O},j}
 \,\text{.}
 \label{eq:Gamma_total_dip}
 \end{equation}
 The amplitudes of the EM fields scattered by the electric and
 magnetic dipoles are proportional to their corresponding
 radiative emission rates $\Gamma_{\text{E1},j}$
 and $\Gamma_{\text{M1},j}$. We write the EM fields
 due to the point electric and magnetic dipoles sources
 as~\cite{JenkinsLongPRB}
 \begin{align}
 {\bf E}_{\text{sc},j}({\bf r}) &=
 b_j\frac{3}{2}\sqrt{\frac{k^3}{12\pi\epsilon_0}}
 \bigg[
 \sqrt{\Gamma_{\text{E1},j}}
 {\bf G}({\bf r} - {\bf r}_j)\cdotp
 {\bf \hat d}_j
 \nonumber
 \\
 &\quad \quad
 - i\sqrt{\Gamma_{\text{M1},j}}
 {\bf G}_\times({\bf r} - {\bf r}_j)\cdotp{\bf \hat m}_j
 \bigg]
 \text{,}
 \label{eq:Esc_dip_scaled}
 \\
 {\bf H}_{\text{sc},j}({\bf r}) &=
 - ib_j\frac{3}{2}\sqrt{\frac{k^3}{12\pi\mu_0}}
 \bigg[
 \sqrt{\Gamma_{\text{M1},j}}
 {\bf G}({\bf r} - {\bf r}_j)\cdotp
 {\bf \hat m}_j
 \nonumber
 \\
 &\quad \quad
 -i \sqrt{\Gamma_{\text{E1},j}}
 {\bf G}_\times({\bf r} - {\bf r}_j)\cdotp{\bf \hat d}_j
 \bigg]
 \text{.}
 \label{eq:Hsc_dip_scaled}
 \end{align}
 
  \subsection{Interacting point dipoles}
 
 The   incident EM field, \esref{eq:Din} and~\eqref{eq:Bin},
 driving the charge oscillations within a
 resonator resulting in the emf~\cite{JenkinsLongPRB}
 $\mathcal{E}_{\text{in},j}^{\text{E1}}$
 and flux~\cite{JenkinsLongPRB}
 $\Phi_{\text{in},j}^{\text{M1}}$,
 follow from \esref{eq:emf}
 and~\eqref{eq:flux}:
 \begin{align}
 \mathcal{E}_{\text{in},j}^{\text{E1}} & =
 \frac{1}{\epsilon_0\sqrt{\omega_jL_j}}
 \int\d^3r\,{\bf p}^{\text{d}}_{j}({\bf r})
 \cdotp{\bf D}_{\text{in}}({\bf r},t)
 \,\text{,}
 \label{eq:emf_in_E1}
 \\
 \Phi_{\text{in},j}^{\text{M1}}
 & =
 \frac{1}{\sqrt{\omega_jL_j}}
 \int\d^3r\,
 {\bf  w}_j({\bf r})
 \cdotp
 {\bf B}_{\text{in}}({\bf r},t)
 \,\text{.}
 \label{eq:Phi_in_M1}
 \end{align}
 The scattered electric field from the $j$th resonator driving the
 polarization source oscillations within
 resonator $i\ne j$, result in  the emf~\cite{JenkinsLongPRB}
 $\mathcal{E}_{i,j}^{\text{sc,E1}}$
 \begin{equation}
 \mathcal{E}_{i,j}^{\text{sc,E1}}=
 \sqrt{\Gamma_{\text{E1},i}\Gamma_{\text{E1},j}}
 \left[\mathcal{G}_{\text{E1}}\right]_{i,j}
 \frac{b_{j}}{\sqrt{2}}
 \,\text{.}
 \label{eq:emf_E1}
 \end{equation}
 The matrix $\mathcal{G}_{\text{E1}}$ determines
 how the geometrical
 properties and orientations of the
 resonators influence the scattered electric field
 contributions to the emf. The diagonal
 elements of $\mathcal{G}_{\text{E1}}$
 are zero, the off-diagonal elements, with
 point electric dipole sources, are
 \begin{equation}
 \big[\mathcal{G}_{\text{E1}}\big]_{i,j}=
 \frac{3}{2}
 {\bf \hat d}_{i}
 \cdotp
 {\bf G}({\bf r}_i-{\bf r}_{j})
 \cdotp
 {\bf \hat d}_{j}
 \,\text{.}
 \label{eq:G_E1}
 \end{equation}
 In a similar manner, the scattered magnetic field from the $j$th
 resonator driving the magnetization source oscillations within
 resonator $i\ne j$, results in the flux~\cite{JenkinsLongPRB}
 $\Phi_{i,j}^{\text{sc,M1}}$, where
 \begin{equation}
 \Phi_{i,j}^{\text{sc,M1}}
  =
 \frac{i}{\omega_j}
 \sqrt{\Gamma_{\text{M1},i}\Gamma_{\text{M1},j}}
 \left[
 \mathcal{G}_{\text{M1}}
 \right]_{i,j}
 \frac{b_{j}}{\sqrt{2}}
 \,\text{.}
 \label{eq:Phi_sc_M1}
 \end{equation}
 The matrix $\mathcal{G}_{\text{M1}}$ is the magnetic counterpart
 of \eref{eq:G_E1}.
 The  diagonal elements  of $\mathcal{G}_{\text{M1}}$ are zero,
 the off-diagonal elements, for point magnetic dipole sources, are
 \begin{equation}
 \left[\mathcal{G}_{\text{M1}}\right]_{i,j}
 =
 \frac{3}{2}
 {\bf \hat m}_{i}
 \cdotp
 {\bf G}({\bf r}_i - {\bf r}_j)
 \cdotp
 {\bf \hat m}_j
 \,\text{.}
 \label{eq:G_M1}
 \end{equation}
 The driving of the polarization (magnetization) sources
 within the $j$th resonator by the magnetic (electric) field
 scattered by resonator $i$ result in additional
 contributions to the emf and flux.
 We call this type of driving ``cross driving". In the
 dipole approximation the cross driving contributions to the
 emf and flux are~\cite{JenkinsLongPRB},
 $\mathcal{E}_{i,j}^{\text{sc,X1}}$ and
 $\Phi_{i,j}^{\text{sc,X1}}$, respectively, where
 \begin{align}
 \mathcal{E}_{i,j}^{\text{sc,X1}} &
 =
 -i\sqrt{\Gamma_{\text{E1},i}\Gamma_{\text{M1},j}}
 \left[
 \mathcal{G}_{\text{X1}}
 \right]_{i,j}
 \frac{b_{j}}{\sqrt{2}}
 \text{,}
 \label{eq:emf_X1}
 \\
 \Phi_{i,j}^{\text{sc,X1}} &
 =
 -\frac{1}{\omega_j}
 \sqrt{\Gamma_{\text{M1},i}\Gamma_{\text{E1},j}}
 \left[
 \mathcal{G}_{\text{X1}}
 \right]_{i,j}^T
 \frac{b_{j}}{\sqrt{2}}
 \text{.}
 \label{eq:flux_X1}
 \end{align}
 The matrix,
 $\mathcal{G}_{\text{X1}} $,
 and its transpose,
 $\mathcal{G}_{\text{X1}}^T$, are the
 cross driving counterparts of \esref{eq:G_E1}
 and~\eqref{eq:G_M1}, the off diagonal elements are
 \begin{equation}
 \left[\mathcal{G}_{\text{X1}}\right]_{i,j}
 =
 \frac{3}{2}
 {\bf \hat m}_{i}\cdotp
 {\bf G}_\times({\bf r}_i - {\bf r}_j)
 \cdotp
 {\bf \hat d}_j
 \, \text{.}
 \label{eq:G_X1}
 \end{equation}
 In the point  dipole approximation,
 the interactions between the
 resonators depend exclusively upon the
 orientation and relative positions of the
 point sources. The coupling matrix $\mathcal{C}$ is
 \begin{equation}
 \mathcal{C} = \Delta - \frac{1}{2}\Upsilon +
 \frac{1}{2}\Big[
 i\mathcal{C}_{\text{E1}} + i\mathcal{C}_{\text{M1}}
 +
 \mathcal{C}_{\text{X1}} + \mathcal{C}_{\text{X1}}^T
 \Big]
 \,\text{,}
 \label{eq:C_E1M1}
 \end{equation}
 where
 \begin{subequations}
 \begin{align}
 \mathcal{C}_{\text{E1}} &
 =
 \Upsilon_{\text{E1}}^{1/2}
 \mathcal{G}_{\text{E1}}
 \Upsilon_{\text{E1}}^{1/2}
 \,\text{,}
 \label{eq:C_E1}
 \\
 \mathcal{C}_{\text{M1}} &
 =
 \Upsilon_{\text{M1}}^{1/2}
 \mathcal{G}_{\text{M1}}
 \Upsilon_{\text{M1}}^{1/2}
 \,\text{,}
 \label{eq:C_M1}
 \\
 \mathcal{C}_{\text{X1}} &
 =
 \Upsilon_{\text{M1}}^{1/2}
 \mathcal{G}_{\text{X1}}
 \Upsilon_{\text{E1}}^{1/2}
 \,\text{.}
 \label{eq:C_X1}
 \end{align}
 \label{eq:C_E1M1X1}
 \end{subequations}
 The diagonal elements of $\mathcal{C}$ contain the
 detuning of the incident EM field from the resonator's
 resonance frequency $\omega_j$ and
 the resonator's total decay rate $\Gamma_j$. The detuning is
 described by the diagonal matrix $\Delta$,
 where~\cite{JenkinsLongPRB}
 \begin{equation}
 \big[\Delta\big]_{j,j} \equiv -i(\omega_j- \Omega_0)\,\text{,}
 \label{eq:Delta}
 \end{equation}
 and the decay rate by the diagonal matrix $\Upsilon$, with
 \begin{equation}
 \big[\Upsilon\big]_{j,j} = \Gamma_{\text{E1},j}
 +
 \Gamma_{\text{M1},j}
 +
 \Gamma_{\text{O},j}
 \,\text{.}
 \label{eq:Upsilon_total_E1}
 \end{equation}
 The radiative decay rates of each resonator are contained in the diagonal
 matrices $\Upsilon_{\text{E1}}$ and $\Upsilon_{\text{M1}}$, where,
 \begin{align}
 \big[\Upsilon_{\text{E1}}\big]_{j,j} &= \Gamma_{\text{E1},j}
 \,\text{,}
 \label{eq:UpsilonE1}
 \\
 \big[\Upsilon_{\text{M1}}\big]_{j,j} &= \Gamma_{\text{M1},j}
 \,\text{.}
 \label{eq:UpsilonM1}
 \end{align}

 %
 \section{Point electric quadrupole approximation}
 \label{sec:E2}

 In \sref{sec:MADyn}, we reviewed the general model for
 interacting resonators and the point electric and magnetic
 dipole approximation of the scattered EM fields.
 In this section,
 we extend the point dipole approximation
 from \sref{subsec:E1M1_general}, to include the point electric quadrupole
 contribution to the scattered EM field and its interaction
 with the other  multipole sources.

 The previously derived interaction matrix
 \eref{eq:C_E1M1} between resonators
 that exhibit point electric and magnetic dipoles is generalized
 for the case of point electric quadrupoles
 \begin{align}
 \mathcal{C} =&  \Delta - \frac{1}{2}\Upsilon
 + \frac{1}{2}\Big[
 i\mathcal{C}_{\text{E1}}
 + i\mathcal{C}_{\text{M1}}
 + \mathcal{C}_{\text{X1}}
 + \mathcal{C}_{\text{X1}}^T
 \nonumber
 \\
 &
 + i\mathcal{C}_{\text{E2}}
 + i\mathcal{C}_{\text{X2e}}
 + i\mathcal{C}_{\text{X2e}}^T
 + \mathcal{C}_{\text{X2m}}
 + \mathcal{C}_{\text{X2m}}^T
 \Big]
 \,\text{.}
 \label{eq:C_E1M1E2}
 \end{align}
 The diagonal elements of \eref{eq:C_E1M1E2}
 contain the detuning $\Delta$ [see \eref{eq:Delta}],
 and the total decay rate $\Upsilon$ for the resonator
 \begin{equation}
 \big[\Upsilon\big]_{j,j} = \Gamma_{\text{E1},j}
 + \Gamma_{\text{M1},j}
 + \Gamma_{\text{E2},j}
 + \Gamma_{\text{O},j}
 \,\text{.}
 \label{eq:Gamma_M1E2_total}
 \end{equation}
 Here, $\Gamma_{\text{E2}}$ denotes the electric quadrupole
 radiative emission rate. The electric and magnetic dipole emission
 rates are $\Gamma_{\text{E1}}$ and $\Gamma_{\text{M1}}$,
 respectively, see \esref{eq:GammaE1} and~\eqref{eq:GammaM1_temp}.
 The  matrices for electric and magnetic dipole interactions,
 $\mathcal{C}_{\text{E1}}$,
 $\mathcal{C}_{\text{M1}}$,
 and $\mathcal{C}_{\text{X1}}$
 are given in \eref{eq:C_E1M1X1}. The
 additional interaction terms are similarly defined,
 \begin{subequations}
 \begin{align}
 \mathcal{C}_{\text{E2}} &
 =
 \Upsilon_{\text{E2}}^{1/2}
 \mathcal{G}_{\text{E2}}
 \Upsilon_{\text{E2}}^{1/2}
 \, \text{,}
 \label{eq:C_E2}
 \\
 \mathcal{C}_{\text{X2e}} &
 =
 \Upsilon_{\text{E1}}^{1/2}\mathcal{G}_{\text{X2e}}
 \Upsilon_{\text{E2}}^{1/2}
 \,\text{,}
 \label{eq:C_X2e}
 \\
 \mathcal{C}_{\text{X2m}} &
 =
 \Upsilon_{\text{M1}}^{1/2}\mathcal{G}_{\text{X2m}}
 \Upsilon_{\text{E2}}^{1/2}
 \,\text{,}
 \label{eq:C_X2m}
 \end{align}
 \label{eq:C_E2X2eX2m}
 \end{subequations}
 and describe:
 electric quadrupole--electric quadrupole;
 electric quadrupole--electric  dipole;
 and
 electric quadrupole--magnetic dipole interactions, respectively.
 The transpose matrices,
 $\mathcal{C}_\text{X2e}^T$ and
 $\mathcal{C}_{\text{X2m}}^T$ are,
 respectively, the
 electric dipole--electric quadrupole and
 magnetic dipole--electric quadrupole interactions. 
 Explicit expressions for: $\mathcal{G}_{\text{E2}}$;
 $\mathcal{G}_{\text{X2e}}$; and
 $\mathcal{G}_{\text{X2m}}$ are given in 
 \esref{eq:G_E2}; \eqref{eq:G_X2e}; and~\eqref{eq:G_X2m},
 respectively, and are derived in this section.

 The electric dipole and magnetic dipole radiative emission rates
 are contained in the diagonal matrices $\Upsilon_{\text{E1}}$
 and $\Upsilon_{\text{M1}}$, respectively, see \esref{eq:UpsilonE1}
 and~\eqref{eq:UpsilonM1}. The electric quadrupole radiative emission
 rate is contained in the equivalent diagonal matrix $\Upsilon_{\text{E2}}$,
 where
 \begin{equation}
 \big[\Upsilon_{\text{E2}}\big]_{j,j}
 = \Gamma_{\text{E2},j}
 \,\text{.}
 \label{eq:UpsilonE2}
 \end{equation}
 In this section, we also derive an explicit expression
 for $\Gamma_{\text{E2}}$.

 \subsection{Interacting point electric quadrupoles}
 \label{subsec:interacting_point_E2}

 We expand the polarization density to include the
 electric quadrupole term ${\bf p}^{\text{q}}_{j}({\bf r})$
 \begin{align}
 {\bf P}_{j}({\bf r},t) &= Q_j(t){\bf p}_{j}({\bf r})
 \,\text{,}
 \nonumber
 \\
 &  = Q_{n}(t)
 \Big[
 {\bf p}^{\text{d}}_{j}({\bf r}) +
 {\bf p}^{\text{q}}_{j}({\bf r}) + \ldots
 \Big]
 \,\text{.}
 \label{eq:p_multipole}
 \end{align}
 Whilst the electric dipole term ${\bf p}^{\text{d}}_{j}({\bf r})$
 is vector quantity, ${\bf p}^{\text{q}}_{j}({\bf r})$ is a tensor.
 The index $\alpha$ of the Cartesian component of the
 electric quadrupole contribution of the $j$th
 resonator is $p_{\alpha,j}^{\text{q}}({\bf r})$,
 where we define~\cite{Zangwill}
 \begin{equation}
 p^{\text{q}}_{\alpha,j}({\bf r})=
 -  \sum_{\beta}A_{\alpha\beta,j}
 \frac{\p}{\p r_\beta}
 \delta({\bf r} - {\bf r}_j)
 \,\text{.}
 \label{eq:p_E2}
 \end{equation}
 Here, $A_{\alpha\beta,j}$ is symmetric
 and traceless with dimensions of area, and
 the indices $\alpha,\,\beta$ refer to the Cartesian coordinates $x,y,z$
 and the summation is over $\beta$.
 The exact form of $A_{\alpha\beta,j}$ depends on the
 geometry of the resonator.

 The scattered electric
 ${\bf E}_{\text{E2},j}({\bf r})$ and magnetic
 ${\bf H}_{\text{E2},j}({\bf r})$
 fields due to the quadrupole moment
 located at ${\bf r}'$ are derived
 from the first terms, respectively, in 
 \esref{eq:Esc} and~\eqref{eq:Bsc}. Here, the spatial profile of the 
 polarization density in \eref{eq:P} has Cartesian component
 $\alpha$ defined in \eref{eq:p_E2}, and we 
 find the Cartesian component $\nu$
 of ${\bf E}_{\text{E2},j}({\bf r})$ 
 and ${\bf H}_{\text{E2,j}}({\bf r})$ are;
 \begin{align}
 E_{\text{E2},\nu,j}({\bf r}) & =
 \frac{Q_jk^3}{4\pi\epsilon_0}
 \sum_{\alpha}
 \int\d^3r'\,
 G_{\nu\alpha}({\bf r} - {\bf r}') p^{\text{q}}_{\alpha,j}({\bf r}')
 \,\text{,}
 \label{eq:E_E2}
 \\
 H_{\text{E2},\nu,j}({\bf r})  &=
 -\frac{cQ_jk^3}{4\pi}
 \sum_{\alpha}
 \int\d^3r'\,
 G_{\times,\nu\alpha}({\bf r} - {\bf r}') p^{\text{q}}_{\alpha,j}({\bf r}')
 \,\text{.}
 \label{eq:H_E2}
 \end{align}
 The radiation kernels $G_{\nu\alpha}({\bf r})$
 and $G_{\times,\nu\alpha}({\bf r})$
 are the tensor components ($\nu,\alpha =x,y,z$)
 of the radiation kernels defined in
 \esref{eq:G} and~\eqref{eq:G_x}, respectively.

 Whilst in the electric and magnetic dipole limit
 the radiation kernels act directly on the moments
 ${\bf p}^{\text{d}}_{j}({\bf r})$
 and ${\bf  w}_j({\bf r})$, respectively,  the
 quadrupole  EM fields, \esref{eq:E_E2} and~\eqref{eq:H_E2},
 are more complicated due to the derivative
 in $p_{\alpha,j}^{\text{q}}({\bf r})$.
 After integrating by parts \esref{eq:E_E2} and~\eqref{eq:H_E2},
 the EM field components $\nu$ are
 \begin{align}
 E_{\text{E2},\nu,j}({\bf r}) & =
 \frac{Q_jk^3}{4\pi\epsilon_0}
 \sum_{\alpha,\beta}
 \frac{\p}{\p r_\beta}
 G_{\nu\alpha}({\bf r} - {\bf r}_j) A_{\alpha\beta,j}
 \,\text{,}
 \label{eq:E_E2_ibp}
 \\
 H_{\text{E2},\nu,j}({\bf r})  &=
 -\frac{cQ_jk^3}{4\pi}
 \sum_{\alpha,\beta}
 \frac{\p }{\p r_\beta}
 G_{\times,\nu\alpha}({\bf r} - {\bf r}_j)A_{\alpha\beta,j}
 \,\text{.}
 \label{eq:H_E2_ibp}
 \end{align}
 The derivatives of the radiation kernel ${\bf G}({\bf r})$
 and cross kernel ${\bf G}_\times({\bf r})$, with respect to the
 Cartesian coordinate $r_{\mu=x,y,z}$ are given
 in \aref{sec:E2_kernels},
 see \esref{eq:Gradient_G} and~\eqref{eq:Gradient_Gx}.

 \Esref{eq:E_E2_ibp} and~\eqref{eq:H_E2_ibp}
 are the full EM field equations evaluated
 at ${\bf r}$ (in Cartesian coordinates), for
 an oscillating electric quadrupole source
 located at ${\bf r}'$. The EM fields are
 determined by contracting \esref{eq:Gradient_G}
 and~\eqref{eq:Gradient_Gx}, acting on the
 quadrupole moment $A_{\alpha\beta,j}$.

 In the electric dipole approximation, it is a
 relatively simple exercise to expand the
 radiation kernel, in powers of $kr$, to obtain
 an expression for the electric dipole radiative decay rate.
 For the electric quadrupole, there is no simple
 expansion for \eref{eq:Gradient_G}. In order
 to determine an expression for the electric
 quadrupole (and other higher order multipoles)
 self-interaction strength and radiative emission
 rate, we find it convenient to compare the
 multipole radiated power~\cite{Jackson,Zangwill}
 to the [rate of change of] energy of an oscillator.

 The radiated power can be obtained by integrating~\cite{Jackson}
 \begin{equation}
 \frac{\d P}{\d \Omega}
 =
 \lim_{r\rightarrow\infty}r^2{\bf \hat r}\cdotp
 \big[{\bf E}({\bf r})\times {\bf H}({\bf r})\big]
 \label{eq:dP_temp}
 \end{equation}
 over a closed spherical surface, where $\d\Omega$
 denotes the solid angle element
 and ${\bf \hat r}$ the vector normal to the surface.
 In the radiation zone, the fields ${\bf E}_\text{rad}({\bf r})$ and
 ${\bf H}_{\text{rad}}({\bf r})$ vary as $1/r$,
 and
 $|{\bf E}_{\text{rad}}({\bf r})|
 =
 c\mu_0|{\bf H}_{\text{rad}}({\bf r})|$. We have
 \begin{equation}
 \frac{\d P}{\d \Omega}
 =
 \frac{r^2}{c\mu_0}|{\bf E}_{\text{rad}}({\bf r})|^2
 \,\text{.}
 \label{eq:dP}
 \end{equation}
 In the limit $kr\gg 1$ we adopt the notation of
 \Rref{Jackson}, and define a quadrupole
 vector  component,
 of the $j$th resonator ${\bf q}_{j}({\bf \hat r})$,
 where
 \begin{equation}
 [{\bf q}_j({\bf \hat r})]_{\alpha}
 = \sum_{\beta=1}^3q_{\alpha\beta,j}\hat{r}_{\beta,j}
 \,\text{.}
 \end{equation}
 Here, $\alpha,\beta$ refer to the Cartesian components,
 ${\bf \hat r}$ is the unit vector in the direction of ${\bf r}$,
 and $q_{\alpha\beta,j}$ is the electric quadrupole moment tensor,
 defined as~\cite{Zangwill}
 \begin{equation}
 q_{\alpha\beta,j} = \frac{1}{2}\int\d^3r\, r_\alpha r_\beta \rho_j({\bf r},t)
 \,\text{.}
 \label{eq:q_alphabeta}
 \end{equation}
 The charge density $\rho_j({\bf r},t)$ in
 \eref{eq:q_alphabeta} is defined in
 \eref{eq:rho}. The electric ${\bf E}_{\text{rad,E2},j}({\bf r})$
 and
 magnetic ${\bf H}_{\text{rad,E2},j}({\bf r})$ radiated fields
 from the $j$th electric quadrupole are~\cite{Jackson}
 \begin{align}
 {\bf E}_{\text{rad,E2},j}({\bf r})
 &
 =
 i\frac{k^3}{4\pi\epsilon_0}\frac{e^{ikr}}{r}
 {\bf \hat r}\times \big[{\bf \hat r} \times {\bf q}_{j}({\bf \hat r})\big]
 \,\text{,}
 \label{eq:E2_rad}
 \\
 {\bf H}_{\text{rad,E2},j}({\bf r})
 &
 =
 i\frac{ck^3}{4\pi}\frac{e^{ikr}}{r}
 {\bf \hat r}\times {\bf q}_{j}({\bf \hat r})
 \,\text{,}
 \label{eq:H2_rad}
 \end{align}
 where $r=|{\bf r} - {\bf r}_j|$. The electric quadrupole
 contribution to the power $P_{\text{E2}}$ is~\cite{Zangwill}
 \begin{equation}
 \frac{\d P_{\text{E2},j}}{\d \Omega}
 =
 \frac{\mu_0c^3k^6}{16\pi^2}
 |{\bf \hat r}\times {\bf q}_j({\bf \hat r})|^2
 \,\text{.}
 \label{eq:P_E2_temp1}
 \end{equation}
 The electric quadrupole radiated power $P_{\text{E2},j}$,
 is the integral of \eref{eq:P_E2_temp1} over all angles~\cite{Zangwill}.
 We find~\cite{Zangwill}  [see \aref{subsec:Power_E2_supp}],
 \begin{equation}
 P_{\text{E2},j} = \frac{\mu_0c^3k^6}{20\pi}
 \sum_{\alpha,\beta}
 \Big[q_{\alpha\beta,j}q_{\alpha\beta,j}
 -
 \frac{1}{3}q_{\alpha\alpha,j}q_{\beta\beta,j}\Big]
 \,\text{,}
 \label{eq:Power_E2}
 \end{equation}
 The quadrupole moment tensors, $q_{\alpha\beta,j}$ and
 $A_{\alpha\beta,j}$, and the dynamic variable $b_j(t)$
 of the $j$th resonator are related through the charge density
 $\rho_j({\bf r},t)$. The electric quadrupole component of the
 charge density, from \esref{eq:rho} and~\eqref{eq:p_E2} is
 \begin{equation}
 \rho({\bf r},t) =
 Q_j(t)\sum_{\alpha\beta,j}\frac{\p}{\p r_\alpha}
 \left[
 A_{\alpha\beta,j}
 \frac{\p }{\p r_\beta}\delta({\bf r} - {\bf r}_j)
 \right]
 \,\text{,}
 \label{eq:rho_E2}
 \end{equation}
 where the summation is over the Cartesian coordinates
 ($\alpha,\beta=x,y,z$). Substituting \eref{eq:rho_E2} into
 \eref{eq:q_alphabeta},
 \begin{equation}
 q_{\alpha\beta,j}
 =
 \frac{1}{2}Q_j(t)\int\d^3r\,
 r_\mu r_\nu
 \frac{\p}{\p r_\alpha}
 \left[
 A_{\alpha\beta,j}
 \frac{\p }{\p r_\beta}\delta({\bf r} - {\bf r}_j)
 \right]
 \,\text{.}
 \label{eq:q_alphabeta_int1}
 \end{equation}
 Integration of \eref{eq:q_alphabeta_int1},
 by parts twice yields
 \begin{equation}
 q_{\alpha\beta,j}
 =
 \frac{1}{2}Q_j(t)\int\d^3r\,
 \left[
 \frac{\p }{\p r_\alpha}
 \frac{\p }{\p r_\beta}
 r_\mu r_\nu
 \right]
 A_{\alpha\beta,j}
 \delta({\bf r} - {\bf r}_j)
 \,\text{.}
 \label{eq:q_alphabeta_int2}
 \end{equation}
 The term in parenthesis in \eref{eq:q_alphabeta_int2},
 simplifies considerably because the derivatives result in
 Kronecker $\delta$ functions,
 \begin{align}
 \frac{\p }{\p r_\alpha}
 \frac{\p }{\p r_\beta}
 r_\mu r_\nu
 &
 =
 \frac{\p }{\p r_\alpha}
 \left[
 \delta_{\beta\mu}r_\nu + \delta_{\beta\nu}r_\mu
 \right]
 \,\text{,}
 \nonumber
 \\
 &
 \quad
 =
 \delta_{\alpha\nu}\delta_{\beta\mu}
 +
 \delta_{\alpha\mu}\delta_{\beta\nu}
 = 2
 \,\text{.}
 \label{eq:kronecker_delta}
 \end{align}
 Writing the charge $Q_j(t)$ in terms of
 the dynamic variable $b_j(t)$ [see \eref{eq:b}], we finally
 have the relationship between $q_{\alpha\beta,j},\,b_j(t)$
 and $A_{\alpha\beta,j}$;
 \begin{equation}
 q_{\alpha\beta,j} =
 \sqrt{\frac{\omega_jC_j}{2}}b_j(t)A_{\alpha\beta,j}
 \,\text{.}
 \end{equation}
 The energy $U_j$
 of an isolated oscillator, from its Hamiltonian,
 is analogous to that of an LC
 circuit~\cite{JenkinsLongPRB}
 \begin{equation}
 U_j(t) = \omega_j \left|b_j\right|^2
 \,\text{.}
 \label{eq:energy}
 \end{equation}
 The electric quadrupole radiated power
 $P_{\text{E2},j}$ of the oscillator, is the
 rate of change of \eref{eq:energy},
 \begin{equation}
 P_{\text{E2},j}
 = -\frac{\d U_j}{\d t}
 =
 \omega_j\Gamma_{\text{E2},j}\left|b_j\right|^2
 \,\text{.}
 \label{eq:power_E2_b}
 \end{equation}
 Here, $\omega_j$ is the resonance frequency and $\Gamma_{\text{E2},j}$
 the  decay rate of the electric quadrupole.
 Comparing \esref{eq:Power_E2} and~\eqref{eq:power_E2_b},
 we obtain the rate at which a resonator radiates energy in the point
 electric  quadrupole approximation as
 \begin{equation}
 \Gamma_{\text{E2},j} =
 \frac{C_jA_{\text{E},j}^2 \omega_j^6}{20\pi\epsilon_0c^5}
 \,\text{,}
 \label{eq:GammaE2}
 \end{equation}
 where we define
 \begin{equation}
 A_{\text{E},j}^2=
 \sum_{\alpha,\beta}
 \left[A_{\alpha\beta,j}A_{\alpha\beta,j} -
 \frac{1}{3}A_{\alpha\alpha,j}A_{\beta\beta,j}\right]
 \,\text{,}
 \label{eq:A_0}
 \end{equation}
 as an effective area of the electric quadrupole. Again, the indices
 $\alpha,\beta$ refer to the Cartesian
 components of the quadrupole moment
 and repeated indices are summed over.

 With the radiative emission rates of the
 electric quadrupole \eref{eq:GammaE2}, and the
 electric and magnetic dipoles \esref{eq:GammaE1}
 and~\eqref{eq:GammaM1_temp}, respectively, we
 can express the normal mode oscillator amplitudes \eref{eq:b}
 in terms of the contributing
 multipole moments
 \begin{align}
 b_j(t)
 & =
 \sqrt{\frac{k^3}{12\pi\epsilon_0}}
 \Bigg[
 Q_j\frac{H_j}{\sqrt{\Gamma_{\text{E1},j}}}
 +
 kQ_j\sqrt{\frac{3}{5}}
 \frac{A_{\text{E},j}}{\sqrt{\Gamma_{\text{E2,j}}}}
 \nonumber
 \\
 &
 \hspace{0.5cm}
 +
 i
 \frac{I_j}{c}\frac{A_{\text{M},j}}{\sqrt{\Gamma_{\text{M1},j}}}
 \Bigg]
 \,\text{.}
 \label{eq:b_multipole}
 \end{align}
 The real part of \eref{eq:b_multipole} comprises
 the electric dipole and electric quadrupole contributions.
 The imaginary part corresponds to the
 magnetic dipole contribution.

 In the point emitter approximation,
 the radiative emission rates
 of the magnetic dipole  and the
 electric quadrupole both depend
 on their respective, effective cross sectional areas $A_{\text{M},j}$ and
 $A_{\text{E},j}$, see
 \esref{eq:GammaM1_temp} and~\eqref{eq:GammaE2},
 respectively.
 For simplicity, we assume that the magnetic dipole and electric quadrupole
 have the same resonance frequency $\omega_j$ [\eref{eq:omega}].
 Comparing \esref{eq:GammaM1_temp} and~\eqref{eq:GammaE2},
 we find $\Gamma_{\text{M1},j}$ and $\Gamma_{\text{E2},j}$
 are of the same order of magnitude, their relative radiation emission rates are
 \begin{equation}
 \frac{\Gamma_{\text{E2},j}}{\Gamma_{\text{M1},j}} =
 \frac{3}{10}
 \frac{A_{\text{E},j}^2}{A_{\text{M},j}^2}
 \,\text{.}
 \label{eq:GammaM1_GammaE2_general}
 \end{equation}

 In \sref{subsec:E1M1_general}, the amplitudes of the
 scattered EM fields were
 proportional to the electric dipole and magnetic dipole
 radiative emission rates. Here, the full
 electric quadrupole EM field amplitudes,
 \esref{eq:E_E2_ibp} and~\eqref{eq:H_E2_ibp},
 are proportional to the electric
 quadrupole decay rate $\Gamma_{\text{E2},j}$.
 We write the scaled EM fields of the $j$th electric quadrupole source as
 \begin{align}
 E_{\text{E2},\nu,j}({\bf r})  &=
 b_j\sqrt{\frac{\wp_0}{\epsilon_0}}
 \sum_{\alpha,\beta}
 \frac{\p}{\p kr_\beta}
 G_{\nu\alpha}({\bf r} - {\bf r}_{j})
 { \hat  A}_{\alpha\beta,j}
 \,\text{,}
 \label{eq:E_E2_scaled}
 \\
 H_{\text{E2},\nu,j}({\bf r})  &=
 -b_j\sqrt{\frac{\wp_0}{\mu_0}}
 \sum_{\alpha,\beta}
 \frac{\p}{\p kr_\beta}
 G_{\times,\nu\alpha}({\bf r} - {\bf r}_{j})
 {\hat A}_{\alpha\beta,j}
 \,\text{,}
 \label{eq:H_E2_scaled}
 \end{align}
 where the constant $\wp_0$ is defined as
 \begin{equation}
 \wp_0 =  \frac{5k^3}{8\pi}\Gamma_{\text{E2},j}
 \,\text{,}
 \label{eq:wp_0}
 \end{equation}
 and ${\hat A}_{\alpha\beta,j}$ is a  tensor which defines the charge
 configuration of the quadrupole moment
 \begin{equation}
 \hat{A}_{\alpha\beta,j} =
 \frac{A_{\alpha\beta,j}}{A_{\text{E},j}}
 \,\text{.}
 \label{eq:A_hat}
 \end{equation}
 The $j$th electric quadrupole is also driven by
 the external electric fields ${\bf E}_{\text{ext},j}({\bf r})$, resulting
 in the induced emf
 \begin{equation}
 \mathcal{E}_{\text{ext},j}^{\text{E2}}
 =
 \frac{1}{\sqrt{\omega_jL_j}}
 \sum_\nu
 \int\d^3r\,
 p_{\nu,j}^{\text{q}}({\bf r})
 E_{\text{ext},\nu,n}({\bf r})
 \,\text{.}
 \label{eq:emf_E2_general}
 \end{equation}
 Here, the mode function $p_{\nu,j}^\text{q}({\bf r})$ is
 defined in \eref{eq:p_E2}.
 For point electric quadrupole sources, the $j$th electric quadrupole
 moment $A_{\alpha\beta,j}$ interacts with gradient of the external
 electric field,
 \begin{equation}
 \mathcal{E}_{\text{ext},j}^{\text{E2}}
 =
 \frac{1}{\sqrt{\omega_jL_j}}
 \sum_{\alpha\beta}
 A_{\alpha\beta,j}
 \bigg[
 \frac{\p}{\p r_\beta}E_{\text{ext},\alpha,j}({\bf r})
 \bigg]
 \,\text{.}
 \label{eq:emf_E2_general_ibp}
 \end{equation}
 The external electric field
 [\eref{eq:Eext}] comprises the
 incident electric field and the different multipole
 scattered fields. These different contributions to the
 external electric field driving the electric quadrupole
 source allow us to decompose the resulting emf
 into different components;
 \begin{align}
 \mathcal{E}_{\text{ext},j}^{\text{E2}}
 =
 \mathcal{E}_{\text{in},j}^{\text{E2}}
 +
 \sum_{i \ne j}
 \Big[
  \mathcal{E}_{i,j}^{\text{sc,X2e}}
 +
  \mathcal{E}_{i,j}^{\text{sc,X2m}}
 +
 \mathcal{E}_{i,j}^{\text{sc,E2}}
 +
 \ldots
 \Big]
 \,\text{.}
 \label{eq:emf_E2_total}
 \end{align}
 In \eref{eq:emf_E2_total}, the incident EM field contribution
 to the emf
 follows from \eref{eq:emf_E2_general_ibp}, with
 the incident displacement field \eref{eq:Din}
 \begin{equation}
 \mathcal{E}_{\text{in},j}^{\text{E2}}  =
 \frac{1}{\epsilon_0\sqrt{\omega_jL_j}}
 \sum_{\alpha\beta}
 A_{\alpha\beta,j}
 \bigg[
 \frac{\p}{\p r_\beta}D_{\text{in},\alpha,j}({\bf r})
 \bigg]
 \,\text{.}
 \label{eq:emf_in_E2}
 \end{equation}
 The contributions $\mathcal{E}_{i,j}^{\text{sc,X2e}}$ and
 $\mathcal{E}_{i,j}^{\text{sc,X2m}}$  are due to
 the interactions of electric and magnetic dipoles, respectively, with
 electric quadrupoles. We discuss these contributions in detail later.
 Here, we provide the
 electric quadrupole driven contribution to the emf
 from two interacting electric
 quadrupoles,  $\mathcal{E}_{i,j}^{\text{sc,E2}}$, the counterpart
 to the emf from two electric dipoles
 [see \eref{eq:emf_E1}]. With the definition
 of the emf, \eref{eq:emf_E2_general},
 we have
 \begin{widetext}
 \begin{equation}
 \mathcal{E}_{i,j}^{\text{sc,E2}}
 =
 -\frac{1}{\sqrt{\omega_jL_j}}
 \frac{Q_jk^3}{4\pi\epsilon_0}
 \sum_\nu\int\d^3r\,
 \left[
 \sum_{\eta}
 A_{\nu\eta,i}
 \frac{\p}{\p r_\eta}\delta({\bf r}- {\bf r}_i)
 \right]
 \left[
 \sum_{\alpha\beta}
 \frac{\p}{\p r_\beta}
 G_{\nu\alpha}({\bf r} - {\bf r}_j)
 A_{\alpha\beta,j}
 \right]
 \,\text{.}
 \label{eq:emf_E2_intermediate_1}
 \end{equation}
 The first term in parenthesis in \eref{eq:emf_E2_intermediate_1}
 is the mode function $p_{\nu,i}^\text{q}({\bf r})$ of the
 $i$th electric quadrupole [see \eref{eq:p_E2}].
 Whilst the second term in parenthesis,
 is the scattered electric field from the $j$th electric quadrupole,
 $E_{\text{E2},\nu,j}({\bf r})$ [see \eref{eq:E_E2_ibp}].
 Integration of \eref{eq:emf_E2_intermediate_1} by parts,
 we have
 \begin{equation}
 \mathcal{E}_{i,j}^{\text{sc,E2}}
 =
 \frac{1}{\sqrt{\omega_jL_j}}
 \frac{Q_jk^3}{4\pi\epsilon_0}
 \sum_{\nu}
 \int\d^3r\,
 A_{\nu\eta,i}
 \delta({\bf r} - {\bf r}_i)
 \sum_{\eta\alpha\beta}
 \bigg[
 \frac{\p}{\p r_\eta}\frac{\p}{\p r_\beta}
 G_{\nu\alpha}({\bf r}_i - {\bf r}_j)
 \bigg]
 A_{\alpha\beta,j}
 \,\text{.}
 \label{eq:emf_E2_intermediate_2}
 \end{equation}
 \end{widetext}
 The integral in \eref{eq:emf_E2_intermediate_2} is
 readily carried out over the $\delta$ function.
 The second derivatives of the radiation kernel with respect
 to the Cartesian coordinate $r_{\mu= x,y,z}$ are given
 in \esref{eq:grad_grad_radiation_kernel_xx}
 and~\eqref{eq:grad_grad_radiation_kernel_xy},
 see \aref{sec:E2_kernels}.
 The electric quadrupole moment $A_{\alpha\beta,j}$, in
 \eref{eq:emf_E2_intermediate_2}, and
 the decay rate $\Gamma_{\text{E2},j}$ are related through the
 effective area $A_{\text{E},j}$ appearing in both
 \esref{eq:GammaE2} and~\eqref{eq:A_hat}.
 This allows us to write \eref{eq:emf_E2_intermediate_2} more
 compactly as
 \begin{equation}
 \mathcal{E}_{i,j}^{\text{sc,E2}} =
 \sqrt{\Gamma_{\text{E2},i}\Gamma_{\text{E2},j}}
 \left[\mathcal{G}_{\text{E2}}\right]_{i,j}
 \frac{b_{j}}{\sqrt{2}}
 \,\text{.}
 \label{eq:emf_E2}
 \end{equation}
 The matrix $\mathcal{G}_{\text{E2}}$ is the
 contribution to $\mathcal{C}_{\text{E2}}$ in \eref{eq:C_E2},
 with off diagonal  components
 \begin{equation}
 \left[\mathcal{G}_{\text{E2}}\right]_{i,j}
 =
 \frac{15}{4}
 \sum_{\nu,\eta,\alpha,\beta}
 {\hat A}_{\nu\eta,i}
 \frac{\p^2}{\p kr_\eta \p kr_\beta}G_{\nu\alpha}({\bf r}_i - {\bf r}_{j})
 {\hat A}_{\alpha\beta,j}
 \,\text{.}
 \label{eq:G_E2}
 \end{equation}
 \Eref{eq:G_E2} is, in general,
 complicated, however, as we show later
 in \sref{subsec:two_pair_general},
 for simple point quadrupole systems,
 \eref{eq:G_E2} simplifies considerably.
 The coupling matrix $\mathcal{C}$ for
 interacting electric quadrupoles only is
 \begin{equation}
 \mathcal{C}= \Delta - \frac{1}{2}\Upsilon
 +
 \frac{i}{2}\mathcal{C}_{\text{E2}}
 \,\text{,}
 \label{eq:Coupling_E2}
 \end{equation}
 where $\mathcal{C}_{\text{E2}}$ is given in \eref{eq:C_E2}.
 The diagonal elements of $\mathcal{C}$ contain the
 detuning $\Delta$ and total decay rate $\Upsilon$.
 \Eref{eq:Delta} gives the
 detuning, the decay rates in the
 electric quadrupole approximation are
 \begin{align}
 \big[\Upsilon\big]_{j,j}& = \Gamma_{\text{E2},j} + \Gamma_{\text{O},j}
 \,\text{.}
 \end{align}

 %
 \subsection{Interacting point electric and magnetic dipoles
 and electric  quadrupoles}
 \label{subsec:M1E2}

 In \sref{sec:E2} we discussed the interactions
 between two point electric quadrupoles
 and in \sref{subsec:E1M1_general} between
 electric and magnetic dipoles. Here we
 introduce the cross coupling of the electric quadrupole
 to the electric and magnetic dipoles.

 \subsubsection{Electric dipole--electric quadrupole interactions}
 \label{subsubsec:E1E2}

 The electric field scattered by an electric dipole is given by the
 first integral in  \eref{eq:Esc_dip_scaled}
 and by an electric quadrupole in \eref{eq:E_E2_scaled}.
 These scattered electric fields drive the
 charge oscillations in other resonators with
 electric quadrupoles and electric dipoles, giving
 rise to the cross driving (dipole-quadrupole) emf
 $\mathcal{E}_{i,j}^{\text{sc,X2e}}$, where
 \begin{align}
 \mathcal{E}_{i,j}^{\text{sc,X2e}} = &
 \Bigg[\sqrt{\Gamma_{\text{E2},i}\Gamma_{\text{E1},j}}
 \left[  \mathcal{G}_{\text{X2e}}\right]_{i,j}
 \nonumber
 \\
 & \quad
 +
 \sqrt{\Gamma_{\text{E1},i}\Gamma_{\text{E2},j}}
 \left[  \mathcal{G}_{\text{X2e}}\right]_{i,j}^T
 \Bigg]
 \frac{b_{j}}{\sqrt{2}}
 \,\text{.}
 \label{eq:emf_E2E1}
 \end{align}
 The matrix $\mathcal{G}_{\text{X2e}}$ has zero diagonal
 elements, the off-diagonal elements are defined by
 \begin{equation}
 \big[\mathcal{G}_{\text{X2e}}\big]_{i,j}
 =
 \sqrt{\frac{15}{2}}
 \sum_{\nu,\eta,\alpha}
 {\hat A}_{\nu\eta,i}
 \frac{\p}{\p kr_\eta }
 G_{\nu\alpha}({\bf r}_i - {\bf r}_{j})
 {\hat d}_{\alpha,j}
 \,\text{.}
 \label{eq:G_X2e}
 \end{equation}
 The interactions between an electric quadrupole
 (electric dipole) with the EM fields from an
 electric dipole   (electric quadrupole) are described by
 $\mathcal{G}_{\text{X2e}}$.
 The derivatives of the radiation kernel are given
 in \eref{eq:Gradient_G} (see \aref{sec:E2_kernels}).

 The interaction matrix $\mathcal{C}_{\text{X2e}}$  in the equation of motion,
 \eref{eq:equmot}, for the cross driving of electric dipoles and
 electric quadrupoles are given in \eref{eq:C_X2e}, where
 \begin{equation}
 \big[\mathcal{C}_{\text{X2e}}\big]_{i,j}
 =
 \sqrt{\Gamma_{\text{E2},i}\Gamma_{\text{E1},j}}
 \big[\mathcal{G}_{\text{X2e}}\big]_{i,j}
 \,\text{.}
 \label{eq:C_X2e_temp}
 \end{equation}

 \subsubsection{Magnetic dipole--electric quadrupole interactions}
 \label{subsubsec:M1E2}

 The electric field from an oscillating magnetic dipole is given by
 the second integral in \eref{eq:Esc_dip_scaled},
 and the magnetic field from an
 electric quadrupole in \eref{eq:H_E2_scaled}. These scattered
 EM fields drive the external electric quadrupole and magnetic dipole
 sources, respectively, resulting in an
 emf $\mathcal{E}^{\text{sc,X2m}}_{i,j}$
 and flux $\Phi^{\text{sc,X2m}}_{i,j}$
 \begin{align}
 \mathcal{E}_{i,j}^{\text{sc,X2m}} = &
 -  \sqrt{\Gamma_{\text{E2},i}\Gamma_{\text{M1},j}}
 \left[  \mathcal{G}_{\text{X2m}}\right]_{i,j}
 \frac{b_{j}}{\sqrt{2}}
 \,\text{,}
 \label{eq:emf_E2M1}
 \\
 \Phi_{i,j}^{\text{sc,X2m}} = &
 -\frac{1}{\omega_j}
 \sqrt{\Gamma_{\text{E2},i}\Gamma_{\text{M1},j}}\big[
 \mathcal{G}_{\text{X2m}}\big]_{i,j}^T
 \frac{b_{j}}{\sqrt{2}}
 \,\text{.}
 \label{eq:flux_M1E2}
 \end{align}
 The terms in  $\mathcal{C}_{\text{X2m}}$
 then follow as in the previous section, where
 the off-diagonal elements of $\mathcal{G}_{\text{X2m}}$ are given by
 \begin{equation}
 \big[\mathcal{G}_{\text{X2m}}\big]_{i,j}
  =
 \sqrt{\frac{15}{2}}
 \sum_{\nu,\eta,\alpha}
 {\hat A}_{\nu\eta,i}
 \frac{\p}{\p kr_\eta }
 G_{\times,\nu\alpha}({\bf r}_i - {\bf r}_{j})
 {\hat m}_{\alpha,j}
 \,\text{,}
 \label{eq:G_X2m}
 \end{equation}
 where the derivatives of the cross kernel
 are given in \eref{eq:Gradient_Gx} (see \aref{sec:E2_kernels}), and
 \begin{equation}
 \big[\mathcal{C}_{\text{X2m}}\big]_{i,j}
 =
 -\sqrt{\Gamma_{\text{E2},i}\Gamma_{\text{M1},j}}
 \big[\mathcal{G}_{\text{X2m}}\big]_{i,j}
 \,\text{.}
 \label{eq:C_X2m_temp}
 \end{equation}
 For simple
 interacting electric quadrupole--magnetic
 dipole systems, \eref{eq:C_X2m_temp}
 simplifies considerably,
 as we show later in
 \sref{subsec:two_pair_general}.

 %
 %

 \section{Examples of simple systems of interacting point emitters}
 \label{sec:point_multipole_resonators}

 Metamaterial arrays typically consist of
 large numbers of subwavelength-spaced
 metamolecules, each of these formed
 by configurations of resonators. Radiative
 interactions between different metamolecules
 can be strong, and when analyzing collective
 interactions in large systems, it may be impractical, or even beyond the computational capacity, to provide a detailed
 intrinsic model of each metamolecule.
 The effective model of point scatterers with
 a multipole expansion of their radiative
 properties can be utilized in the simplification
 of individual metamolecule properties.
 For symmetric and asymmetric split-ring
 resonator metamaterials, in which case
 each metamolecule consists of two
 symmetric or asymmetric resonator arcs,
 the point emitter approximation was previously
 applied separately to each arc~\cite{JenkinsLongPRB,
 JenkinsLineWidthNJP}. In that case it
 was sufficient to represent each circuit
 resonator arc as a point emitter possessing
 electric and magnetic dipoles. The formalism
 was successful in describing collective effects
 in planar asymmetric split-ring metamaterial
 arrays~\cite{JenkinsLineWidthNJP,
 CAIT,
 Jenkinssub}. In analogous systems, it has also
 been used in the development of an electron-beam-driven
 light source from the collective response~\cite{AdamoEtAlPRL2012}.

In order to illustrate and test the point-emitter
 formalism, we introduce models for the
 interactions between effective point emitters
 that not only possess electric and magnetic dipoles,
 but also the electric quadrupole, developed in \sref{sec:E2}.
 After analyzing the elementary case of two point
 electric dipoles, we consider systems
 comprising two parallel pairs of electric point dipoles.
When a parallel pair is symmetrically excited,
 it may be approximated by a single effective
 point emitter possessing an electric dipole
 located at the center of the two dipoles.
For an antisymmetrically excited pair we use
 a single effective point emitter possessing
 both a magnetic dipole and electric
 quadrupole located at the center of the two dipoles.
We denote the decay rates of the
 effective point emitters by $\gamma_\text{s,a}^{(1)}$
 for a symmetrically and antisymmetrically
 excited pair of dipoles, respectively, that
 depend on the separation of the dipoles
 within the pair. For simplicity, we assume
 that the decay rates and the resonance
 frequencies of all point electric dipole are equal.

\subsection{Two parallel point electric dipoles}
 \label{subsec:E1_example}

\begin{figure}[h!]
 \centering
 \includegraphics[]{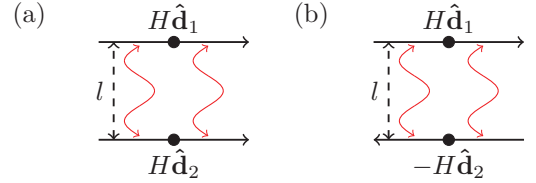}
 \caption[Two interacting parallel point electric dipoles.]
 {\label{Drg:InteractingDipoles}
 Schematic illustration of two
 interacting
 point electric dipoles (dots) with magnitude $H$
 and orientation vectors ${\bf d}_1$ and ${\bf d}_2$,
 separated by a distance $l$. In~(a) we show the symmetric and
 in~(b) the antisymmetric excitation.}
 \end{figure}

 As the first example to illustrate our model
 we take two parallel electric dipoles (\fref{Drg:InteractingDipoles}) located at
 \begin{equation}
 {\bf r}_{1} = \frac{1}{2}\begin{bmatrix} s_1\\  y_1\\0 \end{bmatrix},
 \quad %
 {\bf r}_{2} =  \frac{1}{2}\begin{bmatrix} s_2\\  y_2\\ 0 \end{bmatrix}
 \,\text{,}
 \label{eq:2EmitterLocations}
 \end{equation}
 and specifically set $s_1 = s_2 = 0$  and $|y_1 - y_2| = l$,
 i.e.,  ${\bf r}_1 = -{\bf r}_2 = [0, l/2, 0]$.

 The decay rate of an electric dipole,
 \begin{equation}
 \Gamma^{(1)} =
 \Gamma_\text{O} + \Gamma_{\text{E1}}
 \,\text{,}
 \label{eq:Gamma_1}
 \end{equation}
 depends on the rate of dipole radiation
 and nonradiative losses that we set to
 $\Gamma_{\text{E1}} = 0.83\Gamma^{(1)}$
 and $\Gamma_{\text{O}}=0.17\Gamma^{(1)}$.

 When the driving field is tuned to the
 resonance frequency of the point electric
 dipoles, $\Omega_0=\omega_{0}$, the
 coupling matrix in the equation of motion [\eref{eq:C_E1M1E2},
 with $\Gamma_{\text{M1}} \equiv \Gamma_{\text{E2}} \equiv 0$],
 of a pair of electric dipoles is,
 \begin{equation}
 \mathcal{C} =
 \begin{bmatrix}
 -\dfrac{\Gamma^{(1)}}{2}
 &
 i\dfrac{3}{4}\Gamma_{\text{E1}}
 G_{\text{E1}}({\bf r}_{12})
 \\
 i\dfrac{3}{4}\Gamma_{\text{E1}}
 G_{\text{E1}}(-{\bf r}_{12})
 &
 -\dfrac{\Gamma^{(1)}}{2}
 \end{bmatrix}
 \,\text{,}
 \label{eq:C_E1_two_dipoles}
 \end{equation}
 where
 $\Gamma^{(1)} = \Gamma_\text{O} + \Gamma_{\text{E1}}$
 [see \eref{eq:Gamma_1}],
 ${\bf r}_{12}= {\bf r}_2 - {\bf r}_1$, and
 $ G_{\text{E1}} ({\bf r}_{12})
 =
 G_{\text{E1}} (-{\bf r}_{12})$, which
 from \eref{eq:G}
 \begin{equation}
 G_{\text{E1}} ({\bf r}_{12})
 =
 \frac{i}{3}
 \left[
 2h_0^{(1)}(kl) - h_2^{(1)}(kl)
 \right]
 \,\text{.}
 \label{eq:mathcalG_E1}
 \end{equation}
 \Eref{eq:C_E1_two_dipoles}
 has two eigenmodes of
 current oscillation: a symmetric
 mode (denoted by a subscript `s'),
 where both dipoles current oscillations are in phase,
 i.e., ${\bf \hat d}_1={\bf \hat d}_2$,
 see \fref{Drg:InteractingDipoles}(a); and an antisymmetric
 mode (denoted by a subscript `a'), where the current
 oscillations of the dipoles are out of phase,
 i.e., ${\bf \hat d}_1 = -{\bf \hat d}_2$,
 see \fref{Drg:InteractingDipoles}(b). The
 eigenvectors (${\bf v}_n^{(1)}$) and corresponding
 eigenvalues ($\xi_n^{(1)}$) of the the two
 modes of current oscillation are
 \begin{equation}
 {\bf v}_\text{s}^{(1)} =
 \frac{1}{\sqrt{2}}
 \begin{bmatrix}
 1 \\
 1
 \end{bmatrix}
 \quad
 \text{and}
 \quad
 {\bf v}_{\text{a}}^{(1)}
 =
 \frac{1}{\sqrt{2}}
 \begin{bmatrix}
 1 \\ -1
 \end{bmatrix}
 \,\text{,}
 \label{eq:eigenvectors_two_E1}
 \end{equation}
 and eigenvalues
 \begin{equation}
 \xi_{\text{a,s}}^{(1)} =
 - \frac{\Gamma^{(1)}}{2}
 \pm
 i\frac{3}{4}\Gamma_{\text{E1}}G_{\text{E1}}({\bf r}_{12})
 \,\text{,}
 \label{eq:eigenvalues_two_E1}
 \end{equation}
 respectively.
 The eigenvalues determine the
 mode resonance frequency shifts
 $\delta\omega_{\text{a,s}}^{(1)} =
 -(\Omega_{\text{a,s}}^{(1)}- \Omega_0)
 =\text{Im}(\xi_{\text{a,s}}^{(1)})$,
 and mode decay rates
 $\gamma_{\text{a,s}}^{(1)}
 = 2\text{Re}(\xi_{\text{a,s}}^{(1)})$,
 see  \eref{eq:xi_n}. We will later use
 the symmetric and antisymmetric mode decay rates
 $\gamma_{\text{s}}^{(1)}$ and $\gamma_{\text{a}}^{(1)}$, and
 the corresponding line shifts
 $\delta\omega_{\text{s}}^{(1)}$
 and $\delta\omega_{\text{a}}^{(1)}$,
 as estimates for the total decay rates
 and resonance frequencies of a single
 point emitter possessing either electric
 dipole or both magnetic dipole and electric
 quadrupole moments.

 \begin{figure}[h!]
 \centering
 \includegraphics[]
 {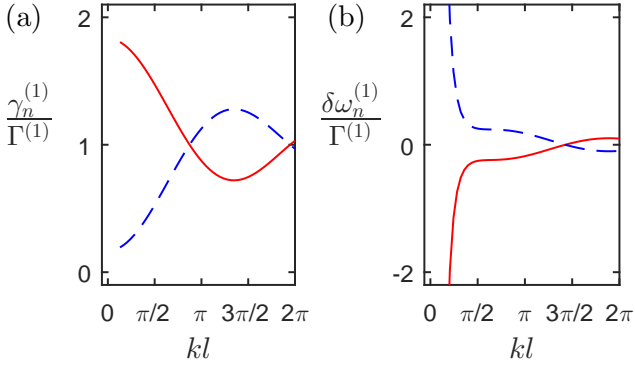}
 \caption[The radiative resonance linewidths and line shifts of two
 parallel point electric dipole resonators.]
 {\label{fig:2_point_E1_shift_width}
 The radiative resonance (a) linewidths
 and (b) line shifts for the collective antisymmetric (blue dashed line) and
 symmetric (red solid line) eigenmodes, as a function of the
 separation $l$, for two parallel point
 electric dipoles. The radiative losses of each
 dipole are $\Gamma_{\text{E1}}=0.83\Gamma^{(1)}$,
 the ohmic losses are
 $\Gamma_{\text{O}}=0.17\Gamma^{(1)}$.
 }
 \end{figure}

 In \fref{fig:2_point_E1_shift_width} we show the
 radiative resonance linewidths and line shifts
 for the collective antisymmetric and
 symmetric eigenmodes.
 As the separation because small $l\rightarrow 0$, the
 linewidth of the antisymmetric mode approaches the
 ohmic loss rate
 ($\gamma_{\text{a}}^{(1)}\rightarrow\Gamma_{\text{O}}$)
 and is subradiant,
 the symmetric mode linewidth approaches
 $\gamma_{\text{s}}^{(1)}\rightarrow 1.8\Gamma^{(1)}$
 and is superradiant. At
 approximately $kl\approx \pi$ (where
 $k=2\pi/\lambda_0$), the
 symmetric and antisymmetric modes become
 subradiant and superradiant, respectively.
 The line shifts of two modes are symmetric about
 $\Omega_0$. At $kl\approx \pi/4$,
 the line shifts diverge with
 $\delta\omega_{\text{s}}^{(1)}$
 red shifted, and
 $\delta\omega_{\text{a}}^{(1)}$ blue shifted,
 from $\Omega_0$.

 \subsection{Effective point emitter for a pair of out-of-phase electric dipoles}
 \label{subsec:antisymmetric_E1}

\begin{figure}[h!]
 \centering
 \includegraphics[]{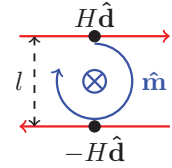}
 \caption[Visualization of the magnetic dipole and electric quadrupole
 moment formed by
 two antisymmetrically excited parallel point electric dipoles.]
 {\label{Drg:M1E2}
 Schematic illustration of the effective magnetic dipole moment
 [pointing into the page] and
 electric quadrupole moment, both located at the center of the cross,
 formed by two parallel antisymmetrically
 excited point electric dipoles (dots), with magnitude $H$ and
 orientation vectors $\pm{\bf \hat d}$, separated
 by a distance $l$.
 }
 \end{figure}

 Two closely-spaced parallel electric dipoles
 have eigenmodes that represent in-phase
 and out-of-phase excitations \eref{eq:eigenvectors_two_E1}.
 The in-phase oscillations of a pair
 of dipoles can be approximated by a
 single electric dipole point emitter. For the antisymmetric, out-of-phase oscillations
 the total electric dipole is weak, but
 the pair exhibits nonvanishing electric quadrupole
 and magnetic dipole moments  (see \fref{Drg:M1E2}).
 We therefore approximate a pair of
 closely-spaced, parallel out-of-phase point
 electric dipoles by a single point emitter,
 possessing a magnetic dipole and
 electric quadrupole moment,
 located between the two electric dipoles.

 We write the decay rate of a point
 emitter corresponding to the pair of out-of-phase electric dipoles as
 \begin{equation}
 \gamma_{\text{a}}^{(1)} =
 \Gamma_{\text{O}} +
 \Gamma_{\text{M1}} +
 \Gamma_{\text{E2}}
 \,\text{,}
 \label{eq:gamma_a_M1E2_approx}
 \end{equation}
 where the antisymmetric collective mode
 linewidth $\gamma_{\text{a}}^{(1)}$ of two
 parallel electric dipoles can be calculated and is shown
 in \fref{fig:2_point_E1_shift_width}.
 Using this formula, we may then derive
 analytical expressions for $\Gamma_{\text{M1}}$ and $\Gamma_{\text{E2}}$.

 Also shown in \fref{fig:2_point_E1_shift_width}, is the
 antisymmetric collective mode line shift $\delta\omega_{\text{a}}^{(1)}$.
 The resonance frequency
 of the magnetic dipole and electric
 quadrupole resonator $\Omega_{\text{a}}^{(1)}$,
 is related to the line shift of the antisymmetric mode
 of two parallel electric dipoles by
 \begin{equation}
 \Omega_{\text{a}}^{(1)}
 =
 (\Omega_0 - \delta\omega_{\text{a}}^{(1)})
 \,\text{.}
 \label{eq:omega_M1E2_approx}
 \end{equation}

 \subsubsection{Magnetic dipole moment
 of two parallel electric dipoles}
 \label{subsubsec:Magnetic_dipole_moment}

 In \sref{subsec:E1M1_general}
 and \Rref{JenkinsLongPRB}, the magnetic dipole moment
 arose solely due to the magnetization ${\bf M}_j({\bf r},t)$.
 We assume the resonator $j$ comprises two  electric dipoles
 located at ${\bf r}_{\pm,j} = [x_j, y_j\pm l_j/2, z_j]$,
 where $l_j\ll \lambda_0$,
 with linear charge and current oscillations along
 their axes.
 An effective magnetization ${\bf M}_{\text{P},j}({\bf r},t)$ is
 present due to the polarization of the two parallel electric dipoles, where
 \begin{equation}
 {\bf M}_{\text{P},j}({\bf r},t)=I_j(t){\bf \bar w}_j({\bf r})
 \,\text{,}
 \label{eq:M_effective}
 \end{equation}
 where ${\bf \bar w}_j({\bf r})$ is the effective current profile
 function.
 The scattered EM fields
 due to the effective magnetic source
 located at ${\bf r}'$, follow from the EM
 fields, \esref{eq:Esc} and~\eqref{eq:Bsc},
 with the effective magnetization
 \eref{eq:M_effective}.

 In the point magnetic dipole approximation,
 the spatial profile function ${\bf \bar w}_j({\bf r})$
 of the effective magnetization is approximated as a
 delta function at the origin
 \begin{align}
 {\bf \bar w}_j({\bf r})
 &
 =
 \frac{1}{2}\int\d^3r\,({\bf r}-{\bf r}_j)\times
 \big[{\bf p}_{+,j}({\bf r}) + {\bf p}_{-,j}({\bf r})\big]
 \nonumber
 \\
 &
 \simeq
 A_{\text{M},j}{\bf \hat m}_j\delta({\bf r} -{\bf r}_j)
 \,\text{.}
 \label{eq:bar_w_n}
 \end{align}
 The effective area $A_{\text{M},j}$, of the point magnetic dipole [see
 \fref{Drg:M1E2}] may
 be approximated by evaluating \eref{eq:bar_w_n} with
 ${\bf p}_{\pm,j}({\bf r}) =
 \pm H{\bf \hat d}\delta({\bf r}- {\bf r}_{\pm,j})$.
 We find the point magnetic dipole moment,
 of the $j$th pair of antisymmetrically excited point
 electric dipoles,
 has an effective area
 \begin{equation}
 A_{\text{M},j} = \frac{l_jH_j}{2}
 \,\text{.}
 \label{eq:A_{M},n}
 \end{equation}
 The magnetic dipole  decay rate is dependent upon the magnitude of the
 electric dipoles $H_j$ which comprise the pair,
 and their separation $l_j$, i.e.,\ their effective cross sectional area.

 If the two electric point dipoles are not
 extremely close to each other, the resonance frequency
 of the antisymmetric mode is close to
 that of the single isolated electric dipole. 
 We use the same resonance frequency
 (\eref{eq:omega}) in both \esref{eq:GammaE1} 
 and~\eqref{eq:GammaM1_temp}, together with the
 effective area of the magnetic dipole \eref{eq:A_{M},n},
 to find the ratio of the
 two emission rates is approximately given by
 \begin{equation}
 \Gamma_{\text{M1}}
 = \frac{\pi^2l_j^2}{\lambda^2}\Gamma_{\text{E1}}
 \,\text{.}
 \label{eq:GammaM1_GammaE1}
 \end{equation}

 \subsubsection{Electric quadrupole moment
 of two parallel electric dipoles}
 \label{subsubsec:Electric_quadrupole_moment}

 The effective area of the point electric
 quadrupole is obtained from a pair of out-of-phase point
 electric dipoles. We compare \eref{eq:q_alphabeta}
 (using the charge density of the electric dipoles, i.e.,
 $\rho_j({\bf r},t) = -Q\nabla\cdotp {\bf p}^{\text{d}}_j({\bf r})$)
 with \eref{eq:q_alphabeta_int2} to obtain the 
 effective area
 \begin{equation}
 A_{\alpha\beta,j} = 
 -\sum_{\pm}\int\d^3r'\, r_\alpha r_\beta
 \frac{\p}{\p r_\gamma}p_{\gamma}^{\text{d},\pm}({\bf r}')
 \,\text{,}
 \label{eq:A_alphabeta_intermediate}
 \end{equation}
 where ${\bf r}'={\bf r} -{\bf r}_j$ and the
 summation is over each of the $\pm$ orientated
 electric dipoles.
 Integrating \eref{eq:A_alphabeta_intermediate} by parts 
 results in Kronecker $\delta$ functions, see e.g., 
 \eref{eq:kronecker_delta}, and we find
 \begin{equation}
 A_{\alpha\beta,j} = \sum_{\pm}\int\d^3r'\,
 \Big[r'_\beta p^{\text{d},\pm}_{\alpha,j}({\bf r'})
 +
 r'_\alpha p^{\text{d},\pm}_{\beta,j}({\bf r'})\Big]
 \,\text{.}
 \label{eq:A_alphabeta}
 \end{equation}
 For an electric dipole  pair separated
 by $l$ along the $y$ axis (\fref{Drg:M1E2}), the dipoles are
 perpendicular to ${\bf \hat y}$,
 i.e.,\ ${\bf \hat d}_{j}^{\pm} = \pm{\bf \hat y}_\perp$.
 By symmetry, all elements of the tensor $A_{\alpha\beta,j}$ are
 zero, with the exception of  $A_{y\hat{d},j} = A_{\hat{d}y,j}$.
 For example, let the electric dipoles
 be aligned along the $x$ axis. Then,
 two antisymmetrically excited point electric dipoles located at
 ${\bf r}_{\pm,j} = [x_j, y_j\pm l_j/2,z_j]$, have mode functions
 ${\bf p}^{\text{d},\pm}_{j}({\bf r})
 =
 \pm H_j{\bf \hat x}\delta({\bf r} - {\bf r}_{\pm,j})$.
 The nonzero components of \eref{eq:A_alphabeta} are
 \begin{equation}
 A_{yx,j} =A_{xy,j}
 =
 \int\d^3r'\,
 \Big[
 r'_yp_{x,j}^{\text{d},+}({\bf r}')
 -
 r'_yp_{x,j}^{\text{d},-}({\bf r}')
 \Big]
  \,\text{.}
 \label{eq:Ayz}
 \end{equation}
 The integral in \eref{eq:Ayz} is carried out over the
 $\delta$ function, and summing over the
 indices $\alpha,\beta$ in \eref{eq:A_0}
 provides the effective area of our point electric quadrupole
 \begin{equation}
 A_{\text{E},j} = \sqrt{2} l_jH_j
 \,\text{.}
 \label{eq:A_{E},n}
 \end{equation}
 The electric quadrupole  radiative emission rate also
 depends on the amplitudes of the point electric dipoles and their separation.
 If the electric dipoles are symmetrically
 excited, then one may readily
 verify the point electric quadrupole moment vanishes.

 In the examples in this section, the radiative emission rates of the
 magnetic dipole and the electric quadrupole both depend the
 effective area of a pair of parallel electric dipoles. We assume that
 resonance frequencies of the magnetic dipole and
 electric quadrupole moments of the $j$th source
 are $\omega_j$, [\eref{eq:omega}]. The
 relative decay rate of the magnetic dipole and electric quadrupole,
 \eref{eq:GammaM1_GammaE2_general}, depends on
 the effective areas, $A_{\text{M},j}$
 and $A_{\text{E},j}$, \esref{eq:A_{M},n} and~\eqref{eq:A_{E},n},
 respectively. We find the relative radiation rates for our
 example are
 \begin{equation}
 \Gamma_{\text{E2}} = \frac{12}{5}\Gamma_{\text{M1}}
 \,\text{.}
 \label{eq:GammaM1_GammaE2}
 \end{equation}
The radiative emission rate of an isolated point electric quadrupole may be
 related to the point electric dipole through \eref{eq:GammaM1_GammaE1}.

 \subsection{Two interacting pairs of point electric dipoles}
 \label{subsec:two_pair_general}

\begin{figure}[h!]
 \centering
 \includegraphics[]{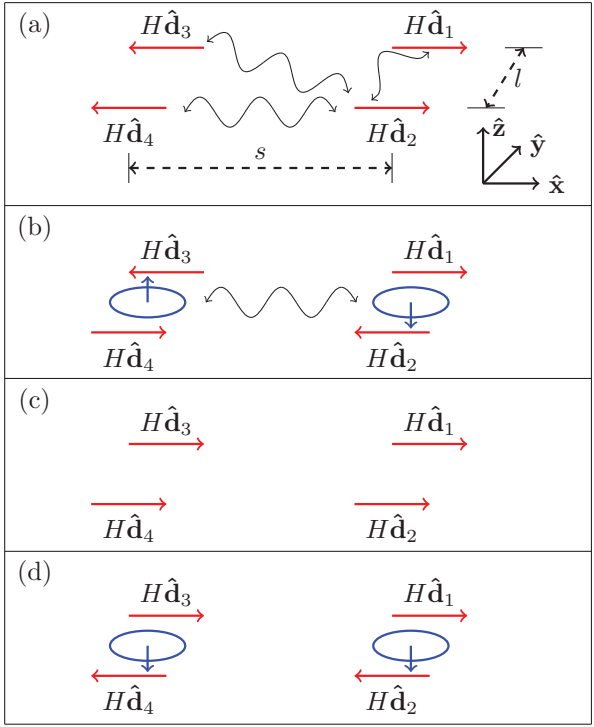}
 \caption[The eigenmodes of two horizontal pairs of
 point electric dipole resonators.]
 {\label{Drg_4E1_Horizontal_ModesCartoon}
 Schematic illustration of the eigenmodes of two horizontal
 pairs of parallel electric dipoles. The
 loops illustrate induced effective magnetic dipoles of each pair.
 The separation between the pairs is denoted
 by $s$ and the separation between the dipoles within each pair by $l$.
 We show the modes: (a)~E1a; (b)~M1E2a; (c)~E1s;
 and (d)~M1E2s.
 }
 \end{figure}

 In this section, we illustrate and test the
 effective point emitter model for the system
 of two pairs of parallel electric point dipoles by describing each pair
 by an effective scatterer possessing an
 electric dipole or a magnetic dipole and electric quadrupole moments.
 We consider two geometries: two horizontal parallel pairs
 [\fref{Drg_4E1_Horizontal_ModesCartoon}(a)]
 and two perpendicular  parallel pairs
 [\fref{Drg_4E1_Perpendicular_ModesCartoon}(a)].

 From the coupling matrix \eref{eq:C_E1_two_dipoles}
 we obtain four collective eigenmodes of current oscillation.
 We classify the modes as (see \fsref{Drg_4E1_Horizontal_ModesCartoon}
 and~\ref{Drg_4E1_Perpendicular_ModesCartoon}):
 antisymmetric electric dipoles (E1a);
 antisymmetric magnetic dipole--electric quadrupoles (M1E2a);
 symmetric electric dipoles (E1s);
 and symmetric magnetic dipole--electric quadrupoles (M1E2s).
 In the E1a and E1s modes each
 parallel pair of resonators forms an effective electric dipole.
 However,
 in the E1a mode the current oscillations  in the different parallel pairs
 are out of phase, and in the E1s mode
 they are in phase. In the M1E2a and M1E2s modes,
 the pairs form effective magnetic dipoles. 

 In the remainder of this section we will directly compare
 the E1a, E1s modes of the four 
 point electric dipoles (denoted by a superscript (1)) to 
 an corresponding modes of an 
 effective electric dipole model (denoted by a superscript (2s)).
 We also compare the M1E2a and M1E2s modes of the four 
 point electric dipoles (also denoted by a superscript (1)) 
 to an effective magnetic dipole-electric quadrupole model 
 (denoted by a superscript (2a)).

 \subsubsection{Description of electric dipoles by two multipole point emitters}
 \label{subsubsec:effective_multipole_interactions}

 In general, the $j$th pair of parallel electric dipoles are
 located at ${\bf r}_{\pm, j}= [x_j,y_j\pm l/2,z_j]$.
 When the separation $l$, between parallel electric dipoles is
 small, the $j$th pair may be approximated by a
 single resonator located at ${\bf r}_{j} = [x_j,y_j,z_j]$.

 When each parallel pair of electric dipoles
 are symmetrically excited [see, e.g.,
 \fref{Drg_4E1_Horizontal_ModesCartoon}(a,c)],
 each pair may be approximated by a
 single point
 electric dipole, located at the center of the pair.
 We introduced the interactions between two point electric
 dipoles in \sref{subsec:E1_example}. We may use similar
 analysis to model the symmetric (E1s and E1a) collective
 modes of two interacting pairs of electric dipoles. The coupling
 matrix in this case is given by \eref{eq:C_E1_two_dipoles};
 with $\Gamma^{(1)}\rightarrow\gamma_\text{s}^{(1)}$ and
 the driving field is tuned to the resonance
 frequency of the symmetrically excited pair of point
 electric dipoles, i.e, $\Omega_0=\Omega_\text{s}^{(1)}$. 
 The eigenvectors of the effective electric dipole interaction matrix
 are 
 \begin{equation}
 {\bf v}^{\text{(2s)}}_{\text{E1s}}
 =
 \frac{1}{\sqrt{2}}
 \begin{bmatrix}
 1 \\ 1
 \end{bmatrix}
 \quad
 \text{and}
 \quad
 {\bf v}^{\text{(2s)}}_{\text{E1a}}
 =
 \frac{1}{\sqrt{2}}
 \begin{bmatrix}
 1 \\ -1
 \end{bmatrix} 
 \,\text{.}
 \end{equation}
 with corresponding eigenvalues $\xi^{\text{(2s)}}_{\text{E1s}}$
 and $\xi^{\text{(2s)}}_{\text{E1a}}$, where
 \begin{align}
 \xi_{\text{E1a,E1s}}^{\text{(2s)}}
 = - \frac{\gamma_\text{s}^{(1)}}{2} \pm 
 \frac{3}{4}\Gamma_{\text{E1}}\mathcal{G}_{\text{E1}}({\bf r}_{12})
 \,\text{.}
 \end{align}
 Here, $\mathcal{G}_{\text{E1}}({\bf r}_{12})$ 
 is defined in \eref{eq:mathcalG_E1} and the argument 
 ${\bf r}_{12}$ depends on the locations of the 
 effective electric dipoles.

 On the other hand, we argued in \sref{subsec:antisymmetric_E1}  how
 a pair of antisymmetrically excited point electric
 dipoles can be approximated
 by a single point emitter possessing both
 magnetic dipole and electric quadrupole
 moments [see, e.g., \fref{Drg_4E1_Horizontal_ModesCartoon}(b,d)].
 For two point emitters located at ${\bf r}_1$ and
 ${\bf r}_2$, with both
 magnetic dipole and electric quadrupole moments, the interaction
 matrix, $\mathcal{C}$, is
 \begin{equation}
 \mathcal{C}
 =
 \Delta'
 -\frac{1}{2}\Upsilon
 +
 \frac{1}{2}
 \Big[
 i\mathcal{C}_{\text{M1}}
 +
 i\mathcal{C}_{\text{E2}}
 +
 \mathcal{C}_{\text{X2}}
 +
 \mathcal{C}_{\text{X2}}^T
 \Big]
 \,\text{.}
 \label{eq:C_example_M1E2}
 \end{equation}
 Similar to our example of two electric dipoles,
 the contributing matrices in \eref{eq:C_example_M1E2} are also
 $2\times2$.
 However, the off-diagonal elements of $\mathcal{C}$
 are more complicated. In the diagonal elements,
 the total decay rate of each
 emitter is $\gamma^{(1)}_\text{a}$,
 given in \eref{eq:gamma_a_M1E2_approx}.
 In general, the resonance frequency
 $\Omega^{(1)}_\text{a}\ne \omega_0$
 [see \eref{eq:omega_M1E2_approx}],
 thus $\Delta'$ is not trivial and contains a frequency shift.
 The matrix
 contributions to $\mathcal{C}$ are then
 \begin{align}
 \Delta' =
 &
 \begin{bmatrix}
 i\delta\omega_{\text{a}}^{(1)} & 0
 \\
 0 &  i\delta\omega_{\text{a}}^{(1)}
 \end{bmatrix}
 \, \text{,}
 \label{eq:Delta_M1E2_example}
 \\
 \Upsilon =
 &
 \begin{bmatrix}
  \gamma^{(1)}_\text{a} & 0
 \\
 0 &   \gamma^{(1)}_{\text{a}}
 \end{bmatrix}
 \, \text{,}
 \label{eq:Upsilon_M1E2_example}
 \\
 \mathcal{C}_{\text{M1}}
 =
 &
 \Gamma_{\text{M1}}
 \begin{bmatrix}
 0
 &
 G_{\text{M1}}({\bf r}_{12})
 \\
 G_{\text{M1}}({\bf r}_{12})
 &
 0
 \end{bmatrix}
 \,\text{,}
 \label{eq:C_M1_example}
 \\
 \mathcal{C}_{\text{E2}}
 =
 &
 \Gamma_{\text{E2}}
 \begin{bmatrix}
 0
 &
 G_{\text{E2}}({\bf r}_{12})
 \\
 G_{\text{E2}}({\bf r}_{12})
 &
 0
 \end{bmatrix}
 \, \text{,}
 \label{eq:C_E2_example}
 \\
 \mathcal{C}_{\text{X2}}
 =
 &
 \sqrt{\Gamma_{\text{M1}}\Gamma_{\text{E2}}}
 \begin{bmatrix}
 0
 &
  G_{\text{X2}}({\bf r}_{12})
 \\
 G_{\text{X2}}({\bf r}_{12})
 &
 0
 \end{bmatrix}
 \, \text{,}
 \label{eq:C_M1E2Cross_example}
 \end{align}
 where ${G}_{\text{M1}}({\bf r}_{12})$,
 ${G}_{\text{E2}}({\bf r}_{12})$ and
 ${G}_{\text{X2}}({\bf r}_{12})$ depend exclusively on
 the orientations and locations of the magnetic dipoles and
 electric quadrupoles and we have utilized
 the symmetry property of the matrices, e.g.,
 $[\mathcal{C}_{\text{M1}}]_{i,j}
 = [\mathcal{C}_{\text{M1}}]_{j,i}$, etc.
 The eigenvectors ${\bf v}_n^{(\text{2a})}$
 of \eref{eq:C_example_M1E2} are
 independent of the resonator locations, and correspond to
 symmetric and antisymmetric oscillations,
 \begin{equation}
 {\bf v}_\text{M1E2s}^{(\text{2a})} =
 \frac{1}{\sqrt{2}}
  \begin{bmatrix}
 1 \\
 1
 \end{bmatrix}
 \quad
 \text{and}
 \quad
 {\bf v}_\text{M1E2a}^{(\text{2a})} =
 \frac{1}{\sqrt{2}}
 \begin{bmatrix}
 1 \\
 -1
 \end{bmatrix}
 \,\text{.}
 \label{eq:eigenvectors_M1E2}
 \end{equation}
 However, the eigenvalues $\xi_{n}^{(\text{2a})}$
 of \eref{eq:C_example_M1E2}
 depend on both the orientations and locations of the
 resonators. In the following section, we
 analyze in detail the point magnetic dipole and
 electric quadrupole interacting systems, utilizing the geometries
 introduced in \fsref{Drg_4E1_Horizontal_ModesCartoon}
 and~\ref{Drg_4E1_Perpendicular_ModesCartoon}.

 \subsubsection{Two horizontal pairs of point electric dipoles}
 \label{subsubsection:two_horizontal_pairs}

 When the point electric dipoles are arranged in horizontal
 pairs, the Cartesian coordinates of the electric dipoles are
 \begin{equation}
 {\bf r}_{1,2} = \frac{1}{2}
 \begin{bmatrix} s\\  \pm l\\ 0 \end{bmatrix}
 \,\text{,} \quad %
 {\bf r}_{3,4} =  \frac{1}{2}
 \begin{bmatrix} -s\\ \pm  l\\ 0 \end{bmatrix}
 \,\text{,}
 \label{eq:locations_4RodInLine}
 \end{equation}
 The effective point emitters of these
 are then located at the center of each pair
 ${\bf r}_1 = -{\bf r}_2 = [s/2, 0, 0]$.

 The interaction terms:
 ${G}_{\text{M1}}({\bf r}_{12})$;
 ${G}_{\text{E2}}({\bf r}_{12})$; and
 ${G}_{\text{X2}}({\bf r}_{12})$, in
 \esref{eq:C_M1_example}--\eqref{eq:C_M1E2Cross_example},
 are given by
 \begin{align}
 G_{\text{M1}}({\bf r}_{12})
 &
 = \frac{i}{2}\left[2h_0^{(1)}(ks) - h_2^{(1)}(ks)\right]
 \,\text{,}
 \label{eq:G_M1_parallel_M1}
 \\
 G_{\text{E2}}({\bf r}_{12})
 &
 = -i\frac{5}{2}\Bigg[\frac{16}{35}h_4^{(1)}(ks)
 - \frac{3}{7}h_2^{(1)}(ks)
 + \frac{2}{5}h_0^{(1)}(ks)
 \Bigg]
 \,\text{,}
 \label{eq:G_E2_parallel_E2}
 \\
 G_{\text{X2}}({\bf r}_{12})
 &
 = -\sqrt{\frac{15}{2}}h_2^{(1)}(ks)
 \,\text{,}
 \label{eq:G_X2_parallel_E2M1}
 \end{align}
 respectively.
 In  this example, the eigenmodes correspond
 antisymmetric excitations
 (${\bf \hat m}_1 = -{\bf \hat m}_2$, and
 $\hat{A}_{\alpha\beta,1} = -\hat{A}_{\alpha\beta,2}$),
 see \fref{Drg_4E1_Horizontal_ModesCartoon}(b), and
 to symmetric excitations of the resonators
 (${\bf \hat m}_1 = {\bf \hat m}_2$, and
 $\hat{A}_{\alpha\beta,1} = \hat{A}_{\alpha\beta,2}$),
 see \fref{Drg_4E1_Horizontal_ModesCartoon}(d).
 These eigenmodes are represented
 by the eigenvectors ${\bf v}_{\text{M1E2a}}^{(\text{2a})}$
 and ${\bf v}_{\text{M1E2s}}^{(\text{2a})}$,
 respectively,
 given in \eref{eq:eigenvectors_M1E2}.
 The eigenvalues $\xi_{\text{M1E2a}}^{(\text{2a})}$ 
 and $\xi_{\text{M1E2s}}^{(\text{2a})}$,
 of \eref{eq:C_example_M1E2} with
 \esref{eq:G_M1_parallel_M1}--\eqref{eq:G_X2_parallel_E2M1},
 are complicated and include contributions
 from $h_4^{(1)}(ks)$, $h_2^{(1)}(ks)$,
 and $h_0^{(1)}(ks)$.
 In the leading order expansion of the
 spherical Hankel functions, the real and imaginary
 parts of $\xi_{\text{M1E2a}}^{(\text{2a})}$ 
 and $\xi_{\text{M1E2s}}^{(\text{2a})}$
 are dominated by $h_0^{(1)}(ks)$
 and $h_4^{(1)}(ks)$, respectively. Specifically, we find
 \begin{subequations}
 \begin{align}
 \gamma_\text{M1E2a}^{(\text{2a})}
 =
 &
 \text{Re}\big(\xi_{\text{M1E2a}}^{\text{(2a)}}\big)
 \approx
 -\frac{\gamma_{\text{a}}^{\text{(1)}}}{2}
 +
 \frac{1}{2}
 \Big[ \Gamma_{\text{M1}}+\Gamma_{\text{E2}}\Big]
 \,\text{,}
 \label{eq:real_eigenvectors_parallel_pair_a}
\\
 \gamma_\text{M1E2s}^{(\text{2a})}
 =
 &
 \text{Re}\big(\xi_{\text{M1E2s}}^{\text{(2a)}}\big)
 \approx
 -\frac{\gamma_{\text{a}}^{\text{(1)}}}{2}
 -
 \frac{1}{2}
 \Big[ \Gamma_{\text{M1}}+\Gamma_{\text{E2}}\Big]
 \,\text{,}
 \label{eq:real_eigenvectors_parallel_pair_s}
 \\
 \delta\omega^{\text{(2a)}}_\text{M1E2a}
 &
 = \text{Im}\big(\xi_{\text{M1E2a}}^{\text{(2a)}}\big)
 \approx
 \delta\omega_\text{a}^{(1)}
 -
 \frac{240}{(ks)^5}\Gamma_{\text{E2}}
 \,\text{,}
 \label{eq:imag_eigenvectors_parallel_pair_a}
 \\
 \delta\omega^{\text{(2a)}}_\text{M1E2s}
 &
 = \text{Im}\big(\xi_{\text{M1E2s}}^{\text{(2a)}}\big)
 \approx
 \delta\omega_\text{a}^{(1)}
 +
 \frac{240}{(ks)^5}\Gamma_{\text{E2}}
 \,\text{.}
 \label{eq:imag_eigenvectors_parallel_pair_s}
 \end{align}
 \end{subequations}
 Here, we have the antisymmetric collective mode decay rate
 $\gamma_\text{M1E2a}^{(\text{2a})}$ is subradiant
 approaching $\Gamma_{\text{O}}$, while
 the symmetric excitation is superradiant with
 $\gamma_{\text{M1E2s}}^{(\text{2a})}$
 approaching $2\gamma^{(1)}_{\text{a}}$.
 When the resonators are close together, the line shifts
 of the collective modes are dominated by $\Gamma_{\text{E2}}$
 and quickly diverge,  with $\delta\omega^{\text{(2a)}}_\text{M1E2a}$
 red shifted,
 and $\delta\omega^{\text{(2a)}}_\text{M1E2s}$ blue shifted,
 from $\delta\omega_\text{a}^{(1)}$.

 \begin{figure}[h!]
 \centering
 \includegraphics[]
 {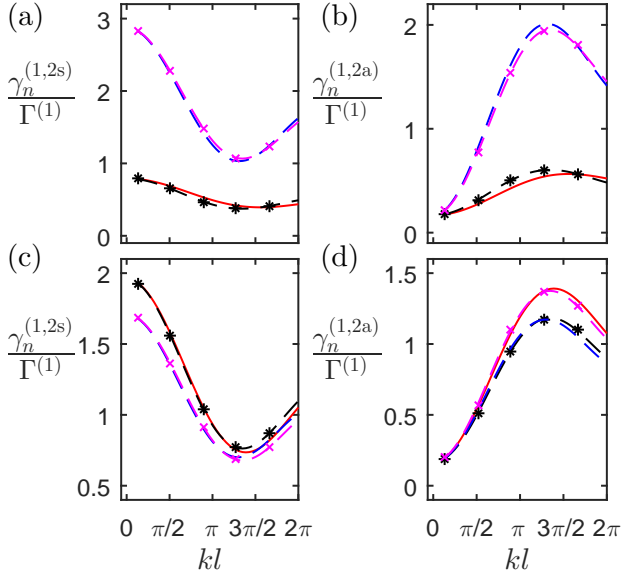}
 \caption[The radiative resonance linewidths of two horizontal
 pairs of point electric dipoles and their corresponding
 effective multipoles as
 a function of $l$.]
 {\label{fig:PointMultipole_horizontal_width_vary_y}
 The radiative resonance
 linewidth $\gamma_n^{(\text{1,2s,2a})}$
 for the collective eigenmodes as a
 function of the separation of dipoles within each pair $l$,
 for two horizontal pairs
 of point electric dipoles, with:~(a) and~(b) $ks=2\pi/3$;
 and~(c) and~(d) $ks=2\pi$.
 We show the linewidth $\gamma_n^{(1)}$
 in the $N=4$
 point electric dipole model, with the different modes
 shown as: E1a--red solid line~(a) and~(c);
 E1s--blue dashed line~(a) and~(c);
 M1E2a--red solid line~(b) and~(d);
 and M1E2s--blue dashed line~(b) and~(d).
 We show the linewidth $\gamma_n^{(\text{2s})}$
 in the $N=2$
 effective electric dipole resonator model:
 antisymmetric excitations--magenta dash circles~(a) and~(c);
 and
 symmetric excitations--black dash squares~(a) and~(c).
 The linewidth $\gamma_n^{(\text{2a})}$
 in the $N=2$ effective magnetic dipole--electric
 quadrupole resonator model:
 antisymmetric excitations--magenta dash circles~(b) and~(d);
 and symmetric excitations--black dash squares~(b) and~(d).
 The radiative losses of
 each electric dipole are
 $\Gamma_{\text{E1}}=0.83\Gamma^{\text{(1)}}$,
 the ohmic losses are
 $\Gamma_{\text{O}}=0.17\Gamma^{\text{(1)}}$.
 }
 \end{figure}

\begin{figure}[h!]
 \centering
 \includegraphics[]
 {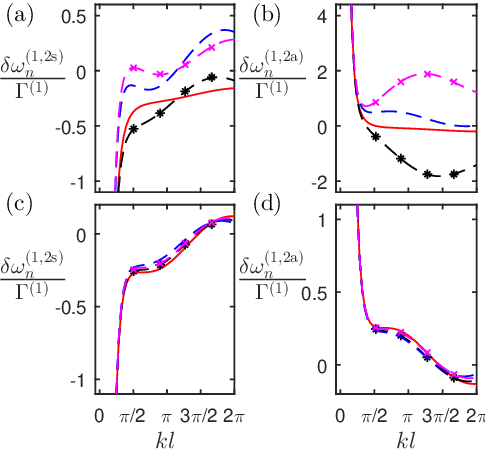}
 \caption[The radiative resonance line shifts of two horizontal
 pairs of point electric dipoles and their corresponding
 effective multipoles as
 a function of $l$.]
 {\label{fig:PointMultipole_horizontal_shift_vary_y}
 The radiative resonance
 line shift $\delta\omega_n^{(\text{1,2s,2a})}$
 for the collective eigenmodes as a
 function of the separation parameter $l$,
 for two horizontal pairs
 of point electric dipoles, with:~(a) and~(b) $ks=2\pi/3$;
 and~(c) and~(d) $ks=2\pi$.
 For plot descriptions, see
 \fref{fig:PointMultipole_horizontal_width_vary_y}
 caption.
 }
 \end{figure}

 In \fsref{fig:PointMultipole_horizontal_width_vary_y}
 and~\ref{fig:PointMultipole_horizontal_shift_vary_y},
 we show how the collective mode linewidths and line shifts,
 respectively, for the $N=4$
 interacting point electric dipoles, vary with the
 separation of dipoles within each pair $l$, when
 $ks=2\pi/3$ and $ks=2\pi$. Also,
 in \fsref{fig:PointMultipole_horizontal_width_vary_y}
 and~\ref{fig:PointMultipole_horizontal_shift_vary_y}, we
 show the collective mode linewidths  and line shifts for the
 effective $N=2$ multipole point emitters.

 When $l$ is varied, the linewidths $\gamma_n^{(\text{2s,2a})}$
 of the $N=2$ effective multipole point emitters  closely approximate
 the corresponding linewidths $\gamma_n^{(1)}$
 of the $N=4$ point electric dipoles;
 both when $ks=2\pi/3$ and $ks=2\pi$. When $s$ is
 small [$ks=2\pi/3$,
 see \fref{fig:PointMultipole_horizontal_width_vary_y}(a)
 and~\ref{fig:PointMultipole_horizontal_width_vary_y}(b)],
 the E1s mode is always superradiant, while
 both the E1a and M1E2a modes are always subradiant.  When
 $s$ is small, the M1E2s mode exhibits both  superradiant and
 subradiant behavior. For large
 $kl\gtrsim \pi/2$, we have superradiant behavior
 $\gamma_{\text{M1E2s}}^{(1)}> \Gamma^{(1)}$
 and
 $kl\lesssim \pi/2$, subradiant behavior;
 $\gamma_{\text{M1E2s}}^{(1)}<\Gamma^{(1)}$.
 At the minimum value
 $\gamma_{\text{E1s}}^{(1)}\simeq 1.1\Gamma^{(1)}$.
 For large $s$
[see \fref{fig:PointMultipole_horizontal_width_vary_y}(c)
 and~\ref{fig:PointMultipole_horizontal_width_vary_y}(d)],
 all the linewidths exhibit both superradiant
 and subradiant behavior.

 While the collective mode linewidths resulting from the
 effective point emitter approximation qualitatively match those of the
 electric dipoles as $l$ varies at both large and small $s$, the
 corresponding line shifts show greater variations. In  particular,
 when $ks=2\pi/3$, see
 \fref{fig:PointMultipole_horizontal_shift_vary_y}(a)
 and~\ref{fig:PointMultipole_horizontal_shift_vary_y}(b),
 the collective line shifts $\delta\omega^{\text{(2a)}}_\text{M1E2a}$
 and $\delta\omega^{(\text{2a})}_\text{M1E2s}$ begin to deviate
 from the corresponding line shifts
 $\delta\omega^{(1)}_\text{M1E2a}$ and
 $\delta\omega^{(1)}_\text{M1E2s}$, when $kl\simeq\pi/4$.
 When $kl\simeq \pi/4$, all the collective mode line shifts
 begin to significantly diverge as $l$ reduces further.
 In contrast, the E1a and E1s line shifts of the point electric dipole
 model qualitatively agree with the corresponding shifts of the
 effective multipole resonator model. When
 $s$ is large and $l$ is varied, there is no significant
 difference in the line shifts, even when $l$ is large. As
 we reduce the separation $l$, the line shifts
 $\delta\omega_\text{E1a}^{(1)}$ and
 $\delta\omega_\text{E1s}^{(1)}$ are red shifted from $\Omega_0$,
 see \fref{fig:PointMultipole_horizontal_shift_vary_y}(a)
 and~\ref{fig:PointMultipole_horizontal_shift_vary_y}(c).
 In contrast, $\delta\omega_{\text{M1E2a}}^{(1)}$
 and $\delta\omega_{\text{M1E2s}}^{(1)}$ are blue shifted,
 see \fref{fig:PointMultipole_horizontal_shift_vary_y}(b)
 and~\ref{fig:PointMultipole_horizontal_shift_vary_y}(d).

\begin{figure}[h!]
 \centering
 \includegraphics[]
 {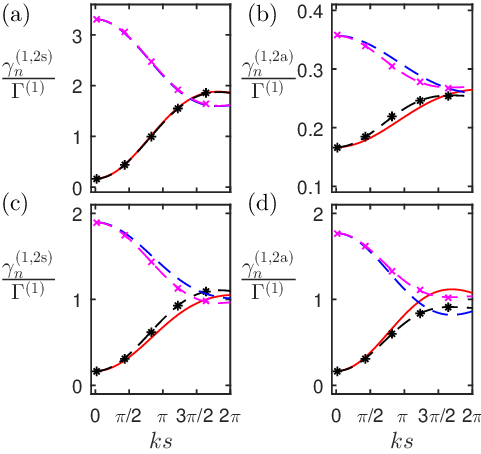}
 \caption[The radiative resonance linewidths of two horizontal
 pairs of point electric dipoles and their corresponding
 effective multipoles as
 a function of $s$.]
 {\label{fig:PointMultipole_horizontal_width_vary_s}
 The resonance
 linewidth $\gamma_n^{(\text{1,2s,2a})}$
 for the collective eigenmodes as a
 function of the separation parameter $s$,
 for two horizontal pairs
 of point electric dipoles, with:~(a) and~(b) $kl=\pi/4$;
 and~(c) and~(d) $kl=2\pi$.
 For plot descriptions,
 see \fref{fig:PointMultipole_horizontal_width_vary_y} caption.
 }
 \end{figure}

  In \fref{fig:PointMultipole_horizontal_width_vary_s}
 we show analogous linewidths when the
 separation $s$ between the pairs is varied.
 The linewidths of the effective model again
 agree well with the full point dipole results.
 The antisymmetric mode linewidth approaches
 the nonradiative loss rate when both $l$ and $s$ are small
 $\gamma_{\text{M1E2s}}^{(1)}=\gamma_\text{M1E2s}^{(\text{2a})}
 \simeq \Gamma_\text{O}$, see
 \fref{fig:PointMultipole_horizontal_width_vary_s}(b).

 \begin{figure}[h!]
 \centering
 \includegraphics[]
 {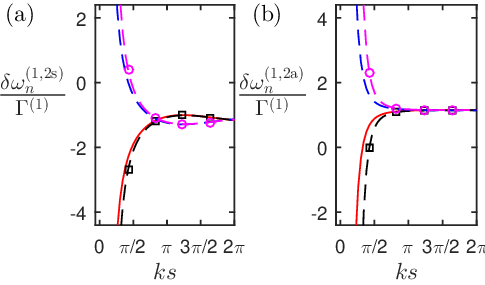}
 \caption[The resonance line shifts of two horizontal
 pairs of point electric dipoles and their corresponding
 effective multipoles as
 a function of $s$.]
 {\label{fig:PointMultipole_horizontal_shift_vary_s}
 The resonance
 line shift $\delta\omega_n^{(\text{1,2s,2a})}$
 for the collective eigenmodes as a
 function of the separation parameter $s$,
 for two horizontal pairs
 of point electric dipoles, with:~(a) and~(b) $kl=\pi/4$.
 For plot descriptions,
 see \fref{fig:PointMultipole_horizontal_shift_vary_y} caption.
 }
 \end{figure}

 In \fref{fig:PointMultipole_horizontal_shift_vary_s},
 we show the line shifts of the different collective modes
 as $s$ varies. The line shifts $\delta\omega_\text{M1E2a}^{(\text{2s})}$
 and $\delta\omega_{\text{M1E2s}}^{\text{(2s)}}$, show
 no significant deviation from the E1a and E1s modes, even
 at small $s$ and $kl=\pi/4$. In contrast,
 the line shifts $\delta\omega_\text{M1E2a}^{\text{(2a)}}$
 and $\delta\omega_\text{M1E2s}^{\text{(2a)}}$ begin
 to deviated from the M1E2a and M1E2s line shifts when
 $ks\simeq \pi/2$ (when $kl=\pi/4$).

 \subsubsection{Two perpendicular pairs of point electric dipoles}
 \label{subsubsection:two_perp_pairs}

\begin{figure}[h!]
 \centering
 \includegraphics[]{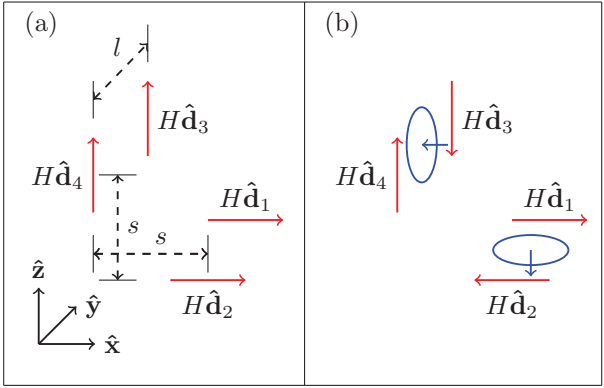}
 \caption[The eigenmodes of four perpendicular electric dipole resonators.]
 {\label{Drg_4E1_Perpendicular_ModesCartoon}
 A schematic illustration of two perpendicular
 pairs of electric dipoles and the induced effective magnetic dipoles.
 The separation between dipoles within each pair is $l$ and the
 position of the
 pair on the $x,z$ axis is determined by $s$.
 We illustrate two of the four modes;
 see \fref{Drg_4E1_Horizontal_ModesCartoon}.
 }
 \end{figure}

 Our second example is two perpendicular pairs
 of point electric dipoles. The Cartesian coordinates
 of each electric dipole are
 \begin{equation}
 {\bf r}_{1,2} =
 \begin{bmatrix} s\\  \pm  \dfrac{l}{2}\\ 0 \end{bmatrix}
 \, \text{,}\quad %
 {\bf r}_{3,4} =
 \begin{bmatrix} 0\\ \pm \dfrac{l}{2}\\ s \end{bmatrix}
 \,\text{.}
 \label{eq:locations_4RodPerpendicular}
 \end{equation}
 In \fref{Drg_4E1_Perpendicular_ModesCartoon}(a)
 and~\ref{Drg_4E1_Perpendicular_ModesCartoon}(c),
 we show the E1a and E1s modes, respectively. In
 \fref{Drg_4E1_Perpendicular_ModesCartoon}(b)
 and~\ref{Drg_4E1_Perpendicular_ModesCartoon}(d),
 we show the M1E2a and M1E2s modes, respectively.
 The locations vectors of the effective resonators with both
 point magnetic dipole and point electric quadrupole
 sources [and effective electric dipole sources]
 are ${\bf r}_1 = [s, 0, 0]$ and ${\bf r}_2 = [0, 0, s]$.

 In this example, the magnetic dipole orientation vectors
 ${\bf \hat m}_{1,2}$ and the unit vector ${\bf \hat k}$ form an
 orthonormal set, i.e.,
 ${\bf \hat m}_1  = \pm {\bf \hat k}\times {\bf \hat m}_2$,
 corresponding to symmetric ($+$) and antisymmetric ($-$)
 oscillations,
 see \fref{Drg_4E1_Perpendicular_ModesCartoon}(d),
 and~\ref{Drg_4E1_Perpendicular_ModesCartoon}(b), respectively.
 Similarly, the electric quadrupole unit tensors, for the
 symmetric ($+$) and antisymmetric ($-$) excitations are
 \begin{equation}
 \hat{A}_{1} =
 \begin{bmatrix}
 0 & 1 & 0
 \\
 1 & 0 & 0
 \\
 0 & 0 & 0
 \end{bmatrix}
 \quad
 \text{and}
 \quad
 \hat{A}_{2} =
 \pm
 \begin{bmatrix}
 0 & 0 & 0
 \\
 0 & 0 & 1
 \\
 0 & 1 & 0
 \end{bmatrix}
 \,\text{.}
 \label{eq:A_perp}
 \end{equation}
 These oscillations are represented through
 the eigenvectors ${\bf v}_{\text{M1E2s}}^\text{(2a)}$ 
 and ${\bf v}_{\text{M1E2a}}^\text{(2a)}$,
 respectively, see \eref{eq:eigenvectors_M1E2}.
 The interaction terms:
 ${G}_{\text{M1}}({\bf r}_{12})$;
 ${G}_{\text{E2}}({\bf r}_{12})$; and
 ${G}_{\text{X2}}({\bf r}_{12})$, in
 \esref{eq:C_M1_example}--\eqref{eq:C_M1E2Cross_example}, respectively,
 in this case  are given by
 \begin{align}
 G_{\text{M1}}({\bf r}_{12})
 &
 = i\frac{3}{4} h_2^{(1)}(\sqrt{2}ks)
 \,\text{,}
 \label{eq:G_M1_perp_M1}
 \\
 G_{\text{E2}}({\bf r}_{12})
 &
 = i\frac{20}{28}\Bigg[h_4^{(1)}(\sqrt{2}ks)
 - \frac{3}{4}h_2^{(1)}(\sqrt{2}ks)
 \Bigg]
 \,\text{,}
 \label{eq:G_E2_perp_E2}
 \\
 G_{\text{X2}}({\bf r}_{12})
 &
 = -\sqrt{\frac{15}{4}}h_2^{(1)}(\sqrt{2}ks)
 \,\text{.}
 \label{eq:G_X2_perp_E2M1}
 \end{align}
 The eigenvalues $\xi_\text{M1E2a}^{\text{(2a)}}$
 and $\xi_\text{M1E2s}^{\text{(2a)}}$
 of \eref{eq:C_example_M1E2}, with
 \esref{eq:G_M1_perp_M1}--\eqref{eq:G_X2_perp_E2M1} are
 complex, involving contributions from $h_4^{(1)}(\sqrt{2}ks)$
 and $h_2^{(1)}(\sqrt{2}ks)$. For analytical expressions
 of $\xi_{\text{M1E2a}}^{\text{(2a)}}$ 
 and $\xi_{\text{M1E2s}}^{\text{(2a)}}$,
 we again consider the leading order expansions of the
 spherical Hankel functions.
 In this limit $\text{Im}(\xi_{n}^{\text{(2a)}})$ is dominated by
 $h_4^{(1)}(\sqrt{2}ks)$ and $\text{Re}(\xi_{n}^{\text{(2a)}})$
 by $h_2^{(1)}(\sqrt{2}ks)$. Specifically, we find
 \begin{subequations}
 \begin{align}
 \delta\omega^{\text{(2a)}}_\text{M1E2a}
 =
 &\,
 \text{Im}\big(\xi_{\text{M1E2a}}^{\text{(2a)}}\big)
 \approx
 \delta\omega_{\text{a}}^{(1)}
 +
 \frac{1125}{(\sqrt{2}ks)^5}\Gamma_{\text{E2}}
 \,\text{,}
 \label{eq:imag_eigenvalue_perp_pair_a}
 \\
 \delta\omega^{\text{(2a)}}_\text{M1E2s}
 =
 &\,
 \text{Im}\big(\xi_{\text{M1E2s}}^{\text{(2a)}}\big)
 \approx
 \delta\omega_{\text{a}}^{(1)}
 -
 \frac{1125}{(\sqrt{2}ks)^5}\Gamma_{\text{E2}}
 \,\text{,}
 \label{eq:imag_eigenvalue_perp_pair_s}
 \\
 \gamma^{\text{(2a)}}_\text{M1E2a}
 =
 &\,
 \text{Re}\big(\xi_{\text{M1E2a}}^{\text{(2a)}}\big)
 \approx
 - \frac{\gamma^{(1)}_{\text{a}}}{2}
 +
 \bigg[
 \frac{\Gamma_{\text{M1}}}{10}
 +
 \frac{\Gamma_{\text{E2}}}{28}
 \nonumber
 \\
 &\quad
 +
 \sqrt{\frac{15}{4}}\sqrt{\Gamma_{\text{M1}}\Gamma_{\text{E2}}}
 \bigg]
 (\sqrt{2}ks)^2
 \,\text{,}
 \label{eq:real_eigenvalue_perp_pair_a}
 \\
 \gamma^{\text{(2a)}}_\text{M1E2s}
 =
 &\,
 \text{Re}\big(\xi_{\text{M1E2s}}^{\text{(2a)}}\big)
 \approx
 - \frac{\gamma^{(1)}_{\text{a}}}{2}
 -
 \bigg[
 \frac{\Gamma_{\text{M1}}}{10}
 +
 \frac{\Gamma_{\text{E2}}}{28}
 \nonumber
 \\
 &\quad
 +
 \sqrt{\frac{15}{4}}\sqrt{\Gamma_{\text{M1}}\Gamma_{\text{E2}}}
 \bigg]
 (\sqrt{2}ks)^2
 \,\text{.}
 \label{eq:real_eigenvalue_perp_pair_s}
 \end{align}
 \end{subequations}
 When the dipoles are close and perpendicular,
 they interact only weakly,
 $\gamma_{\text{M1E2a}}^{\text{(2a)}}\approx
 \gamma_{\text{M1E2s}}^{\text{(2a)}}\approx \gamma^{(1)}_\text{a}$.
 The line shifts of the modes diverge as $ks \rightarrow 0$ and are dominated
 by $\Gamma_{\text{E2}}$, with
 $\delta\omega_{\text{M1E2a}}^{\text{(2a)}}$
 blue shifted and $\delta\omega_{\text{M1E2s}}^{\text{(2a)}}$ red
 shifted from $\delta\omega_\text{a}^{(1)}$.

 \begin{figure}[h!]
 \centering
 \includegraphics[]
 {Figure_10.eps}
 \caption[The radiative resonance linewidths of two perpendicular
 pairs of point electric dipoles and their corresponding
 effective multipoles as
 a function of $l$.]
 {\label{fig:PointMultipole_perp_width_vary_y}
 The radiative resonance
 linewidth $\gamma$
 for the collective eigenmodes as a
 function of the separation parameter $l$,
 for two perpendicular pairs
 of point electric dipoles, with:~(a) and~(b) $ks=2\pi/5$.
 For plot descriptions, see
 \fref{fig:PointMultipole_horizontal_width_vary_y} caption
 }
 \end{figure}

 \begin{figure}[h!]
 \centering
 \includegraphics[]
 {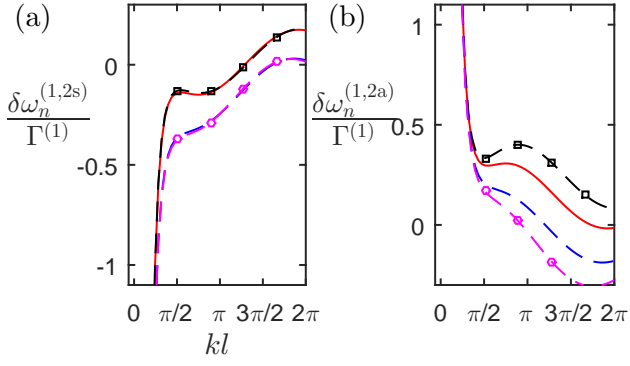}
 \caption[The radiative resonance line shifts of two perpendicular
 pairs of point electric dipoles and their corresponding
 effective multipoles as
 a function of $l$.]
 {\label{fig:PointMultipole_perp_shift_vary_y}
 The radiative resonance
 line shift $\delta\omega$
 for the collective eigenmodes as a
 function of the separation parameter $l$,
 for two perpendicular pairs
 of point electric dipoles, with:~(a) and~(b) $ks=2\pi/5$.
 For plot descriptions, see
 \fref{fig:PointMultipole_horizontal_shift_vary_y} caption
 }
 \end{figure}

 In \fsref{fig:PointMultipole_perp_width_vary_y}
 and~\ref{fig:PointMultipole_perp_shift_vary_y}, we
 show how the collective mode linewidths and
 line shifts, respectively, for situations similar to those of the parallel dipoles of
\fsref{fig:PointMultipole_horizontal_width_vary_y}
 and~\ref{fig:PointMultipole_horizontal_shift_vary_y},
 When $ks=2\pi/5$, and $l$ is varied, the linewidths
 of all the collective modes of perpendicular
 pairs closely resemble the linewidths of the corresponding horizontal
 pairs, see \fsref{fig:PointMultipole_perp_width_vary_y}(a,b)
 and~\ref{fig:PointMultipole_horizontal_width_vary_y}(c,d).
 There is no significant difference in the linewidths between
 the two different models, even when $s$ is small.

\begin{figure}[h!]
 \centering
 \includegraphics[]
 {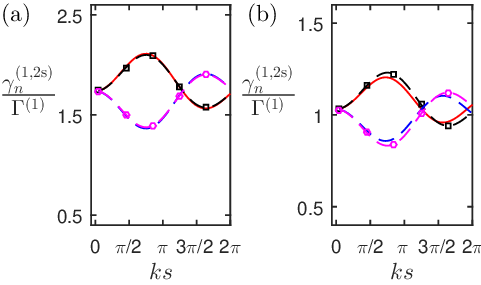}
 \caption[The resonance linewidths of two perpendicular
 pairs of point electric dipoles and their corresponding
 effective multipoles as
 a function of $s$.]
 {\label{fig:PointMultipole_perp_width_vary_s}
 The resonance
 linewidth $\gamma$
 for the collective eigenmodes as a
 function of the separation parameter $s$,
 for two perpendicular pairs
 of point electric dipoles with~(a) $kl=\pi/4$
 and~(b) $kl=2\pi$.
 For plot descriptions, see
 \fref{fig:PointMultipole_horizontal_width_vary_y} caption.
 }
 \end{figure}

  When we vary $s$
 (\fref{fig:PointMultipole_perp_width_vary_s}), the perpendicular dipoles
 collective mode linewidths
 exhibit different behavior to those of
 horizontal dipoles, displaying
 characteristic oscillations as a function of the separation $s$.
 The effective multipole model provides a
 good approximation of the corresponding
 point electric dipole model linewidths.
 The line shifts of the perpendicular pairs have very similar
 characteristics to horizontal pairs, both when $s$ is large
 and when $s$ is small.

 \subsection{The response of an effective
 point emitter model to external fields}

 In this section, we compare the response
 of the four point electric dipole system with
 that of the effective two point emitter model
 under external driving, when we approximate
 the effective point emitter model with only one eigenmode.
 We consider the antisymmetric excitations, in
 which case the point emitter model exhibits
 the magnetic dipole and electric quadrupole moments. The driven
 dipoles radiate and induce excitations on
 the nearby dipoles, resulting in a strongly coupled system.

 We solve the equation of motion \eref{eq:equmot}
 in a  steady-state ($\dot{\bf b}=0$)
 for horizontal pairs of electric dipoles.
 We focus on the case when the external
 EM field drives one pair of dipoles only
 and propagates in the direction normal to the pair.
 For simplicity, we assume that the field
 perfectly couples to the antisymmetric excitation of the pair.
 In the point electric dipole system, we drive
 the pair 12,  formed by the dipoles $n=1,2$.
 The driving by  incident fields,
 ${\bf F}_{\text{in}}$ in
 \eref{eq:equmot}, that takes the form
 \begin{equation}
 {\bf F}_{\text{in}} =
 \frac{F_0}{\sqrt{2}}
 \begin{bmatrix}
 1 \\ -1 \\ 0 \\ 0
 \end{bmatrix}
 \,\text{.}
 \label{eq:f_drivingM1E2}
 \end{equation}
 We only take the antisymmetric mode
 for the $N=2$ effective point emitter system that
 exhibits magnetic dipole and
 electric quadrupole moments.  The incident driving takes the form
 \begin{equation}
 {\bf F}_{\text{in}} =
 F_{0}
 \begin{bmatrix}
 1\\
 0
 \end{bmatrix}
 \,\text{.}
 \end{equation}

 The coupling matrix $\mathcal{C}$ in
 \eref{eq:equmot} is non-Hermitian, but we
 can define an occupation measure
 for a particular eigenmode ${\bf v}_n$ in an excitation ${\bf b}$ by
 \begin{equation}
{O}_n({\bf b})
 \equiv
 \left|{\bf v}_n^T{\bf b}\right|^2
 \,\text{.}
 \label{eq:b_occupation}
 \end{equation}
 For the four electric dipoles, we project the excitation onto the basis
 \begin{equation}
 {\bf v}_{12}^{(\pm)}
 =
 \frac{1}{\sqrt{2}}
 \begin{bmatrix}
 1 \\ \pm 1 \\ 0 \\ 0
 \end{bmatrix}
 \,\text{,}
 \quad
  {\bf v}_{34}^{(\pm)} =
 \frac{1}{\sqrt{2}}
 \begin{bmatrix}
 0 \\ 0 \\ 1 \\ \pm 1
 \end{bmatrix}
 \,\text{.}
 \label{eq:E1NewBasis}
 \end{equation}
 Here, the superscript $(\pm)$ indicates
 symmetric/antisymmetric excitations of the
 dipole pair.
 In the  effective two-emitter model we use the basis
 \begin{equation}
 {\bf v}_1^{(-)}
 =
 \begin{bmatrix}
 1\\0
 \end{bmatrix}
 \,\text{,}
 \qquad
 {\bf v}_2^{(-)}
 =
 \begin{bmatrix}
 0 \\ 1
 \end{bmatrix}
 \,\text{.}
 \label{eq:M1E2NewBasis}
 \end{equation}

\begin{figure}
 \centering
 \includegraphics[]{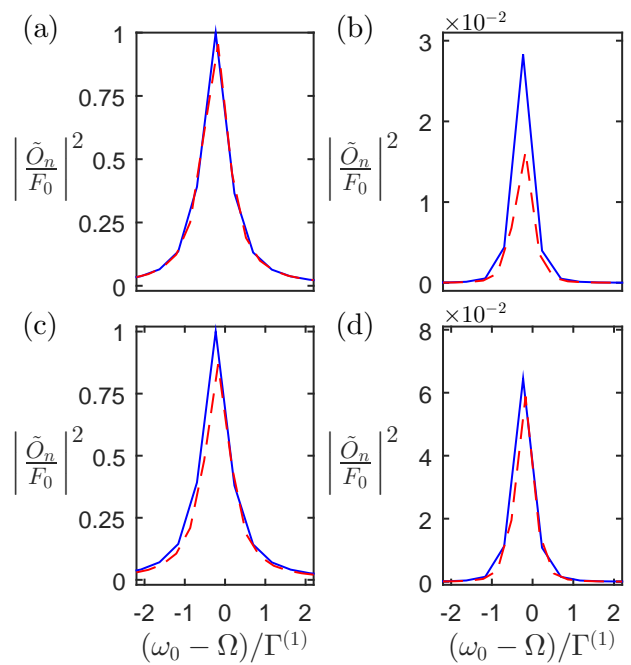}
 \caption{
 The excitation spectra of the antisymmetric
 modes of the electric dipole pairs (a,c) 1 and 2;
 (b,d) 3 and 4, as a function of
 the detuning of the incident frequency $\Omega$
 from the resonance frequency $\omega_0$ of a
 single point electric dipole. The resonance is at
 $-(\omega_0 - \Omega) \simeq 0.25 \Gamma^{(1)}$.
 The full four point electric dipole model (blue solid line)
 and the corresponding effective
 two-point-emitter model, exhibiting  magnetic
 dipole and electric quadrupole moments (red dashed line).
 In~(a,b)  $ks = 4\pi/3$ and $kl = \pi/3$,
 (c,d)   $ks = 8\pi/11$ and $kl = 8\pi/7$.
 }
 \label{fig:occupations}
 \end{figure}

 In \fref{fig:occupations}, we show the excitation spectra of the
 antisymmetric modes for the two models.
 For our choice of the driving in \eref{eq:f_drivingM1E2},
 the symmetric excitations
 of the electric dipoles are negligible.
 We find that, despite the inclusion of
 only one mode, the effective model agrees
 with the four-dipole case, provided that neither $s$ nor $l$ is too small.
 For small $s$ the geometry of the
 configuration starts becoming important,
 while for small $l$, the contribution of the
 symmetric excitations would need to be included.
 For $ks = 4\pi/3$ and $kl = \pi/3$ (a,b),
 the effective model underestimates the
 excitation of the non-driven pair, while
 for $ks = 8\pi/9$ and $kl = 4\pi/9$ (c,d) the agreement is better.

 %
 \section{Conclusions}
 \label{sec:conclusions}

 We have developed a formalism for
 effective point scatterer models that
 goes beyond the electric and magnetic
 dipole approximations, and also includes the more complicated
 electric quadrupole contributions to
 the interactions between the resonators.
 The resulting theory can then be expressed
 as a coupled set of dynamical equations for the resonators
 and EM fields. The interactions between
 the resonators result in collective eigenmodes
 with associative collective resonance frequency shifts and linewidths.

 There is a clear motivation for introducing
 discrete models for the studies of
 EM field responses. For closely-spaced
 resonant emitters the EM-field-mediated
 interactions can be strong. The
 combination of recurrent scattering~\cite{Wiersma_recurrent,
 Ruostekoski1997a,
 Morice1995a} -- a process in which a wave
 is scattered more than once by the
 same emitter -- and position-dependent radiative
 coupling between the emitters can lead
 to a correlated EM-field response. Even in a randomly
 distributed ensemble of emitters, such
 correlations have been shown to result in a qualitative failure
 of standard homogeneous-medium
 electrodynamics that, by construction,
 is a mean-field approximation~\cite{Javanainen2014a,
 JavanainenMFT}.
 In large planar arrays of resonators,
 on the other hand, the collective effects
 can manifest themselves despite the
 presence of nonradiative losses,
 resulting, e.g., in a correlated excitation
 of a subradiant mode that can extend over
 the entire lattice, including over
 1000 metamolecules~\cite{Jenkinssub}.

 Here, we have tested and illustrated the
 effective theory using  simple point scatterer models
 where we only include one mode of
 the corresponding two-point-dipole scatterer.
 The effective models could be extended to
 the studies of large arrays in which case they
 can provide considerable numerical
 simplifications, or also to the studies of
 more complex multipole resonatators~\cite{LiuEtAlNatMat2009,
 Lovera,
 FanCapasso,
 GiessenOligomers,
 Frimmer,
 Dregely,
 Watson2016}.

 \begin{acknowledgments}
 We acknowledge discussions with Vassili Fedotov and Nikolay Zheludev, and financial support from
 the EPSRC and the Leverhulme Trust.
 \end{acknowledgments}

 %
 \appendix
 %
 \section{Electric quadrupole radiation kernel}
 \label{sec:E2_kernels}

 In \sref{sec:E2}, we calculated the
 EM fields scattered from a point electric quadrupole source, and
 the interaction between these scattered EM fields and other
 electric quadrupoles, electric dipoles and magnetic dipoles. The
 scattered EM fields and the resulting emf and flux terms
 contain derivatives of the radiation kernels
 \esref{eq:G} and~\eqref{eq:G_x}.
 In this Appendix, we give the corresponding
 derivatives of ${\bf G}({\bf r})$
 and ${\bf G}_\times({\bf r})$.

 The scattered electric ${\bf E}_{\text{E2},j}$ and
 magnetic ${\bf H}_{\text{E2},j}$ fields from the $j$th
 electric quadrupole are given in
 \esref{eq:E_E2_ibp} and~\eqref{eq:H_E2_ibp}. These
 equations contain rank three tensors which are the gradients of
 ${\bf G}({\bf r})$ and ${\bf G}_\times({\bf r})$, given by
 \begin{widetext}
 \begin{align}
 \frac{\p }{\p kr_\mu}{\bf G}({\bf r})
 &
 =
 i\Bigg[
 \left[
 \frac{1}{5}\frac{r_\mu}{r}{\bf I}
 +
 \frac{1}{5} \frac{{\bf r}{\bf \hat r}_\mu + {\bf \hat r}_\mu{\bf r}}{r}
 - \frac{r_\mu}{r}\frac{{\bf r}{\bf r}}{r^2}
 \right]
 h_3^{(1)}(kr)
 -
  \left[
 \frac{12}{15}\frac{r_\mu}{r}{\bf I}
 -
 \frac{1}{5} \frac{{\bf r}{\bf \hat r}_\mu + {\bf \hat r}_\mu{\bf r}}{r}
 \right]
 h_1^{(1)}(kr)
 \Bigg]
 \,\text{,}
 \label{eq:Gradient_G}
 \\
 \frac{\p}{\p kr_\mu}{\bf G}_\times({\bf r})
 &
 =
 \frac{1}{r^2}\left[
 r_\eta{\bf \hat r}_\nu
 \left(
 {\bf r}_\mu + {\bf r}_\nu + {\bf r}_\eta
 \right)
 -
 r_\nu({\bf \hat r}_\eta
 \left(
 {\bf r}_\mu + {\bf r}_\nu + {\bf r}_\eta
 \right)
 \right]
 h_2^{(1)}(kr)
 -
 \frac{1}{r}
 \left[
 {\bf \hat r}_\nu{\bf \hat r}_\eta
 -
 {\bf \hat r}_\eta{\bf \hat r}_\nu
 \right]
 h_1^{(1)}(kr)
 \,\text{.}
 \label{eq:Gradient_Gx}
 \end{align}
 \end{widetext}
 Here: $r_\mu$ is Cartesian component $\mu=x,y,z$ of
 the vector ${\bf r}$; ${\bf r}{\bf r}$ is the outer product of
 ${\bf r}$ with itself; ${\bf I}$ is the identity matrix;
 and ${\bf \hat r}_\mu{\bf r}$ is the outer product
 of the unit vector ${\bf \hat r}_\mu$ in the Cartesian direction
 $\mu=x,y,z$ with the vector ${\bf r}$;
 and $h_n^{(1)}(kr)$ are the $n$th order spherical Hankel functions
 of the first kind, defined by
 \begin{align}
 h_1^{(1)}(x)
 & =
 -
 \left[
 1 + \frac{i}{x}
 \right]
 \frac{e^{ix}}{x}
 \,\text{,}
 \label{eq:h_1}
 \\
 h_3^{(1)}(x) & =
 \left[
 1 + \frac{6i}{x} - \frac{15}{x^2} - \frac{15i}{x^3}
 \right]
 \frac{e^{ix}}{x}
 \,\text{,}
 \label{eq:h_3}
 \end{align}
 $h_2^{(1)}(x)$ is defined in \eref{eq:h_2}.
 In the cross kernel derivatives,
 \eref{eq:Gradient_Gx}, the
 component $\p r_{\mu}G_{\times,\mu\nu}({\bf r})=0$
 for $\nu=\mu,\nu,\eta$.

 The interaction between two separate
 electric quadrupoles $i$ and $j$ results in an effective emf
 $\mathcal{E}_{i,j}^{\text{sc,E2}}$, see
 \esref{eq:emf_E2_intermediate_2}--\eqref{eq:G_E2}. \Eref{eq:G_E2}
 describes the interaction matrix $\mathcal{G}_{\text{E2}}$
 whose off-diagonal elements represent the interactions between two
 electric quadrupoles, taking into account their relative locations and
 orientations only. $\mathcal{G}_\text{E2}$ is a contraction
 of the quadrupole moment tensors $\hat{A}_{\alpha\beta,m}$ and
 $\hat{A}_{\mu\nu,n}$ and a rank four tensor.
 The rank four tensor contains second order derivatives of ${\bf G}({\bf r})$,
 \begin{widetext}
 \begin{align}
 \frac{\p^2}{\p kr_\mu\p kr_\mu}&\bigg|_{r_\mu = r_\mu}{\bf G}({\bf r})
 =
 i
 \Bigg\{
 \left[
 \frac{r_\mu^2}{r^4} {\bf r}{\bf r}
 -
 \frac{1}{7r^2}
 \left(
 {\bf r}{\bf r} + r_\mu^2{\bf I}
 -
 4{\bf \hat r}_\mu{\bf \hat r}_\mu
 -
 2\left[{\bf r}_\mu{\bf r}_\nu + {\bf r}_\nu{\bf r}_\mu\right]
 \right)
 +
 \frac{1}{35}
 \left(
 {\bf I}
 +
 2{\bf \hat r}_\mu{\bf \hat r}_\mu
 \right)\right]
 h_4^{(1)}(kr)
 \nonumber
 \\
 &
 -
 \left[
 \frac{1}{7r^2}
 \left[
 {\bf r}{\bf r}
 +
 2\left({\bf r}_\mu{\bf r}_\nu + {\bf r}_\nu{\bf r}_\mu\right)
 +
 r_\mu^2{\bf I}
 +
 3{\bf r}_\mu{\bf r}_\mu
 \right]
 +
 \frac{5}{21}
 \left({\bf I} + 4{\bf \hat r}_\mu{\bf \hat r}_\mu\right)
 \right]
 h_2^{(1)}(kr)
 -
 \frac{4}{15}
 \left[{\bf I} + 2{\bf \hat r}_\mu{\bf \hat r}_\mu
 \right]
 h_0^{(1)}(kr)
 \Bigg\}
 \,\text{.}
 \label{eq:grad_grad_radiation_kernel_xx}
 \end{align}
 \begin{align}
 \frac{\p^2}{\p kr_\mu\p kr_\nu}&\bigg|_{r_\mu\ne r_\nu}
 {\bf G}({\bf r})
 =
 i\Bigg\{
 \left[\frac{r_\mu r_\nu}{r^4}{\bf r}{\bf r}
 -
 \frac{1}{7r^2}
 \left[
 r_\mu
 \left(
 {\bf r}{\bf \hat r}_\nu + {\bf \hat r}_\nu{\bf r}
 \right)
 +
 r_\nu
 \left(
 {\bf r}{\bf \hat r}_\mu + {\bf \hat r}_\mu{\bf r}
 \right)
  +
 r_\mu r_\nu{\bf I}
 \right]
 +
 \frac{1}{35}
 \left(
 {\bf \hat r}_\mu{\bf \hat r}_\nu
 +
 {\bf \hat r}_\nu{\bf \hat r}_\mu
 \right)
 \right]
 h_4^{(1)}(kr)
 \nonumber
 \\
 &
 -
 \left[
 \frac{1}{7r^2}
 \left(
 r_\mu
 \left[
 {\bf r}{\bf \hat r}_\nu + {\bf \hat r}_\nu{\bf r}
 \right]
 +
 r_\nu
 \left[
 {\bf r}{\bf \hat r}_\mu + {\bf \hat r}_\mu{\bf r}
 - 6r_\mu{\bf I}
 \right]
 \right)
 -
 \frac{2}{21}
\left(
 {\bf \hat r}_\mu{\bf \hat r}_\nu
 +
 {\bf \hat r}_\nu{\bf \hat r}_\mu
 \right)
 \right]
 h_2^{(1)}(kr)
 + \frac{1}{15}
 \left[
 {\bf \hat r}_\mu{\bf \hat r}_\nu
 +
 {\bf \hat r}_\nu{\bf \hat r}_\mu
 \right]
 h_0^{(1)}(kr)
 \Bigg\}
 \,\text{.}
 \label{eq:grad_grad_radiation_kernel_xy}
 \end{align}
 \end{widetext}
 The spherical Hankel functions $h_0^{(1)}(kr)$ and
 $h_2^{(1)}(kr)$ are defined in \esref{eq:h_0} and~\eqref{eq:h_2},
 respectively, while
 \begin{equation}
 h_4^{(1)}(x)
 =
 -i
 \left[
 1 + \frac{10i}{x} - \frac{45}{x^2} - \frac{105i}{x^3} + \frac{105}{x^4}
 \right]
 \frac{e^{ix}}{x}
 \,\text{.}
 \label{eq:h_4}
 \end{equation}

 \subsection{Electric quadrupole radiated power}
 \label{subsec:Power_E2_supp}

 In \sref{sec:E2}, we obtain an expression for the electric quadrupole
 radiative emission rate $\Gamma_{\text{E2},j}$, by calculating
 the radiated power $P_\text{E2}$, see \esref{eq:P_E2_temp1}
 and~\eqref{eq:Power_E2}. To arrive at \eref{eq:Power_E2}, we
 had to evaluate the integral of $|{\bf \hat r}\times{\bf q}_j({\bf \hat r})|^2$,
 over all angles.
 To do so,
 we note the Cartesian coordinate identity~\cite{Zangwill}
 \begin{align}
 |{\bf \hat r}\times {\bf q}_j({\bf \hat r})|^2
 &
 =
 \sum_{\alpha\beta\eta}
 q_{\alpha\beta,j}\hat{r}_{\beta}q_{\alpha\eta,n}^\ast\hat{r}_{\eta}
 \nonumber
 \\
 & \quad\quad
 -
 \sum_{\alpha\beta\eta\nu}
 \hat{r}_{\alpha}q_{\alpha\beta,j}\hat{r}_{\beta}
 \hat{r}_{\eta}q_{\eta\nu,n}^\ast\hat{r}_{\nu}
 \,\text{.}
 \label{eq:E2_identity}
 \end{align}
 The different ${\hat r}_\alpha$'s
 are direction cosines which obey the identities~\cite{Zangwill}
 \begin{align}
 \int\d\Omega\,{\hat r}_\beta{\hat r}_{\eta}
 &
 =
 \frac{4\pi}{3}\delta_{\beta\eta}
 \,\text{,}
 \label{eq:id_int1}
 \\
 \int\d\Omega\, {\hat r}_{\alpha} {\hat r}_{\beta} {\hat r}_{\eta}{\hat r}_{\nu}
 &
 =
 \frac{4\pi}{15}
 \Big[
 \delta_{\alpha\beta}\delta_{\eta\nu}
 +
 \delta_{\alpha\eta}\delta_{\beta\nu}
 +
 \delta_{\alpha\nu}\delta_{\beta\eta}
 \Big]
 \,\text{.}
 \label{eq:id_int2}
 \end{align}
 Evaluating the integrals in \esref{eq:id_int1} and~\eqref{eq:id_int2},
 and summing over the
 Cartesian indices $x$, $y$ and $z$, hence results in
 \eref{eq:Power_E2}.


\begin{thebibliography}{38}%
\makeatletter
\providecommand \@ifxundefined [1]{%
 \@ifx{#1\undefined}
}%
\providecommand \@ifnum [1]{%
 \ifnum #1\expandafter \@firstoftwo
 \else \expandafter \@secondoftwo
 \fi
}%
\providecommand \@ifx [1]{%
 \ifx #1\expandafter \@firstoftwo
 \else \expandafter \@secondoftwo
 \fi
}%
\providecommand \natexlab [1]{#1}%
\providecommand \enquote  [1]{``#1''}%
\providecommand \bibnamefont  [1]{#1}%
\providecommand \bibfnamefont [1]{#1}%
\providecommand \citenamefont [1]{#1}%
\providecommand \href@noop [0]{\@secondoftwo}%
\providecommand \href [0]{\begingroup \@sanitize@url \@href}%
\providecommand \@href[1]{\@@startlink{#1}\@@href}%
\providecommand \@@href[1]{\endgroup#1\@@endlink}%
\providecommand \@sanitize@url [0]{\catcode `\\12\catcode `\$12\catcode
  `\&12\catcode `\#12\catcode `\^12\catcode `\_12\catcode `\%12\relax}%
\providecommand \@@startlink[1]{}%
\providecommand \@@endlink[0]{}%
\providecommand \url  [0]{\begingroup\@sanitize@url \@url }%
\providecommand \@url [1]{\endgroup\@href {#1}{\urlprefix }}%
\providecommand \urlprefix  [0]{URL }%
\providecommand \Eprint [0]{\href }%
\providecommand \doibase [0]{http://dx.doi.org/}%
\providecommand \selectlanguage [0]{\@gobble}%
\providecommand \bibinfo  [0]{\@secondoftwo}%
\providecommand \bibfield  [0]{\@secondoftwo}%
\providecommand \translation [1]{[#1]}%
\providecommand \BibitemOpen [0]{}%
\providecommand \bibitemStop [0]{}%
\providecommand \bibitemNoStop [0]{.\EOS\space}%
\providecommand \EOS [0]{\spacefactor3000\relax}%
\providecommand \BibitemShut  [1]{\csname bibitem#1\endcsname}%
\let\auto@bib@innerbib\@empty
\bibitem [{\citenamefont {Szabo}\ \emph {et~al.}(2010)\citenamefont {Szabo},
  \citenamefont {Park}, \citenamefont {Hedge},\ and\ \citenamefont
  {Li}}]{5565504}%
  \BibitemOpen
  \bibfield  {author} {\bibinfo {author} {\bibfnamefont {Z.}~\bibnamefont
  {Szabo}}, \bibinfo {author} {\bibfnamefont {Gi-Ho}\ \bibnamefont {Park}},
  \bibinfo {author} {\bibfnamefont {R.}~\bibnamefont {Hedge}}, \ and\ \bibinfo
  {author} {\bibfnamefont {Er-Ping}\ \bibnamefont {Li}},\ }\bibfield  {title}
  {\enquote {\bibinfo {title} {A unique extraction of metamaterial parameters
  based on kramers - kronig relationship},}\ }\href@noop {} {\bibfield
  {journal} {\bibinfo  {journal} {IEEE Trans. Microwave Theory Tech.}\ }\textbf
  {\bibinfo {volume} {58}},\ \bibinfo {pages} {2646--2653} (\bibinfo {year}
  {2010})}\BibitemShut {NoStop}%
\bibitem [{\citenamefont {Koschny}\ \emph {et~al.}(2004)\citenamefont
  {Koschny}, \citenamefont {Kafesaki}, \citenamefont {Economou},\ and\
  \citenamefont {Soukoulis}}]{KoschnyEtAlPRL2004}%
  \BibitemOpen
  \bibfield  {author} {\bibinfo {author} {\bibfnamefont {T.}~\bibnamefont
  {Koschny}}, \bibinfo {author} {\bibfnamefont {M.}~\bibnamefont {Kafesaki}},
  \bibinfo {author} {\bibfnamefont {E.~N.}\ \bibnamefont {Economou}}, \ and\
  \bibinfo {author} {\bibfnamefont {C.~M.}\ \bibnamefont {Soukoulis}},\
  }\bibfield  {title} {\enquote {\bibinfo {title} {Effective medium theory of
  left-handed materials},}\ }\href@noop {} {\bibfield  {journal} {\bibinfo
  {journal} {Phys. Rev. Lett.}\ }\textbf {\bibinfo {volume} {93}},\ \bibinfo
  {pages} {107402} (\bibinfo {year} {2004})}\BibitemShut {NoStop}%
\bibitem [{\citenamefont {Pendry}\ \emph {et~al.}(1999)\citenamefont {Pendry},
  \citenamefont {Holden}, \citenamefont {Robbins},\ and\ \citenamefont
  {Stewart}}]{PendryEtAlIEEE1999}%
  \BibitemOpen
  \bibfield  {author} {\bibinfo {author} {\bibfnamefont {J.~B.}\ \bibnamefont
  {Pendry}}, \bibinfo {author} {\bibfnamefont {A.~J.}\ \bibnamefont {Holden}},
  \bibinfo {author} {\bibfnamefont {D.~J.}\ \bibnamefont {Robbins}}, \ and\
  \bibinfo {author} {\bibfnamefont {W.~J.}\ \bibnamefont {Stewart}},\
  }\bibfield  {title} {\enquote {\bibinfo {title} {Magnetism from conductors
  and enhanced nonlinear phenomena},}\ }\href@noop {} {\bibfield  {journal}
  {\bibinfo  {journal} {IEEE Transactions on Microwave Theory and Techniques}\
  }\textbf {\bibinfo {volume} {47}},\ \bibinfo {pages} {2075} (\bibinfo {year}
  {1999})}\BibitemShut {NoStop}%
\bibitem [{\citenamefont {Belov}\ and\ \citenamefont
  {Simovski}(2005)}]{BelovSimovskiPRE2005}%
  \BibitemOpen
  \bibfield  {author} {\bibinfo {author} {\bibfnamefont {Pavel~A.}\
  \bibnamefont {Belov}}\ and\ \bibinfo {author} {\bibfnamefont {Constantin~R.}\
  \bibnamefont {Simovski}},\ }\bibfield  {title} {\enquote {\bibinfo {title}
  {Homogenization of electromagnetic crystals formed by uniaxial resonant
  scatterers},}\ }\href@noop {} {\bibfield  {journal} {\bibinfo  {journal}
  {Phys. Rev. E}\ }\textbf {\bibinfo {volume} {72}},\ \bibinfo {pages} {026615}
  (\bibinfo {year} {2005})}\BibitemShut {NoStop}%
\bibitem [{\citenamefont {Liu}\ \emph {et~al.}(2007)\citenamefont {Liu},
  \citenamefont {Cui}, \citenamefont {Huang}, \citenamefont {Zhao},\ and\
  \citenamefont {Smith}}]{LiuEtAlPRE2007}%
  \BibitemOpen
  \bibfield  {author} {\bibinfo {author} {\bibfnamefont {Ruopeng}\ \bibnamefont
  {Liu}}, \bibinfo {author} {\bibfnamefont {Tie~Jun}\ \bibnamefont {Cui}},
  \bibinfo {author} {\bibfnamefont {Da}~\bibnamefont {Huang}}, \bibinfo
  {author} {\bibfnamefont {Bo}~\bibnamefont {Zhao}}, \ and\ \bibinfo {author}
  {\bibfnamefont {David~R.}\ \bibnamefont {Smith}},\ }\bibfield  {title}
  {\enquote {\bibinfo {title} {Description and explanation of electromagnetic
  behaviors in artificial metamaterials based on effective medium theory},}\
  }\href@noop {} {\bibfield  {journal} {\bibinfo  {journal} {Phys. Rev. E}\
  }\textbf {\bibinfo {volume} {76}},\ \bibinfo {pages} {026606} (\bibinfo
  {year} {2007})}\BibitemShut {NoStop}%
\bibitem [{\citenamefont {Jenkins}\ and\ \citenamefont
  {Ruostekoski}(2012{\natexlab{a}})}]{JenkinsLongPRB}%
  \BibitemOpen
  \bibfield  {author} {\bibinfo {author} {\bibfnamefont {S.~D.}\ \bibnamefont
  {Jenkins}}\ and\ \bibinfo {author} {\bibfnamefont {J.}~\bibnamefont
  {Ruostekoski}},\ }\bibfield  {title} {\enquote {\bibinfo {title} {Theoretical
  formalism for collective electromagnetic response of discrete metamaterial
  systems},}\ }\href@noop {} {\bibfield  {journal} {\bibinfo  {journal} {Phys.
  Rev. B}\ }\textbf {\bibinfo {volume} {86}},\ \bibinfo {pages} {085116}
  (\bibinfo {year} {2012}{\natexlab{a}})}\BibitemShut {NoStop}%
\bibitem [{\citenamefont {de~Vries}\ \emph {et~al.}(1998)\citenamefont
  {de~Vries}, \citenamefont {van Coevorden},\ and\ \citenamefont
  {Lagendijk}}]{devries98}%
  \BibitemOpen
  \bibfield  {author} {\bibinfo {author} {\bibfnamefont {Pedro}\ \bibnamefont
  {de~Vries}}, \bibinfo {author} {\bibfnamefont {David~V.}\ \bibnamefont {van
  Coevorden}}, \ and\ \bibinfo {author} {\bibfnamefont {Ad}~\bibnamefont
  {Lagendijk}},\ }\bibfield  {title} {\enquote {\bibinfo {title} {Point
  scatterers for classical waves},}\ }\href {\doibase
  10.1103/RevModPhys.70.447} {\bibfield  {journal} {\bibinfo  {journal} {Rev.
  Mod. Phys.}\ }\textbf {\bibinfo {volume} {70}},\ \bibinfo {pages} {447--466}
  (\bibinfo {year} {1998})}\BibitemShut {NoStop}%
\bibitem [{\citenamefont {M{\"u}hlschlegel}\ \emph {et~al.}(2005)\citenamefont
  {M{\"u}hlschlegel}, \citenamefont {Eisler}, \citenamefont {Martin},
  \citenamefont {Hecht},\ and\ \citenamefont {Pohl}}]{optical_antennas}%
  \BibitemOpen
  \bibfield  {author} {\bibinfo {author} {\bibfnamefont {P.}~\bibnamefont
  {M{\"u}hlschlegel}}, \bibinfo {author} {\bibfnamefont {H.-J.}\ \bibnamefont
  {Eisler}}, \bibinfo {author} {\bibfnamefont {O.~J.~F.}\ \bibnamefont
  {Martin}}, \bibinfo {author} {\bibfnamefont {B.}~\bibnamefont {Hecht}}, \
  and\ \bibinfo {author} {\bibfnamefont {D.~W.}\ \bibnamefont {Pohl}},\
  }\bibfield  {title} {\enquote {\bibinfo {title} {Resonant optical
  antennas},}\ }\href {\doibase 10.1126/science.1111886} {\bibfield  {journal}
  {\bibinfo  {journal} {Science}\ }\textbf {\bibinfo {volume} {308}},\ \bibinfo
  {pages} {1607--1609} (\bibinfo {year} {2005})},\ \Eprint
  {http://arxiv.org/abs/http://science.sciencemag.org/content/308/5728/1607.full.pdf}
  {http://science.sciencemag.org/content/308/5728/1607.full.pdf} \BibitemShut
  {NoStop}%
\bibitem [{\citenamefont {Fromm}\ \emph {et~al.}(2004)\citenamefont {Fromm},
  \citenamefont {Sundaramurthy}, \citenamefont {Schuck}, \citenamefont {Kino},\
  and\ \citenamefont {Moerner}}]{Fromm}%
  \BibitemOpen
  \bibfield  {author} {\bibinfo {author} {\bibfnamefont {David~P.}\
  \bibnamefont {Fromm}}, \bibinfo {author} {\bibfnamefont {Arvind}\
  \bibnamefont {Sundaramurthy}}, \bibinfo {author} {\bibfnamefont {P.~James}\
  \bibnamefont {Schuck}}, \bibinfo {author} {\bibfnamefont {Gordon}\
  \bibnamefont {Kino}}, \ and\ \bibinfo {author} {\bibfnamefont {W.~E.}\
  \bibnamefont {Moerner}},\ }\bibfield  {title} {\enquote {\bibinfo {title}
  {Gap-dependent optical coupling of single €``bowtie''€ nanoantennas resonant
  in the visible},}\ }\href {\doibase 10.1021/nl049951r} {\bibfield  {journal}
  {\bibinfo  {journal} {Nano Letters}\ }\textbf {\bibinfo {volume} {4}},\
  \bibinfo {pages} {957--961} (\bibinfo {year} {2004})},\ \Eprint
  {http://arxiv.org/abs/http://dx.doi.org/10.1021/nl049951r}
  {http://dx.doi.org/10.1021/nl049951r} \BibitemShut {NoStop}%
\bibitem [{\citenamefont {Evlyukhin}\ \emph {et~al.}(2010)\citenamefont
  {Evlyukhin}, \citenamefont {Reinhardt}, \citenamefont {Seidel}, \citenamefont
  {Luk'yanchuk},\ and\ \citenamefont {Chichkov}}]{Evlyukhin_nanop}%
  \BibitemOpen
  \bibfield  {author} {\bibinfo {author} {\bibfnamefont {Andrey~B.}\
  \bibnamefont {Evlyukhin}}, \bibinfo {author} {\bibfnamefont {Carsten}\
  \bibnamefont {Reinhardt}}, \bibinfo {author} {\bibfnamefont {Andreas}\
  \bibnamefont {Seidel}}, \bibinfo {author} {\bibfnamefont {Boris~S.}\
  \bibnamefont {Luk'yanchuk}}, \ and\ \bibinfo {author} {\bibfnamefont
  {Boris~N.}\ \bibnamefont {Chichkov}},\ }\bibfield  {title} {\enquote
  {\bibinfo {title} {Optical response features of si-nanoparticle arrays},}\
  }\href {\doibase 10.1103/PhysRevB.82.045404} {\bibfield  {journal} {\bibinfo
  {journal} {Phys. Rev. B}\ }\textbf {\bibinfo {volume} {82}},\ \bibinfo
  {pages} {045404} (\bibinfo {year} {2010})}\BibitemShut {NoStop}%
\bibitem [{\citenamefont {Wang}\ \emph {et~al.}(2011)\citenamefont {Wang},
  \citenamefont {Cao}, \citenamefont {Chen},\ and\ \citenamefont
  {Gu}}]{wang_nanoshells}%
  \BibitemOpen
  \bibfield  {author} {\bibinfo {author} {\bibfnamefont {Meng}\ \bibnamefont
  {Wang}}, \bibinfo {author} {\bibfnamefont {Min}\ \bibnamefont {Cao}},
  \bibinfo {author} {\bibfnamefont {Xin}\ \bibnamefont {Chen}}, \ and\ \bibinfo
  {author} {\bibfnamefont {Ning}\ \bibnamefont {Gu}},\ }\bibfield  {title}
  {\enquote {\bibinfo {title} {Subradiant plasmon modes in multilayer
  metal-€“dielectric nanoshells},}\ }\href {\doibase 10.1021/jp205736d}
  {\bibfield  {journal} {\bibinfo  {journal} {The Journal of Physical Chemistry
  C}\ }\textbf {\bibinfo {volume} {115}},\ \bibinfo {pages} {20920--20925}
  (\bibinfo {year} {2011})},\ \Eprint
  {http://arxiv.org/abs/http://dx.doi.org/10.1021/jp205736d}
  {http://dx.doi.org/10.1021/jp205736d} \BibitemShut {NoStop}%
\bibitem [{\citenamefont {Jenkins}\ and\ \citenamefont
  {Ruostekoski}(2012{\natexlab{b}})}]{JenkinsLineWidthNJP}%
  \BibitemOpen
  \bibfield  {author} {\bibinfo {author} {\bibfnamefont {S.~D.}\ \bibnamefont
  {Jenkins}}\ and\ \bibinfo {author} {\bibfnamefont {J.}~\bibnamefont
  {Ruostekoski}},\ }\bibfield  {title} {\enquote {\bibinfo {title} {Cooperative
  resonance linewidth narrowing in a planar metamaterial},}\ }\href@noop {}
  {\bibfield  {journal} {\bibinfo  {journal} {New Journal of Physics}\ }\textbf
  {\bibinfo {volume} {14}},\ \bibinfo {pages} {103003} (\bibinfo {year}
  {2012}{\natexlab{b}})}\BibitemShut {NoStop}%
\bibitem [{\citenamefont {Jenkins}\ and\ \citenamefont
  {Ruostekoski}(2013)}]{CAIT}%
  \BibitemOpen
  \bibfield  {author} {\bibinfo {author} {\bibfnamefont {S.~D.}\ \bibnamefont
  {Jenkins}}\ and\ \bibinfo {author} {\bibfnamefont {J.}~\bibnamefont
  {Ruostekoski}},\ }\bibfield  {title} {\enquote {\bibinfo {title}
  {Metamaterial transparency induced by cooperative electromagnetic
  interactions},}\ }\href@noop {} {\bibfield  {journal} {\bibinfo  {journal}
  {Phys. Rev. Lett.}\ }\textbf {\bibinfo {volume} {111}},\ \bibinfo {pages}
  {147401} (\bibinfo {year} {2013})}\BibitemShut {NoStop}%
\bibitem [{\citenamefont {Jenkins}\ \emph {et~al.}(2016)\citenamefont
  {Jenkins}, \citenamefont {Ruostekoski}, \citenamefont {Papasimakis},
  \citenamefont {Savo},\ and\ \citenamefont {Zheludev}}]{Jenkinssub}%
  \BibitemOpen
  \bibfield  {author} {\bibinfo {author} {\bibfnamefont {S.~D.}\ \bibnamefont
  {Jenkins}}, \bibinfo {author} {\bibfnamefont {J.}~\bibnamefont
  {Ruostekoski}}, \bibinfo {author} {\bibfnamefont {N.}~\bibnamefont
  {Papasimakis}}, \bibinfo {author} {\bibfnamefont {S.}~\bibnamefont {Savo}}, \
  and\ \bibinfo {author} {\bibfnamefont {N.~I.}\ \bibnamefont {Zheludev}},\
  }\href@noop {} {\enquote {\bibinfo {title} {Many-body subradiant excitations
  in metamaterial arrays: Experiment and theory, eprint arxiv:1611.01509},}\ }
  (\bibinfo {year} {2016})\BibitemShut {NoStop}%
\bibitem [{\citenamefont {Smith}\ \emph {et~al.}(2000)\citenamefont {Smith},
  \citenamefont {Padilla}, \citenamefont {Vier}, \citenamefont
  {{Nemat-Nasser}},\ and\ \citenamefont {Schultz}}]{SmithEtAlPRL2000}%
  \BibitemOpen
  \bibfield  {author} {\bibinfo {author} {\bibfnamefont {D.~R.}\ \bibnamefont
  {Smith}}, \bibinfo {author} {\bibfnamefont {W.~J.}\ \bibnamefont {Padilla}},
  \bibinfo {author} {\bibfnamefont {D.~C.}\ \bibnamefont {Vier}}, \bibinfo
  {author} {\bibfnamefont {S.~C.}\ \bibnamefont {{Nemat-Nasser}}}, \ and\
  \bibinfo {author} {\bibfnamefont {S.}~\bibnamefont {Schultz}},\ }\bibfield
  {title} {\enquote {\bibinfo {title} {Composite medium with simultaneously
  negative permeability and permittivity},}\ }\href@noop {} {\bibfield
  {journal} {\bibinfo  {journal} {Phys. Rev. Lett.}\ }\textbf {\bibinfo
  {volume} {84}},\ \bibinfo {pages} {4184} (\bibinfo {year}
  {2000})}\BibitemShut {NoStop}%
\bibitem [{\citenamefont {Fedotov}\ \emph {et~al.}(2007)\citenamefont
  {Fedotov}, \citenamefont {Rose}, \citenamefont {Prosvirnin}, \citenamefont
  {Papasimakis},\ and\ \citenamefont {Zheludev}}]{FedotovEtAlPRL2007}%
  \BibitemOpen
  \bibfield  {author} {\bibinfo {author} {\bibfnamefont {V.~A.}\ \bibnamefont
  {Fedotov}}, \bibinfo {author} {\bibfnamefont {M.}~\bibnamefont {Rose}},
  \bibinfo {author} {\bibfnamefont {S.~L.}\ \bibnamefont {Prosvirnin}},
  \bibinfo {author} {\bibfnamefont {N.}~\bibnamefont {Papasimakis}}, \ and\
  \bibinfo {author} {\bibfnamefont {N.~I.}\ \bibnamefont {Zheludev}},\
  }\bibfield  {title} {\enquote {\bibinfo {title} {Sharp trapped-mode
  resonances in planar metamaterials with a broken structural symmetry},}\
  }\href@noop {} {\bibfield  {journal} {\bibinfo  {journal} {Phys. Rev. Lett.}\
  }\textbf {\bibinfo {volume} {99}},\ \bibinfo {pages} {147401} (\bibinfo
  {year} {2007})}\BibitemShut {NoStop}%
\bibitem [{\citenamefont {Adamo}\ \emph {et~al.}(2012)\citenamefont {Adamo},
  \citenamefont {Ou}, \citenamefont {So}, \citenamefont {Jenkins},
  \citenamefont {{De Angelis}}, \citenamefont {{MacDonald}}, \citenamefont {{Di
  Fabrizio}}, \citenamefont {Ruostekoski},\ and\ \citenamefont
  {Zheludev}}]{AdamoEtAlPRL2012}%
  \BibitemOpen
  \bibfield  {author} {\bibinfo {author} {\bibfnamefont {G.}~\bibnamefont
  {Adamo}}, \bibinfo {author} {\bibfnamefont {J.~Y.}\ \bibnamefont {Ou}},
  \bibinfo {author} {\bibfnamefont {J.~K.}\ \bibnamefont {So}}, \bibinfo
  {author} {\bibfnamefont {S.~D.}\ \bibnamefont {Jenkins}}, \bibinfo {author}
  {\bibfnamefont {F.}~\bibnamefont {{De Angelis}}}, \bibinfo {author}
  {\bibfnamefont {K.~F.}\ \bibnamefont {{MacDonald}}}, \bibinfo {author}
  {\bibfnamefont {E.}~\bibnamefont {{Di Fabrizio}}}, \bibinfo {author}
  {\bibfnamefont {J.}~\bibnamefont {Ruostekoski}}, \ and\ \bibinfo {author}
  {\bibfnamefont {N.~I.}\ \bibnamefont {Zheludev}},\ }\bibfield  {title}
  {\enquote {\bibinfo {title} {Electron-beam-driven collective-mode
  metamaterial light source},}\ }\href@noop {} {\bibfield  {journal} {\bibinfo
  {journal} {Phy. Rev. Lett.}\ }\textbf {\bibinfo {volume} {109}},\ \bibinfo
  {pages} {217401} (\bibinfo {year} {2012})}\BibitemShut {NoStop}%
\bibitem [{\citenamefont {Jenkins}\ and\ \citenamefont
  {Ruostekoski}(2012{\natexlab{c}})}]{JenkinsRuostekoskiPRB2012b}%
  \BibitemOpen
  \bibfield  {author} {\bibinfo {author} {\bibfnamefont {S.~D.}\ \bibnamefont
  {Jenkins}}\ and\ \bibinfo {author} {\bibfnamefont {J.}~\bibnamefont
  {Ruostekoski}},\ }\bibfield  {title} {\enquote {\bibinfo {title} {Resonance
  linewidth and inhomogeneous broadening in a metamaterial array},}\
  }\href@noop {} {\bibfield  {journal} {\bibinfo  {journal} {Phys. Rev. B}\
  }\textbf {\bibinfo {volume} {86}},\ \bibinfo {pages} {085116} (\bibinfo
  {year} {2012}{\natexlab{c}})}\BibitemShut {NoStop}%
\bibitem [{\citenamefont {Liu}\ \emph {et~al.}(2009)\citenamefont {Liu},
  \citenamefont {Langguth}, \citenamefont {Weiss}, \citenamefont {K\"{a}stel},
  \citenamefont {Fleischhauer}, \citenamefont {Pfau},\ and\ \citenamefont
  {Giessen}}]{LiuEtAlNatMat2009}%
  \BibitemOpen
  \bibfield  {author} {\bibinfo {author} {\bibfnamefont {Na}~\bibnamefont
  {Liu}}, \bibinfo {author} {\bibfnamefont {Lutz}\ \bibnamefont {Langguth}},
  \bibinfo {author} {\bibfnamefont {Thomas}\ \bibnamefont {Weiss}}, \bibinfo
  {author} {\bibfnamefont {J\"{u}rgen}\ \bibnamefont {K\"{a}stel}}, \bibinfo
  {author} {\bibfnamefont {Michael}\ \bibnamefont {Fleischhauer}}, \bibinfo
  {author} {\bibfnamefont {Tilman}\ \bibnamefont {Pfau}}, \ and\ \bibinfo
  {author} {\bibfnamefont {Harald}\ \bibnamefont {Giessen}},\ }\bibfield
  {title} {\enquote {\bibinfo {title} {Plasmonic analogue of
  electromagnetically induced transparency at the {Drude} damping limit},}\
  }\href@noop {} {\bibfield  {journal} {\bibinfo  {journal} {Nat. Mater.}\
  }\textbf {\bibinfo {volume} {8}},\ \bibinfo {pages} {758--762} (\bibinfo
  {year} {2009})}\BibitemShut {NoStop}%
\bibitem [{\citenamefont {Lovera}\ \emph {et~al.}(2013)\citenamefont {Lovera},
  \citenamefont {Gallinet}, \citenamefont {Nordlander},\ and\ \citenamefont
  {Martin}}]{Lovera}%
  \BibitemOpen
  \bibfield  {author} {\bibinfo {author} {\bibfnamefont {Andrea}\ \bibnamefont
  {Lovera}}, \bibinfo {author} {\bibfnamefont {Benjamin}\ \bibnamefont
  {Gallinet}}, \bibinfo {author} {\bibfnamefont {Peter}\ \bibnamefont
  {Nordlander}}, \ and\ \bibinfo {author} {\bibfnamefont {Olivier~J.F.}\
  \bibnamefont {Martin}},\ }\bibfield  {title} {\enquote {\bibinfo {title}
  {Mechanisms of fano resonances in coupled plasmonic systems},}\ }\href
  {\doibase 10.1021/nn401175j} {\bibfield  {journal} {\bibinfo  {journal} {ACS
  Nano}\ }\textbf {\bibinfo {volume} {7}},\ \bibinfo {pages} {4527--4536}
  (\bibinfo {year} {2013})}\BibitemShut {NoStop}%
\bibitem [{\citenamefont {Fan}\ \emph {et~al.}(2010)\citenamefont {Fan},
  \citenamefont {Wu}, \citenamefont {Bao}, \citenamefont {Bao}, \citenamefont
  {Bardhan}, \citenamefont {Halas}, \citenamefont {Manoharan}, \citenamefont
  {Nordlander}, \citenamefont {Shvets},\ and\ \citenamefont
  {Capasso}}]{FanCapasso}%
  \BibitemOpen
  \bibfield  {author} {\bibinfo {author} {\bibfnamefont {Jonathan~A.}\
  \bibnamefont {Fan}}, \bibinfo {author} {\bibfnamefont {Chihhui}\ \bibnamefont
  {Wu}}, \bibinfo {author} {\bibfnamefont {Kui}\ \bibnamefont {Bao}}, \bibinfo
  {author} {\bibfnamefont {Jiming}\ \bibnamefont {Bao}}, \bibinfo {author}
  {\bibfnamefont {Rizia}\ \bibnamefont {Bardhan}}, \bibinfo {author}
  {\bibfnamefont {Naomi~J.}\ \bibnamefont {Halas}}, \bibinfo {author}
  {\bibfnamefont {Vinothan~N.}\ \bibnamefont {Manoharan}}, \bibinfo {author}
  {\bibfnamefont {Peter}\ \bibnamefont {Nordlander}}, \bibinfo {author}
  {\bibfnamefont {Gennady}\ \bibnamefont {Shvets}}, \ and\ \bibinfo {author}
  {\bibfnamefont {Federico}\ \bibnamefont {Capasso}},\ }\bibfield  {title}
  {\enquote {\bibinfo {title} {Self-assembled plasmonic nanoparticle
  clusters},}\ }\href {\doibase 10.1126/science.1187949} {\bibfield  {journal}
  {\bibinfo  {journal} {Science}\ }\textbf {\bibinfo {volume} {328}},\ \bibinfo
  {pages} {1135--1138} (\bibinfo {year} {2010})}\BibitemShut {NoStop}%
\bibitem [{\citenamefont {Hentschel}\ \emph {et~al.}(2011)\citenamefont
  {Hentschel}, \citenamefont {Dregely}, \citenamefont {Vogelgesang},
  \citenamefont {Giessen},\ and\ \citenamefont {Liu}}]{GiessenOligomers}%
  \BibitemOpen
  \bibfield  {author} {\bibinfo {author} {\bibfnamefont {Mario}\ \bibnamefont
  {Hentschel}}, \bibinfo {author} {\bibfnamefont {Daniel}\ \bibnamefont
  {Dregely}}, \bibinfo {author} {\bibfnamefont {Ralf}\ \bibnamefont
  {Vogelgesang}}, \bibinfo {author} {\bibfnamefont {Harald}\ \bibnamefont
  {Giessen}}, \ and\ \bibinfo {author} {\bibfnamefont {Na}~\bibnamefont
  {Liu}},\ }\bibfield  {title} {\enquote {\bibinfo {title} {Plasmonic
  oligomers: The role of individual particles in collective behavior},}\ }\href
  {\doibase 10.1021/nn103172t} {\bibfield  {journal} {\bibinfo  {journal} {ACS
  Nano}\ }\textbf {\bibinfo {volume} {5}},\ \bibinfo {pages} {2042--2050}
  (\bibinfo {year} {2011})}\BibitemShut {NoStop}%
\bibitem [{\citenamefont {Frimmer}\ \emph {et~al.}({2012})\citenamefont
  {Frimmer}, \citenamefont {Coenen},\ and\ \citenamefont
  {Koenderink}}]{Frimmer}%
  \BibitemOpen
  \bibfield  {author} {\bibinfo {author} {\bibfnamefont {Martin}\ \bibnamefont
  {Frimmer}}, \bibinfo {author} {\bibfnamefont {Toon}\ \bibnamefont {Coenen}},
  \ and\ \bibinfo {author} {\bibfnamefont {A.~Femius}\ \bibnamefont
  {Koenderink}},\ }\bibfield  {title} {\enquote {\bibinfo {title} {{Signature
  of a Fano Resonance in a Plasmonic Metamolecule's Local Density of Optical
  States}},}\ }\href {\doibase {10.1103/PhysRevLett.108.077404}} {\bibfield
  {journal} {\bibinfo  {journal} {{Phys. Rev. Lett.}}\ }\textbf {\bibinfo
  {volume} {{108}}},\ \bibinfo {pages} {{077404}} (\bibinfo {year}
  {{2012}})}\BibitemShut {NoStop}%
\bibitem [{\citenamefont {Dregely}\ \emph {et~al.}(2011)\citenamefont
  {Dregely}, \citenamefont {Hentschel},\ and\ \citenamefont
  {Giessen}}]{Dregely}%
  \BibitemOpen
  \bibfield  {author} {\bibinfo {author} {\bibfnamefont {Daniel}\ \bibnamefont
  {Dregely}}, \bibinfo {author} {\bibfnamefont {Mario}\ \bibnamefont
  {Hentschel}}, \ and\ \bibinfo {author} {\bibfnamefont {Harald}\ \bibnamefont
  {Giessen}},\ }\bibfield  {title} {\enquote {\bibinfo {title} {Excitation and
  tuning of higher-order fano resonances in plasmonic oligomer clusters},}\
  }\href {\doibase 10.1021/nn202876k} {\bibfield  {journal} {\bibinfo
  {journal} {ACS Nano}\ }\textbf {\bibinfo {volume} {5}},\ \bibinfo {pages}
  {8202--8211} (\bibinfo {year} {2011})}\BibitemShut {NoStop}%
\bibitem [{\citenamefont {Watson}\ \emph {et~al.}(2016)\citenamefont {Watson},
  \citenamefont {Jenkins}, \citenamefont {Ruostekoski}, \citenamefont
  {Fedotov},\ and\ \citenamefont {Zheludev}}]{Watson2016}%
  \BibitemOpen
  \bibfield  {author} {\bibinfo {author} {\bibfnamefont {Derek~W.}\
  \bibnamefont {Watson}}, \bibinfo {author} {\bibfnamefont {Stewart~D.}\
  \bibnamefont {Jenkins}}, \bibinfo {author} {\bibfnamefont {Janne}\
  \bibnamefont {Ruostekoski}}, \bibinfo {author} {\bibfnamefont {Vassili~A.}\
  \bibnamefont {Fedotov}}, \ and\ \bibinfo {author} {\bibfnamefont
  {Nikolay~I.}\ \bibnamefont {Zheludev}},\ }\bibfield  {title} {\enquote
  {\bibinfo {title} {Toroidal dipole excitations in metamolecules formed by
  interacting plasmonic nanorods},}\ }\href@noop {} {\bibfield  {journal}
  {\bibinfo  {journal} {Phys. Rev. B}\ }\textbf {\bibinfo {volume} {93}},\
  \bibinfo {pages} {125420} (\bibinfo {year} {2016})}\BibitemShut {NoStop}%
\bibitem [{\citenamefont {Fedotov}\ \emph {et~al.}(2010)\citenamefont
  {Fedotov}, \citenamefont {Papasimakis}, \citenamefont {Plum}, \citenamefont
  {Bitzer}, \citenamefont {Walther}, \citenamefont {Kuo}, \citenamefont
  {Tsai},\ and\ \citenamefont {Zheludev}}]{FedotovEtAlPRL2010}%
  \BibitemOpen
  \bibfield  {author} {\bibinfo {author} {\bibfnamefont {V.~A.}\ \bibnamefont
  {Fedotov}}, \bibinfo {author} {\bibfnamefont {N.}~\bibnamefont
  {Papasimakis}}, \bibinfo {author} {\bibfnamefont {E.}~\bibnamefont {Plum}},
  \bibinfo {author} {\bibfnamefont {A.}~\bibnamefont {Bitzer}}, \bibinfo
  {author} {\bibfnamefont {M.}~\bibnamefont {Walther}}, \bibinfo {author}
  {\bibfnamefont {P.}~\bibnamefont {Kuo}}, \bibinfo {author} {\bibfnamefont
  {D.~P.}\ \bibnamefont {Tsai}}, \ and\ \bibinfo {author} {\bibfnamefont
  {N.~I.}\ \bibnamefont {Zheludev}},\ }\bibfield  {title} {\enquote {\bibinfo
  {title} {Spectral collapse in ensembles of metamolecules},}\ }\href@noop {}
  {\bibfield  {journal} {\bibinfo  {journal} {Phys. Rev. Lett.}\ }\textbf
  {\bibinfo {volume} {104}},\ \bibinfo {pages} {223901} (\bibinfo {year}
  {2010})}\BibitemShut {NoStop}%
\bibitem [{\citenamefont {Papasimakis}\ \emph {et~al.}(2009)\citenamefont
  {Papasimakis}, \citenamefont {Fedotov}, \citenamefont {Fu}, \citenamefont
  {Tsai},\ and\ \citenamefont {Zheludev}}]{papasimakis2009}%
  \BibitemOpen
  \bibfield  {author} {\bibinfo {author} {\bibfnamefont {N.}~\bibnamefont
  {Papasimakis}}, \bibinfo {author} {\bibfnamefont {V.~A.}\ \bibnamefont
  {Fedotov}}, \bibinfo {author} {\bibfnamefont {Y.~H.}\ \bibnamefont {Fu}},
  \bibinfo {author} {\bibfnamefont {D.~P.}\ \bibnamefont {Tsai}}, \ and\
  \bibinfo {author} {\bibfnamefont {N.~I.}\ \bibnamefont {Zheludev}},\
  }\bibfield  {title} {\enquote {\bibinfo {title} {Coherent and incoherent
  metamaterials and order-disorder transitions},}\ }\href@noop {} {\bibfield
  {journal} {\bibinfo  {journal} {Phys. Rev. B}\ }\textbf {\bibinfo {volume}
  {80}},\ \bibinfo {pages} {041102(R)} (\bibinfo {year} {2009})}\BibitemShut
  {NoStop}%
\bibitem [{\citenamefont {Sentenac}\ and\ \citenamefont
  {Chaumet}(2008)}]{SentenacPRL2008}%
  \BibitemOpen
  \bibfield  {author} {\bibinfo {author} {\bibfnamefont {Anne}\ \bibnamefont
  {Sentenac}}\ and\ \bibinfo {author} {\bibfnamefont {Patrick~C.}\ \bibnamefont
  {Chaumet}},\ }\bibfield  {title} {\enquote {\bibinfo {title} {Subdiffraction
  light focusing on a grating substrate},}\ }\href {\doibase
  10.1103/PhysRevLett.101.013901} {\bibfield  {journal} {\bibinfo  {journal}
  {Phys. Rev. Lett.}\ }\textbf {\bibinfo {volume} {101}},\ \bibinfo {pages}
  {013901} (\bibinfo {year} {2008})}\BibitemShut {NoStop}%
\bibitem [{\citenamefont {Lemoult}\ \emph {et~al.}(2010)\citenamefont
  {Lemoult}, \citenamefont {Lerosey}, \citenamefont {{de Rosny}},\ and\
  \citenamefont {Fink}}]{LemoultPRL10}%
  \BibitemOpen
  \bibfield  {author} {\bibinfo {author} {\bibfnamefont {Fabrice}\ \bibnamefont
  {Lemoult}}, \bibinfo {author} {\bibfnamefont {Geoffroy}\ \bibnamefont
  {Lerosey}}, \bibinfo {author} {\bibfnamefont {Julien}\ \bibnamefont {{de
  Rosny}}}, \ and\ \bibinfo {author} {\bibfnamefont {Mathias}\ \bibnamefont
  {Fink}},\ }\bibfield  {title} {\enquote {\bibinfo {title} {Resonant
  metalenses for breaking the diffraction barrier},}\ }\href {\doibase
  10.1103/PhysRevLett.104.203901} {\bibfield  {journal} {\bibinfo  {journal}
  {Phys. Rev. Lett.}\ }\textbf {\bibinfo {volume} {104}},\ \bibinfo {pages}
  {203901} (\bibinfo {year} {2010})}\BibitemShut {NoStop}%
\bibitem [{\citenamefont {Trepanier}\ \emph {et~al.}(2013)\citenamefont
  {Trepanier}, \citenamefont {Zhang}, \citenamefont {Mukhanov},\ and\
  \citenamefont {Anlage}}]{Anlageprx}%
  \BibitemOpen
  \bibfield  {author} {\bibinfo {author} {\bibfnamefont {M.}~\bibnamefont
  {Trepanier}}, \bibinfo {author} {\bibfnamefont {Daimeng}\ \bibnamefont
  {Zhang}}, \bibinfo {author} {\bibfnamefont {Oleg}\ \bibnamefont {Mukhanov}},
  \ and\ \bibinfo {author} {\bibfnamefont {Steven~M.}\ \bibnamefont {Anlage}},\
  }\bibfield  {title} {\enquote {\bibinfo {title} {Realization and modeling of
  metamaterials made of rf superconducting quantum-interference devices},}\
  }\href {\doibase 10.1103/PhysRevX.3.041029} {\bibfield  {journal} {\bibinfo
  {journal} {Phys. Rev. X}\ }\textbf {\bibinfo {volume} {3}},\ \bibinfo {pages}
  {041029} (\bibinfo {year} {2013})}\BibitemShut {NoStop}%
\bibitem [{\citenamefont {Yang}\ \emph {et~al.}(2014)\citenamefont {Yang},
  \citenamefont {Kravchenko}, \citenamefont {Briggs},\ and\ \citenamefont
  {Valentine}}]{dielectricmeta}%
  \BibitemOpen
  \bibfield  {author} {\bibinfo {author} {\bibfnamefont {Yuanmu}\ \bibnamefont
  {Yang}}, \bibinfo {author} {\bibfnamefont {Ivan~I.}\ \bibnamefont
  {Kravchenko}}, \bibinfo {author} {\bibfnamefont {Dayrl~P.}\ \bibnamefont
  {Briggs}}, \ and\ \bibinfo {author} {\bibfnamefont {Jason}\ \bibnamefont
  {Valentine}},\ }\bibfield  {title} {\enquote {\bibinfo {title}
  {All-dielectric metasurface analogue of electromagnetically induced
  transparency},}\ }\href {http://dx.doi.org/10.1038/ncomms6753} {\bibfield
  {journal} {\bibinfo  {journal} {Nature Communications}\ }\textbf {\bibinfo
  {volume} {5}},\ \bibinfo {pages} {5753 EP --} (\bibinfo {year}
  {2014})}\BibitemShut {NoStop}%
\bibitem [{\citenamefont {Jackson}(1999)}]{Jackson}%
  \BibitemOpen
  \bibfield  {author} {\bibinfo {author} {\bibfnamefont {John~David}\
  \bibnamefont {Jackson}},\ }\href@noop {} {\emph {\bibinfo {title} {Classical
  Electrodynamics}}},\ \bibinfo {edition} {3rd}\ ed.\ (\bibinfo  {publisher}
  {Wiley, New York},\ \bibinfo {year} {1999})\BibitemShut {NoStop}%
\bibitem [{\citenamefont {Zangwill}(2013)}]{Zangwill}%
  \BibitemOpen
  \bibfield  {author} {\bibinfo {author} {\bibfnamefont {Andrew}\ \bibnamefont
  {Zangwill}},\ }\href@noop {} {\emph {\bibinfo {title} {Modern
  Electrodynamics}}}\ (\bibinfo  {publisher} {Cambridge University Press},\
  \bibinfo {address} {Cambridge},\ \bibinfo {year} {2013})\BibitemShut
  {NoStop}%
\bibitem [{\citenamefont {Wiersma}\ \emph {et~al.}(1995)\citenamefont
  {Wiersma}, \citenamefont {van Albada}, \citenamefont {van Tiggelen},\ and\
  \citenamefont {Lagendijk}}]{Wiersma_recurrent}%
  \BibitemOpen
  \bibfield  {author} {\bibinfo {author} {\bibfnamefont {Diederik~S.}\
  \bibnamefont {Wiersma}}, \bibinfo {author} {\bibfnamefont {Meint~P.}\
  \bibnamefont {van Albada}}, \bibinfo {author} {\bibfnamefont {Bart~A.}\
  \bibnamefont {van Tiggelen}}, \ and\ \bibinfo {author} {\bibfnamefont
  {Ad}~\bibnamefont {Lagendijk}},\ }\bibfield  {title} {\enquote {\bibinfo
  {title} {Experimental evidence for recurrent multiple scattering events of
  light in disordered media},}\ }\href {\doibase 10.1103/PhysRevLett.74.4193}
  {\bibfield  {journal} {\bibinfo  {journal} {Phys. Rev. Lett.}\ }\textbf
  {\bibinfo {volume} {74}},\ \bibinfo {pages} {4193--4196} (\bibinfo {year}
  {1995})}\BibitemShut {NoStop}%
\bibitem [{\citenamefont {Ruostekoski}\ and\ \citenamefont
  {Javanainen}(1997)}]{Ruostekoski1997a}%
  \BibitemOpen
  \bibfield  {author} {\bibinfo {author} {\bibfnamefont {Janne}\ \bibnamefont
  {Ruostekoski}}\ and\ \bibinfo {author} {\bibfnamefont {Juha}\ \bibnamefont
  {Javanainen}},\ }\bibfield  {title} {\enquote {\bibinfo {title} {Quantum
  field theory of cooperative atom response: Low light intensity},}\
  }\href@noop {} {\bibfield  {journal} {\bibinfo  {journal} {Phys. Rev. A}\
  }\textbf {\bibinfo {volume} {55}},\ \bibinfo {pages} {513--526} (\bibinfo
  {year} {1997})}\BibitemShut {NoStop}%
\bibitem [{\citenamefont {Morice}\ \emph {et~al.}(1995)\citenamefont {Morice},
  \citenamefont {Castin},\ and\ \citenamefont {Dalibard}}]{Morice1995a}%
  \BibitemOpen
  \bibfield  {author} {\bibinfo {author} {\bibfnamefont {O.}~\bibnamefont
  {Morice}}, \bibinfo {author} {\bibfnamefont {Y.}~\bibnamefont {Castin}}, \
  and\ \bibinfo {author} {\bibfnamefont {J.}~\bibnamefont {Dalibard}},\
  }\bibfield  {title} {\enquote {\bibinfo {title} {Refractive index of a dilute
  bose gas},}\ }\href@noop {} {\bibfield  {journal} {\bibinfo  {journal} {Phys.
  Rev. A}\ }\textbf {\bibinfo {volume} {51}},\ \bibinfo {pages} {3896--3901}
  (\bibinfo {year} {1995})}\BibitemShut {NoStop}%
\bibitem [{\citenamefont {Javanainen}\ \emph {et~al.}(2014)\citenamefont
  {Javanainen}, \citenamefont {Ruostekoski}, \citenamefont {Li},\ and\
  \citenamefont {Yoo}}]{Javanainen2014a}%
  \BibitemOpen
  \bibfield  {author} {\bibinfo {author} {\bibfnamefont {Juha}\ \bibnamefont
  {Javanainen}}, \bibinfo {author} {\bibfnamefont {Janne}\ \bibnamefont
  {Ruostekoski}}, \bibinfo {author} {\bibfnamefont {Yi}~\bibnamefont {Li}}, \
  and\ \bibinfo {author} {\bibfnamefont {Sung-Mi}\ \bibnamefont {Yoo}},\
  }\bibfield  {title} {\enquote {\bibinfo {title} {Shifts of a resonance line
  in a dense atomic sample},}\ }\href {\doibase 10.1103/PhysRevLett.112.113603}
  {\bibfield  {journal} {\bibinfo  {journal} {Phys. Rev. Lett.}\ }\textbf
  {\bibinfo {volume} {112}},\ \bibinfo {pages} {113603} (\bibinfo {year}
  {2014})}\BibitemShut {NoStop}%
\bibitem [{\citenamefont {Javanainen}\ and\ \citenamefont
  {Ruostekoski}(2016)}]{JavanainenMFT}%
  \BibitemOpen
  \bibfield  {author} {\bibinfo {author} {\bibfnamefont {Juha}\ \bibnamefont
  {Javanainen}}\ and\ \bibinfo {author} {\bibfnamefont {Janne}\ \bibnamefont
  {Ruostekoski}},\ }\bibfield  {title} {\enquote {\bibinfo {title} {Light
  propagation beyond the mean-field theory of standard optics},}\ }\href
  {\doibase 10.1364/OE.24.000993} {\bibfield  {journal} {\bibinfo  {journal}
  {Opt. Express}\ }\textbf {\bibinfo {volume} {24}},\ \bibinfo {pages}
  {993--1001} (\bibinfo {year} {2016})}\BibitemShut {NoStop}%
\end{thebibliography}
\end{document}